\documentclass[twocolumn]{aastex63}
\usepackage{amsmath}
\usepackage{graphics}
\usepackage{epstopdf}
\usepackage{enumerate}

\usepackage{savesym}
\savesymbol{tablenum}
\usepackage{siunitx}
\restoresymbol{SIX}{tablenum}

\usepackage{hyperref}

\newcommand{\del}{\partial}

\newcommand{\addsf}[1]{{\color{black} #1}} 

\received{}
\revised{\today}
\accepted{}

\shorttitle{Mass ejeciton and nucleosynthesis in NS-NS mergers leaving hypermassive NSs}

\shortauthors{Fujibayashi et al.}

\begin{document}

\title{Comprehensive study of mass ejection and nucleosynthesis in binary neutron star mergers leaving short-lived massive neutron stars}

\correspondingauthor{Sho Fujibayashi}
\email{sho.fujibayashi@aei.mpg.de}

\author[0000-0001-6467-4969]{Sho Fujibayashi}
\affiliation{Max-Planck-Institut f\"ur Gravitationsphysik (Albert-Einstein-Institut), Am M\"uhlenberg 1, D-14476 Potsdam-Golm, Germany}

\author[0000-0003-4988-1438]{Kenta Kiuchi}
\affiliation{Max-Planck-Institut f\"ur Gravitationsphysik (Albert-Einstein-Institut), Am M\"uhlenberg 1, D-14476 Potsdam-Golm, Germany}
\affiliation{Center for Gravitational Physics and Quantum Information, Yukawa Institute for Theoretical Physics, Kyoto University, Kyoto, 606-8502, Japan}

\author[0000-0002-4759-7794]{Shinya Wanajo}
\affiliation{Max-Planck-Institut f\"ur Gravitationsphysik (Albert-Einstein-Institut), Am M\"uhlenberg 1, D-14476 Potsdam-Golm, Germany}
\affiliation{Interdisciplinary Theoretical and Mathematical Science (iTHEMS) Research Group, RIKEN, Wako, Saitama, 351-0198, Japan}

\author[0000-0003-3179-5216]{Koutarou Kyutoku}
\affiliation{Department of Physics, Kyoto University, Kyoto, 606-8502, Japan}
\affiliation{Center for Gravitational Physics and Quantum Information, Yukawa Institute for Theoretical Physics, Kyoto University, Kyoto, 606-8502, Japan}
\affiliation{Interdisciplinary Theoretical and Mathematical Science (iTHEMS) Research Group, RIKEN, Wako, Saitama, 351-0198, Japan}

\author[0000-0002-2648-3835]{Yuichiro Sekiguchi}
\affiliation{Center for Gravitational Physics and Quantum Information, Yukawa Institute for Theoretical Physics, Kyoto University, Kyoto, 606-8502, Japan}
\affiliation{Department of Physics, Toho University, Funabashi, Chiba 274-8510, Japan}

\author[0000-0002-4979-5671]{Masaru Shibata}
\affiliation{Max-Planck-Institut f\"ur Gravitationsphysik (Albert-Einstein-Institut), Am M\"uhlenberg 1, D-14476 Potsdam-Golm, Germany}
\affiliation{Center for Gravitational Physics and Quantum Information, Yukawa Institute for Theoretical Physics, Kyoto University, Kyoto, 606-8502, Japan}

\begin{abstract}
By performing general relativistic hydrodynamics simulations with an approximate neutrino-radiation transfer, the properties of ejecta in dynamical and post-merger phases are investigated for the cases in which the remnant massive neutron star collapses into a black hole in $\lesssim \SI{20}{ms}$ after the onset of the merger.
The dynamical mass ejection is investigated in three-dimensional simulations.
The post-merger mass ejection is investigated in two-dimensional axisymmetric simulations with viscosity using the three-dimensional post-merger systems as the initial conditions.
We show that the typical neutron-richness of the dynamical ejecta is higher for the merger of more asymmetric binaries; hence, heavier $r$-process nuclei are dominantly synthesized.
The post-merger ejecta are shown to have only a mild neutron-richness, which results in the production of lighter $r$-process nuclei, irrespective of binary mass ratios.
Because of the larger disk mass, the post-merger ejecta mass is larger for more asymmetric binary mergers.
Thus, the post-merger ejecta can compensate for the underproduced lighter $r$-process nuclei for asymmetric merger cases.
As a result, by summing up both ejecta components, the solar residual $r$-process pattern is reproduced within the average deviation of a factor of three, irrespective of the binary mass ratio.
Our result also indicates that the (about a factor of a few) light-to-heavy abundance scatter observed in $r$-process-enhanced stars can be attributed to variation in the binary mass ratio and total mass.
Implications of our results associated with the mass distribution of compact neutron star binaries and the magnetar scenario of short gamma-ray bursts are discussed.
\end{abstract} 


\section{Introduction}
\label{sec:intro}
Binary neutron star mergers expel a fraction of neutron star matter into space, which drives a variety of electromagnetic counterparts of gravitational-wave observation.
The electromagnetic counterparts 
associated with the mass ejection of the merger have rich information on the activities of the system during and after the merger~(e.g., \citealt{Abbott2017oct1,Abbott2017oct2}).
Therefore the theoretical studies on the mass ejection and resulting electromagnetic counterparts have become one of the most critical topics in the era of multi-messenger astronomy composed of gravitational-wave and electromagnetic observations.

A fraction of neutron-star matter can be ejected dynamically by the tidal interaction and shock heating in the violent merger phase with a timescale of $\lesssim \SI{10}{ms}$.
Pioneering simulations for binary neutron star mergers with an approximate neutrino treatment in Newtonian gravity~\citep{Rosswog1999jan,korobkin2012nov} as well as those in an approximate framework of general-relativistic gravity but without weak interaction~\citep{Goriely2011sep,Bauswein2013a} have shown the ejection of very neutron-rich matter (in terms of the electron fraction, $Y_\mathrm{e}<0.1$).
However, in this decade, both including the general relativistic gravity, which results in a more violent merger process \addsf{due to the effectively stronger gravitational potential than that of Newtonian case (\citealt{Takahara1984nov}, \citealt{VanRiper1988mar}, and \citealt{Bruenn2001oct} in the context of stellar core collapse)} and hence in higher matter temperature, and taking weak interaction processes into account have significantly altered our understanding on the neutron-richness of the ejecta~\citep{Sekiguchi2015a,Palenzuela2015a,Sekiguchi2016a,Foucart2016b,Radice2018dec,Kullmann2022feb}.
These studies have shown that the dynamical ejecta have a broad distribution of neutron richness ($Y_\mathrm{e}\approx 0.05$--0.4), and as a result, a wide variety of heavy nuclei are now considered to be synthesized via the rapid neutron capture ($r$-) process~\citep{Wanajo2014a,Goriely2015a,Radice2018dec,Kullmann2022feb}.

The increasing number of the observations of metal-poor stars gives us an interesting constraint on the astrophysical site of $r$-process.
A number of metal-poor stars with enhanced $r$-process elements (more than a three times higher value of the Eu/Fe ratio than that of the solar ratio) have elemental abundances very similar to that of the solar-system $r$-process residuals~(hereafter solar $r$-residuals; e.g., \citealt{Cowan2021jan} and references therein).
\addsf{The agreement is particularly excellent for heavy $r$-process elements with $Z \ge 56$ (Ba and heavier except for actinides Th and U), although a factor of a  few variations for lighter elements with $Z = 38$--48 (Sr to Cd) are observed \citep[e.g.,][]{Holmbeck2020aug}.}
Since the elemental abundances of such \addsf{$r$-process-enhanced} stars are expected to preserve those of single nucleosynthetic events \addsf{\citep[e.g.,][]{Wanajo2021aug,Hirai2022jun}}, this observational fact indicates that each $r$-process event has to provide the abundance pattern similar to that of the solar $r$-residuals \addsf{within a factor of a few deviation (except for actinides)}.

The post-merger system is generally composed of a disk surrounding a remnant compact object, either a massive neutron star or a black hole, unless a black hole is promptly formed from the merger of nearly equal-mass systems~(e.g., \citealt{Hotokezaka2013a}). 
Then, a post-merger mass ejection can be triggered from the remnant disk by magnetohydrodynamical processes.
The magnetic field in the disk is amplified by magnetic winding and magnetorotational instability. This produces a turbulent state with magnetically induced effective viscosity, which governs the evolution of the disk and drives the mass ejection with a timescale of $\gtrsim\SI{1}{s}$~\citep{Fernandez2013a,Metzger2014a,Just2015a,Siegel2018a,Fernandez2019a,Fujibayashi2020a,Fujibayashi2020b,Just2022}.

Our previous work with viscous radiation-hydrodynamics simulation in numerical relativity~\citep{fujibayashi2018a,Fujibayashi2020c} showed that in the presence of a long-lived massive neutron star as a remnant of the merger, the mass of the post-merger ejecta can be $\sim 30$--40\% of that of the disk, which amounts to $\gtrsim0.1M_\odot$ and can be much larger than that of the dynamical ejecta.
In addition, the post-merger ejecta are likely to mainly synthesize the light species of $r$-process nuclei reflecting its moderate neutron-richness ($Y_\mathrm{e}\sim0.3$) that results from the weak interaction in the slowly expanding disk matter during the viscous evolution~(see also \citealt{Just2022}).
The resulting nuclear abundance pattern (from the sum of the dynamical and post-merger ejecta) thus deviates from that of the solar $r$-residuals.
\addsf{If binary neutron star mergers are the main production site of the $r$-process nuclei, our findings indicate that the mergers leaving long-lived remnant neutron stars should be a minority}.

In addition to the turbulence, a global magnetic field could be developed associated with the mass outflow from the disk.
Such magnetic fields also work for the mass ejection from the disk through magnetocentrifugal force and magnetic pressure~\citep[e.g.,][]{Blandford1982jun}.
If a dynamo action works in the remnant system and a massive neutron star survives for the timescale longer than the timescale of magnetic-field amplification inside it, the mass ejection from the post-merger system can be more violent because of the presence of the global magnetic field anchored on the neutron star surface~\citep{Shibata2021b}.
By this effect, the kinetic energy of the ejecta may be enhanced, and as a result, a very bright radio emission may be induced in a timescale of years by its interaction with an interstellar medium. However, such a radio transient has not been observed in untargeted surveys or follow-up observations of short gamma-ray bursts~\citep{Horesh2016mar}.
This also indicates that a binary neutron star merger resulting in a long-lived massive neutron star as its remnant may not be a frequent event~\citep{Kawaguchi2022feb}.

As summarized above, it is expected that the formation of hypermassive neutron stars that leads to a black hole in a short timescale after the merger (see, e.g., \citealt{Baumgarte2000jan} and \citealt{Shibata2016a} for the definition of the hypermassive neutron star) is likely to be the major channel of neutron star mergers.
In this channel, the system composed of a spinning black hole with a disk is a canonical remnant and thus the theoretical study for the mass ejection from the black hole-disk system is one of the most important subjects associated with the multi-messenger astronomy. 
In the last decade, the mass ejection from the disk surrounding a black hole has been investigated in a number of studies using the equilibrium configurations of disks with a certain fixed mass as initial conditions~\citep{Fernandez2013a,Metzger2014a,Just2015a,Lippuner2017a,Siegel2018a,Fernandez2019a,Miller2019a,Fernandez2020a,Christie2019dec,Fujibayashi2020a,Fujibayashi2020b,Just2022,Fahlman2022apr}.
These studies have provided us with an important qualitative picture for the post-merger mass ejection.
However, because the properties of the disk and the resulting post-merger ejecta depend quantitatively on the properties of the merger remnant, the post-merger mass ejection has to be investigated self-consistently throughout the merger and post-merger phases. Also, it is important to follow the post-merger evolution for a long timescale of $\sim 10$\,s, because the post-merger mass ejection proceeds over such a long timescale.

\addsf{
Several magnetohydrodynamics simulations for post-merger systems have been performed using the results of 3D simulations for binary neutron star merger as their initial conditions~\citep{Moesta2020oct}, or solving magnetohydrodynamics equations in the entire simulation~\citep{Combi2022}.
Although these simulations are self-consistent in terms of modeling the angular momentum transport as well as the hydrodynamical profile of a post-merger system, it is computationally very expensive to perform such simulations for several seconds.
As a result, their simulation times are limited to several tens of milliseconds, which are much shorter than post-merger mass ejection timescales.
}

In this work, we investigate binary neutron star mergers in which the hypermassive neutron stars with the lifetime of 3--20\,ms (before collapsing into black holes) are formed.
We first perform three-dimensional (3D) neutrino-radiation hydrodynamics simulations for the merger phase until the merger remnant settles into a hypermassive neutron star and a disk in a quasi-steady state.
We then perform  long-term (10-second long) axisymmetric (two-dimensional; 2D) simulations for the post-merger phase throughout the black hole formation, the disk evolution, and the post-merger mass ejection employing the angle-averaged data of 3D merger remnants as initial conditions.
Our aim with this setting is to self-consistently obtain the properties of the dynamical and post-merger ejecta.
 
During the formation of a black hole, a fraction of the matter surrounding the pre-collapse hypermassive neutron star falls into the black hole, leading to a decrease of the disk mass.
In addition, a large fraction of the disk matter falls into the black hole during the viscous evolution of the disk; this effect is suppressed in the presence of remnant neutron stars.
As a consequence, the mass of the post-merger ejecta from black hole-disk systems tends to be smaller than that in the presence of a remnant massive neutron star~\citep[for studies on the effect of the presence of a massive neutron star with simplified setups, see][]{Metzger2014a,Lippuner2017a}.
This implies that the nucleosynthetic outcomes could be very different between the cases with the formation of hypermassive and long-lived  neutron stars.

Motivated by this consideration, we explore the $r$-process nucleosynthesis for the case that hypermassive neutron stars are formed.
In particular, we investigate the dependence of the properties of ejecta on the mass ratio of binary neutron stars.
Specifically, we employ the binary with the total mass of $M_\mathrm{tot}=2.7M_\odot$ and vary the mass ratio (defined as the ratio of lower mass to higher mass $M_2/M_1$) from 0.8 (1.2-1.5$M_\odot$) to 1.0 (1.35-1.35$M_\odot$).
We also consider a higher-mass case of each mass $1.25$ and $1.55M_\odot$ ($M_2/M_1 = 0.81$) with $M_\mathrm{ tot}=2.8M_\odot$.
This range approximately covers the mass ratio of compact binary neutron stars found in our Galaxy~\citep[although J1913+1102, whose constituent masses are approximately 1.62$M_\odot$ and 1.27$M_\odot$, has a slightly smaller mass ratio, 0.78;][]{Ferdman2020} and inferred total mass of the merged binary neutron star in GW170817~\citep{abbott2017a}.

This article is organized as follows.
In \S~\ref{sec:method}, our methods of 3D and 2D simulations are described briefly.
Then, the result of our hydrodynamics simulations and nucleosynthesis calculations are presented in \S~\ref{sec:result}.
Compiling the previous result~\citep{Fujibayashi2020c} with the present one, we summarize the relation of the lifetime of the remnant massive neutron star and the corresponding nuclear abundance pattern in \S~\ref{sec:discussion}.
\S~\ref{sec:summary} is devoted to the summary.
Throughout this paper, $G$, $c$, and $k_\mathrm{B}$ denote the gravitational constant, speed of light, and Boltzmann's constant, respectively.

\section{Numerical Method} \label{sec:method}
\subsection{Numerical procedure}
As in our latest work~\citep{Fujibayashi2020c}, we first perform 3D general relativistic neutrino-radiation hydrodynamics simulations for the mergers of binary neutron stars.
We continue the simulations until \addsf{the remnant hypermassive neutron star becomes gravitationally unstable.}
Then, we map the hydrodynamical profile of the remnant onto the axisymmetric profile by averaging quantities over the azimuthal angle around the rotational axis (defined as the axis passing through the point of the minimum value of the conformal factor).
The axisymmetric profile is then evolved with our 2D general relativistic neutrino-radiation viscous-hydrodynamics code. Specifically, we map the 3D profile to the 2D one at 0.5--\SI{1}{ms} before the black-hole formation.
We check that the black-hole formation time and the remaining disk mass in our 2D simulations are consistent with those in the 3D simulations (see \S~\ref{subsubsec:diskmass}).

We note that the collapse to a black hole in the 3D simulations is induced primarily by the angular momentum transport effect from the central region to the surrounding matter by the gravitational torque due to the presence of a non-axisymmetric structure of the remnant hypermassive neutron star.
In the 2D simulation, this effect is absent but by the viscous effect, the angular momentum transport effect is effectively taken into account (see also \S~\ref{subsubsec:diskmass}).
\addsf{We also note that the mapping from 3D to 2D profiles erases the non-axisymmetric $Y_\mathrm{e}$ profile of dynamical ejecta.
This results in the different mass distributions of $Y_\mathrm{e}$ for the dynamical ejecta between those from the mapped 2D data and the original 3D data.
Thus, we use the tracer particles obtained from the 3D data for the nucleosynthesis calculations of the dynamical ejecta component (see \S~\ref{subsec:tp}).
}

\begin{table*}[t]
\caption{List of the models and the key results for 3D simulations.
$t_\mathrm{BH}$ denotes the post-merger time at the black hole formation and $\chi_\mathrm{BH}$ is the dimensionless spin of the black hole. The results for simulations with $\Delta x_{i_\mathrm{max}}=\SI{150}{m}$ are listed.
The mass and dimensionless spin of the black hole as well as the disk and ejecta masses are measured at $t=t_\mathrm{BH}+30$\,ms.
}
\begin{center}
\begin{tabular}{lcccccccccc}

\hline \hline

Model & $M_1/M_\odot$ & $M_2/M_\odot$ & $M_2/M_1$ & $t_\mathrm{BH}$ (ms) & $M_\mathrm{BH}/M_\odot$ & $\chi_\mathrm{ BH}$ & $M_\mathrm{disk}\, (10^{-2}M_\odot)$ & $M_\mathrm{dyn}\, (10^{-2}M_\odot)$ & $\langle Y_\mathrm{e,dyn}\rangle$
\\\hline
SFHo135-135 &1.35 & 1.35 & 1.00 & 13 & 2.56 & 0.67 & 1.2  & 0.69 & 0.23\\
SFHo130-140 &1.40 & 1.30 & 0.93 & 16 & 2.55 & 0.67 & 3.0  & 0.46 & 0.24\\
SFHo125-145 &1.45 & 1.25 & 0.86 & 17 & 2.54 & 0.66 & 3.6  & 0.54 & 0.16\\
SFHo120-150 &1.50 & 1.20 & 0.80 & 18 & 2.50 & 0.64 & 6.5  & 0.37 & 0.13\\
SFHo125-155 &1.55 & 1.25 & 0.81 & 3  & 2.68 & 0.75 & 4.3  & 0.86 & 0.09\\
\hline
\end{tabular}
\end{center}
\label{tab:model-3d}
\end{table*}

\subsection{Models}
In this work, we focus on the case that the remnant massive neutron star collapses into a black hole in a short timescale in contrast to our previous work \citep{Fujibayashi2020c}, in which we studied the cases that the massive neutron stars survive for a long time (more than seconds). We employ the models with the total gravitational mass of the binaries to be 2.7$M_\odot$ and consider four different mass ratios: $M_2/M_1 = 0.8$--1.0.
Some of the initial conditions are the same as those in~\cite{Sekiguchi2016a,Shibata2017a}, in which the method for computing initial data is described.
To check that our conclusion does not depend on the total mass of the system, we also employ a more massive model for which the total mass is $2.8M_\odot$ (SFHo125-155).
All these 3D models are listed in Table~\ref{tab:model-3d}. 

For both 3D and 2D simulations, we use the same tabulated equation of state (EOS) referred to as SFHo~\citep{steiner2013a} with an extension to lower-density ($\rho<\SI{1.66e3}{g/cm^3}$) and lower-temperature ($k_\mathrm{B}T<\SI{0.1}{MeV}$) domains using the Timmes (Helmholtz) EOS~\citep{timmes2000a}.
The detailed procedure of the extension is described in Appendix~A of \cite{Hayashi2021}.
For the readers who are interested in our handling for the heating by the nuclear burning in the assumption of the nuclear statistical equilibrium (NSE), we also refer to Appendix~B of \cite{Hayashi2021}.

\subsection{3D simulation}
The 3D simulations are performed with the latest version of our radiation-hydrodynamics code in numerical relativity~\citep{Sekiguchi2010a,Sekiguchi2015a,Sekiguchi2016a,Fujibayashi2020c}, 
which solves Einstein's equation with a version of puncture-Baumgarte-Shapiro-Shibata-Nakamura (BSSN) formalism with a Z4c constraint violation propagation scheme~\citep{shibata1995a,baumgarte1999,Baker:2005vv,Campanelli:2005dd,marronetti2008,Hilditch2013a}. 
The Riemann solver is updated to the one referred to as HLLC \citep[for relativistic hydrodynamics; ][]{Mignone2005nov}, the details of which is described in \cite{Kiuchi2022}.
The neutrino radiation transport is treated with a version of the leakage scheme incorporating a moment-based transport scheme.
In this scheme, neutrinos are separated into two components; ``trapped" and ``streaming" neutrinos.
The trapped neutrinos are assumed to be thermalized with the same local temperature as the fluid and considered as a part of the fluid, i.e., they comove with the fluid.
The trapped neutrinos leak out from the fluid in a rate depending on their diffusion timescale and become a streaming component.
The streaming neutrinos are solved with a truncated-moment formalism with the so-called M1-closure for estimating higher moments (see \citealt{Thorne1981a,shibata2011a}).
The absorption of streaming neutrinos is considered in an approximate manner, while the pair-annihilation of them is not taken into account in this study.
The detail of the treatment for our approximate neutrino transfer is described in \cite{Sekiguchi2011}, \cite{Sekiguchi2012}, and \cite{fujibayashi2017a}.

The simulations are performed using a fixed-mesh-refinement (FMR) algorithm assuming the plane symmetry with respect to the $z=0$ plane.
Each refinement level has the same half-cubic box region with the uniform grid spacing and the $i$-th level has a grid spacing of $\Delta x_i = 2\Delta x_{i+1}$ ($i=$1, 2, \dots, $i_\mathrm{max}-1$) with $\Delta x_{i_\mathrm{ max}}$ the input parameter.
The $i$-th level has the computational domain of $[-L_i:L_i]\times[-L_i:L_i]\times[0:L_i]$ with $L_i=N\Delta x_i$.
We set $i_\mathrm{max}=13$, $\Delta x_{i_\mathrm{max}}=\SI{150}{m}$, and $N=258$ for our fiducial model, with which $L_1 \approx \SI{1.6e5}{km}$.
Several simulations with the different grid resolution of $\Delta x_{i_\mathrm{max}}=100,$ 200, and \SI{250}{m} are performed to check the dependence of the results on the grid resolution.
For the resolutions of $\Delta x_{i_\mathrm{max}}=100,$ 200, and \SI{250}{m}, we set $i_\mathrm{max}=13$ and $N=369$, 193, and 137, with which $L_1\approx\SI{1.5e5}{},$ \SI{1.6e5}{}, and \SI{1.4e5}{km}, respectively.

\begin{table*}[t]
\caption{Setup and several key results of 2D simulations (see the text for details).
The post-merger ejecta mass is defined by the sum of the mass of tracer particles that experience the temperature larger than \SI{10}{GK}.
}
\begin{center}
\begin{tabular}{lccccccc}

\hline \hline

Model & $\Delta x_0$ (m) & $N$ & $L$ (km) & $\alpha_\mathrm{vis} H_\mathrm{tur}$ (m) & $M_\mathrm{post}\,(10^{-2}M_\odot)$& $M_\mathrm{post}/M_\mathrm{dyn}$\\
\hline
SFHo135-135    & 70 $\rightarrow$ 200 & 937 $\rightarrow$ 689 & 9237 $\rightarrow$ 8908 & 400 & 0.22 & 0.32 \\
SFHo130-140    & 70 $\rightarrow$ 200 & 937 $\rightarrow$ 689 & 9237 $\rightarrow$ 8908 & 400 & 0.53 & 1.19\\
SFHo125-145    & 70 $\rightarrow$ 200 & 937 $\rightarrow$ 689 & 9237 $\rightarrow$ 8908 & 400 & 0.69 & 1.26\\
SFHo120-150    & 70 $\rightarrow$ 200 & 937 $\rightarrow$ 689 & 9237 $\rightarrow$ 8908 & 400 & 1.33 & 3.58\\
SFHo125-155    & 70 $\rightarrow$ 200 & 937 $\rightarrow$ 689 & 9237 $\rightarrow$ 8908 & 400 & 0.83 & 0.99\\
\hline
SFHo120-150-lr & 70 $\rightarrow$ 300 & 937 $\rightarrow$ 625 & 9237 $\rightarrow$ 9055 & 400 & -- & --\\
\hline
\end{tabular}
\end{center}
\label{tab:model}
\end{table*}

Because high-resolution shock-capturing schemes cannot treat the vacuum state, we need to set a spurious but tenuous atmosphere outside the neutron stars.
In these simulations, we set the constant atmosphere density profile to be $\rho_\mathrm{atm} = 10^3\,\mathrm{g/cm^3}$ for $r\le L_{13}$.
We also assume the power-law profile of the atmosphere density $\rho_\mathrm{atm} \propto 1/r^3$ for $r>L_{13}$.
The floor value of the atmosphere density is determined to be $\approx 0.166\,\mathrm{g/cm^3}$.
The atmosphere temperature is set to be $10^{-3}\,\mathrm{MeV}/k_\mathrm{B}$.

We also employ the \textit{reflux} prescription at the FMR boundary to ensure the baryon mass conservation. With the help of it, the violation of the baryon mass conservation is kept to be low, $\alt 10^{-7}M_\odot$, in all the simulations.
This enables us to investigate the fast-moving component of the ejecta, the mass of which is very small (see \S~\ref{subsec:3d}).

Simulations for the fiducial models are performed until the bulk of the ejecta reaches $\sim \SI{5000}{km}$ and cools down sufficiently (less than $\SI{1}{GK}=10^9\si{K}$) for the post-process nucleosynthesis calculations.
It takes $\sim 40$--\SI{50}{ms} after merger until such a state is achieved. Here, we note that the 3D simulations are performed even after the formation of the black hole.
However, the mapping to the 2D simulations is done just prior to the black-hole formation, as we already mentioned.

\subsection{2D simulation}
Following our previous work~\citep{Fujibayashi2020c}, a general relativistic neutrino-radiation viscous-hydrodynamics code is employed for the present simulations.
As in the 3D simulations, Einstein's equation is solved with a version of the puncture-BSSN formalism with a Z4c scheme.
To impose the axial symmetry, a cartoon method is used~\citep{Alcubierre2001a,Shibata2000a}.
The scheme for the neutrino transport is also the same as in the 3D simulations.
The shear viscous effect is taken into account using a simplified version of the Israel-Stuart formalism~\citep{Israel1979a} as described in~\cite{shibata2017b}.
The kinematic viscous parameter is modeled as $\nu_\mathrm{vis}=\alpha_\mathrm{vis}c_\mathrm{s}H_\mathrm{tur}$, where $c_\mathrm{s}$ is the sound speed and $H_\mathrm{tur}$ is the length-scale of the turbulence generated hypothetically as a result of magnetohydrodynamical instabilities.
As in our previous study~\citep{Fujibayashi2020c}, we set  $H_\mathrm{tur}=\SI{10}{km}$ and $\alpha_\mathrm{vis}=0.04$, which actually results in a constant viscous-length scale of $\alpha_\mathrm{vis} H_\mathrm{tur}=\SI{400}{m}$ (see Table~\ref{tab:model}).

The grid structure is the same as in the 2D simulations recently performed with the same code~\citep{Fujibayashi2020c}, in which the cylindrical coordinates $(R,z)$ are employed.
In the inner cylindrical region of $R < \SI{15}{km}$ and $z < \SI{15}{km}$, a uniform grid with the grid spacing of $\Delta x_0$ is set, while in the outer region, a non-uniform grid is set with an increase rate of the grid spacing of 1.01.
The grid number $N$ and the location of the outer boundaries along each axis (denoted by $L$) are listed in Table~\ref{tab:model}.
We assume the plane symmetry with respect to the $z=0$ plane.

We employ the snapshots of the fiducial 3D models with $\Delta x_{i_\mathrm{max}}=\SI{150}{m}$ for preparing the initial conditions of the 2D simulations.
After mapping the 3D data to the 2D one, we first perform high-resolution simulations with the innermost grid spacing of $\Delta x_0 = \SI{70}{m}$.
Such a high resolution is particularly important to simulate the formation and evolution of the black hole in a good accuracy.
In our previous work for a spinning black hole surrounded by a disk~\citep{Fujibayashi2020a}, we have found that, to accurately follow the evolution of the black hole with the dimensionless spin of $\sim 0.8$ for 1\,s, the finest grid spacing should be $ \Delta x_0 \lesssim 0.018 GM_{\rm BH}/c^2 \approx \SI{69}{m} (M_{\rm BH}/2.6M_\odot)$, where $M_{\rm BH}$ is the mass of the black hole.

After the black-hole formation, the self-gravity of the disk is minor compared to the gravitation exerted by the black hole.
In particular, after the viscous evolution of the disk, the mass of the disk becomes less than 1\% of the black-hole mass, and in such a stage, the self-gravitational effect of the disk can be safely ignored.
Thus, we stop the time evolution of the gravitational field for the phase in which the rest mass outside the apparent horizon is below 1\% of the mass of the black hole.
After stopping the time evolution of the gravitational field, the black-hole evolution does not have to be followed with a high resolution any longer.
Thus, we carry out a regridding at the same time.
In the regridding process, all the geometrical and radiation viscous-hydrodynamics quantities are mapped onto a coarser grid that has $\Delta x_0=\SI{200}{m}$ with the same increase rate of grid spacing, 1.01, and a uniform grid spacing region, $R \le \SI{15}{km}$ and $z \le \SI{15}{km}$.
We do not observe any artificial behaviors in the subsequent time evolution due to the regridding process.

Before the regridding, the grid number and the size of the computational domain are set to $N=937$ and $L=\SI{9237}{km}$, respectively.
After the regridding, they are changed to $N=689$ and $L=\SI{8908}{km}$, respectively, for models with the fiducial resolution, and $N=625$ and $L=\SI{9055}{km}$ for the lower resolution model 120-150-lr. 

\subsection{Tracer-particle method}
\label{subsec:tp}
To derive the $Y_\mathrm{e}$ distribution (and some other quantities) of the ejecta and to perform nucleosynthesis calculations, we apply our post-process tracer-particle method for the results of 3D and 2D simulations as described in~\cite{Fujibayashi2020c}.
In particle tracing for 3D simulations, the unbound matter is detected in the last snapshot of each simulation using the so-called geometrical criterion $u_t<-1$, where $u_t$ is the time component of the four-velocity of the fluid.
Several thousands of tracer particles are distributed in the region for which the ejecta criterion is satisfied and each particle is assumed to be the representative of neighboring unbound matter.
Then, the tracer particles are evolved backward in time to obtain the time evolution of the thermodynamical quantities along the trajectory of each particle.

In 2D particle tracing, on the other hand, the tracer particles are distributed on 32 points with polar angles in the range of $\theta=[0:\pi/2]$ along the arc with the radius of $r_\mathrm{ext}=\SI{8000}{km}$.
The procedure is repeated every time step of $\Delta t_\mathrm{set}$, which is controlled adaptively based on the average velocity of the fluid at $r_\mathrm{ext}$ so that the particles are distributed uniformly in space.
The mass of each particle is set based on the mass flux of its initial position as $\Delta m={r_\mathrm{ext}}^2\Delta \Omega \rho u^r\sqrt{-g}\Delta t_\mathrm{set}$, where $\Delta \Omega$ is the solid angle element, $g$ is the determinant of the metric, $\rho$ is the rest-mass density, and $u^k$ is the spatial component of the four-velocity of the fluid.
With this definition, the total mass of the particles agrees with the ejecta mass defined in \S~\ref{subsubsec:post-merger} (see Eqs.~\eqref{eq:mej_esc} and \eqref{eq:mdot-baryon-mass}) within the accuracy of 5 percent.
The tracer particles are then evolved backward in time as in the 3D case.

We note that, although the 2D simulations contain the dynamical ejecta at the beginning of the simulation inside their computational domain, the electron fraction distributions of the dynamical ejecta obtained in the 2D simulations are different from the corresponding 3D ones. The reason for this is that the asymmetric structure of $Y_\mathrm{e}$ distribution around the $z$-axis in the 3D simulations is erased after the averaging and mapping onto the 2D computational domain.
Thus, to obtain a self-consistent $Y_\mathrm{e}$ distribution, we replace the tracer particles of the dynamical ejecta component in the 2D simulations with those in the corresponding 3D simulations.
Here, the tracer particles of the dynamical ejecta in the 2D simulations are defined as those having the maximum temperature along its trajectory less than \SI{10}{GK}, i.e., the component already located far from the central region at the beginning of the 2D simulation.

\section{Result} \label{sec:result}

\begin{figure*}
\epsscale{1.17}
\includegraphics[width=0.49\textwidth]{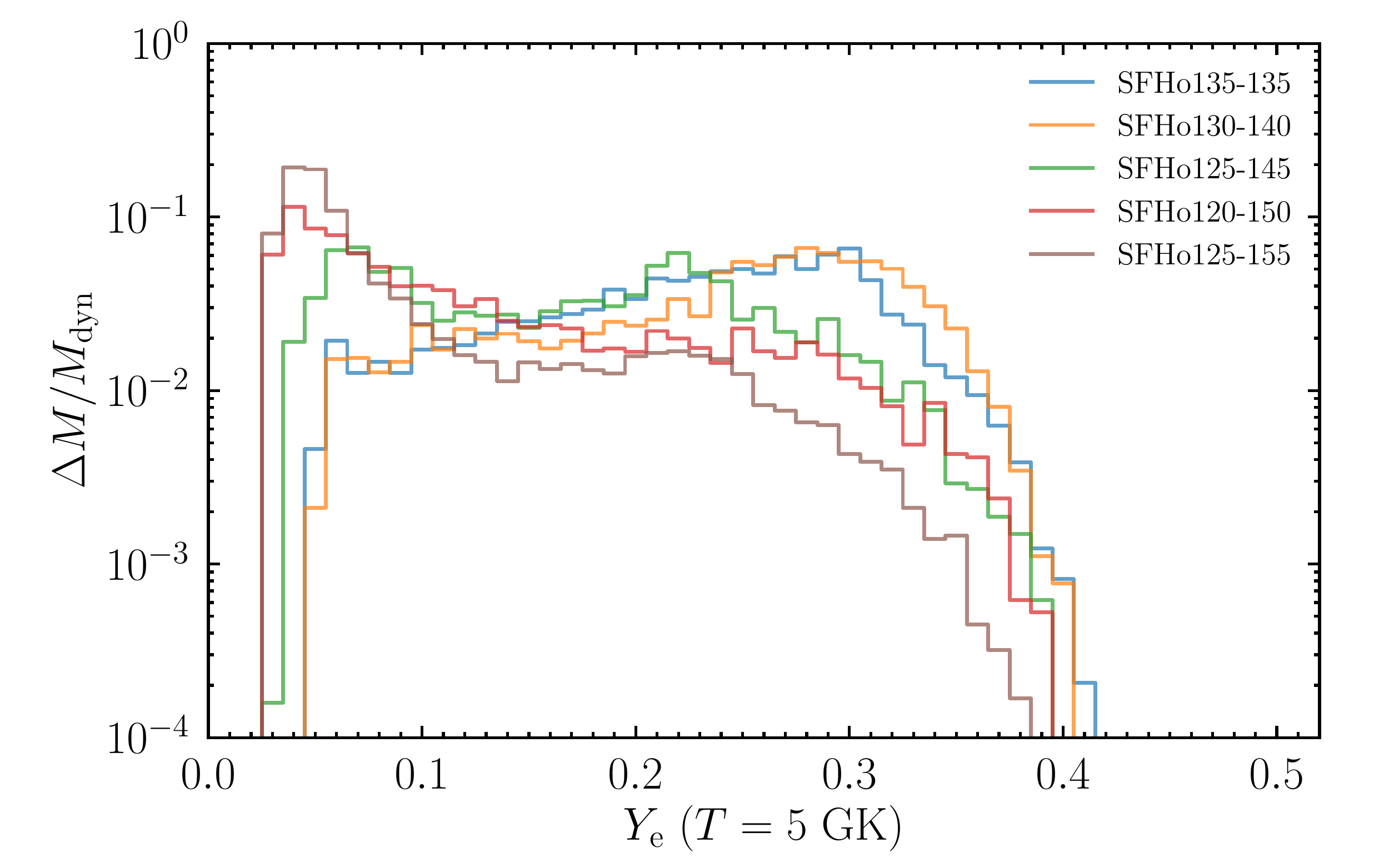}
\includegraphics[width=0.49\textwidth]{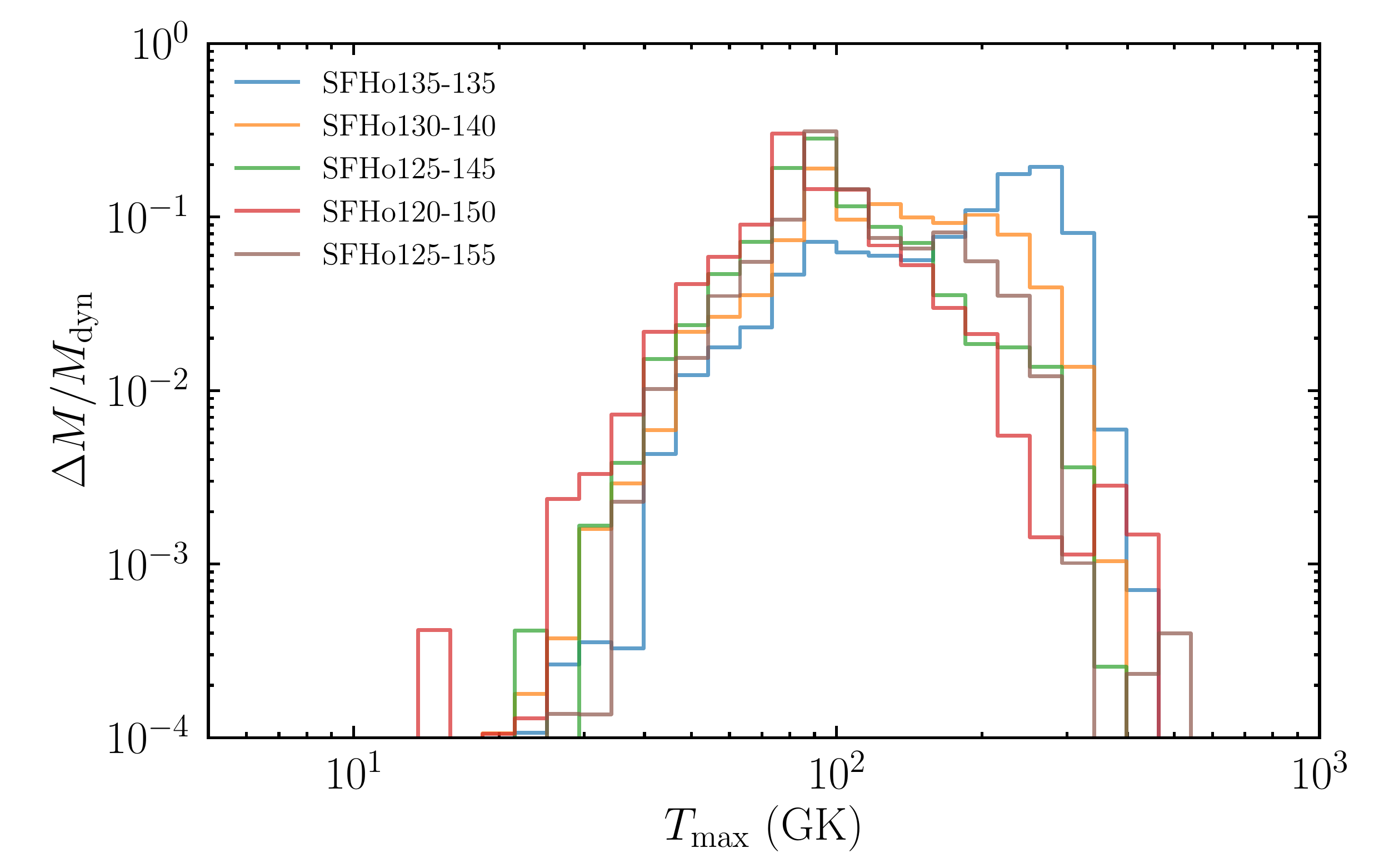}\\
\includegraphics[width=0.49\textwidth]{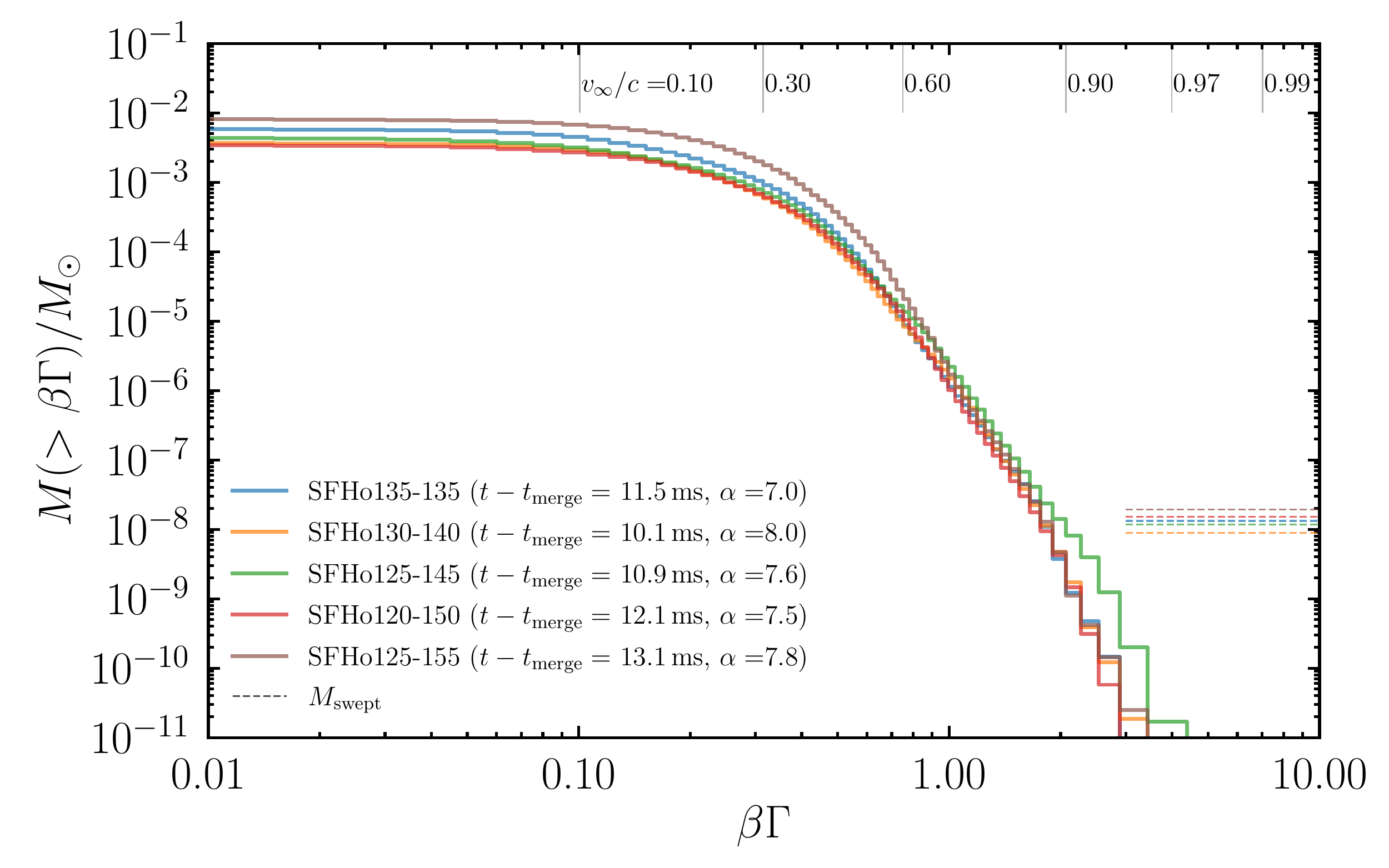}
\includegraphics[width=0.49\textwidth]{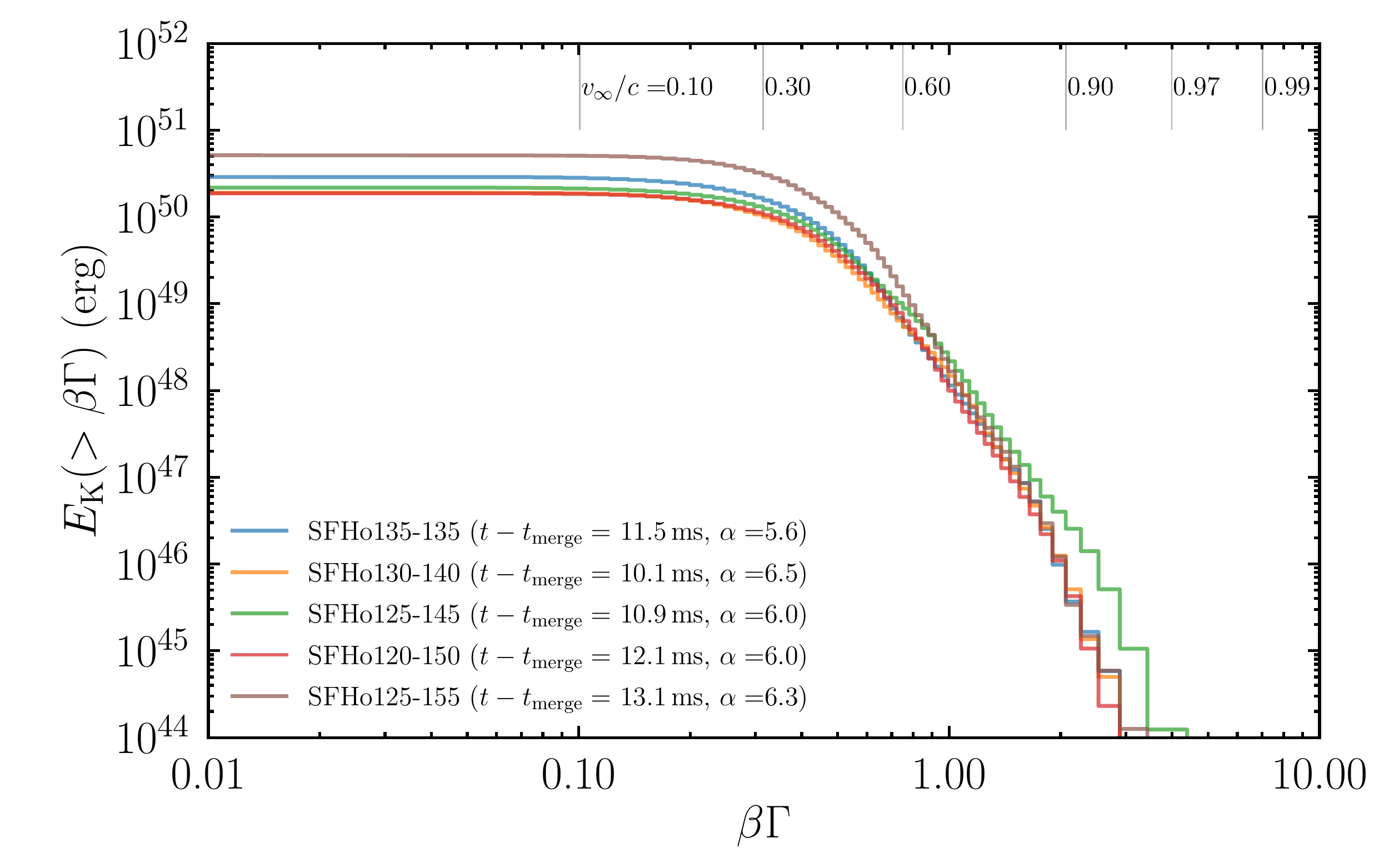}
\caption{
Top: Mass histograms of as functions of $Y_\mathrm{e}$ at $T=\SI{5}{GK}$ (left) and of maximum temperature that each tracer particle experiences (right) calculated from the tracer particles obtained in the 3D simulations.
Bottom: Cumulative mass (left) and kinetic energy (right) histograms of the asymptotic specific momentum $\beta\Gamma$ of the dynamical ejecta obtained using the snapshot at which the highest velocity for a given time slice becomes maximum in the entire time history (the corresponding post-merger time is shown in the parenthesis).
The vertical lines with numbers denote the corresponding asymptotic velocity $v_\infty/c := \sqrt{1-(-u_t)^{-2}}$.
In the caption, $\alpha$ denotes the slope of the distribution ($M(>\beta\Gamma)\propto (\beta\Gamma)^{-\alpha}$ or $E_\mathrm{K}(>\beta\Gamma)\propto (\beta\Gamma)^{-\alpha}$) around $\beta\Gamma=1$.
The horizontal dotted lines in the bottom-left panel denote the mass of the atmosphere swept by the highest-velocity matter until the time at which the histogram is generated.
This indicates the maximum level of the pollution by the atmosphere.
}
\label{fig:hist_dyn}

\end{figure*}

\subsection{3D simulations}
\label{subsec:3d}

We describe the results of the 3D simulations paying a particular attention to their dependence on the mass ratio.
We note that the results are qualitatively the same as in \cite{Sekiguchi2016a,Shibata2017a} but the present simulations are performed with a better grid resolution, with a better hydrodynamics scheme, and with an improved treatment of the weak interaction.
For each model, the binaries merge after approximately five orbits in the present choice of the initial conditions.

For all the models with $M_\mathrm{ tot}=2.7M_\odot$, after the merger, a hypermassive neutron star is temporarily formed and survives for $t_\mathrm{ BH}\approx15$--$\SI{20}{ms}$ before collapsing into a black hole.
For the massive model, SFHo125-155, a hypermassive neutron star is temporarily formed as well, but the lifetime is shorter: $t_\mathrm{BH} \approx \SI{3}{ms}$.
The post-merger time of the black-hole formation, which is listed in Table~\ref{tab:model-3d}, depends weakly on the mass ratio of the binary: A more asymmetric merger results in a slightly longer-lived hypermassive neutron star, because the formed neutron star is less massive due to the formation of more massive disk (see the text below) and also the merger sets in at a larger orbital separation, i.e., less energy and angular momentum are emitted by gravitational radiation before the onset of merger.

For the equal-mass case (i.e., SFHo135-135), the ejecta is driven primarily by the shock heating at the contact surface of merging two neutron stars and additionally by a subsequent tidal torque exerted by the deformed remnant massive neutron star.
As a result of the shock heating, the temperature increases and the electron degeneracy decreases, and consequently, its $Y_\mathrm{e}$ increases from the originally low values up to $\approx 0.4$ due to the positron capture, although the ejecta that does not experience the shock heating preserves a low value of $Y_\mathrm{e}\lesssim0.1$.

For more asymmetric cases (e.g., SFHo120-150 and 125-155), on the other hand, the tidal interaction becomes more important for the mass-ejection channel.
Since the shock-heating effect is relatively minor and the temperature enhancement is not as efficient as in the symmetric case, the electron-positron pair production is relatively inactive and the positron capture does not occur in a sufficiently short timescale.
Thus, the values of $Y_\mathrm{e}$ for the majority of the ejecta are preserved to be their original (i.e., low) values determined by the neutrino-less beta-equilibrium (that is $\approx 0.03$--0.04 at lowest for the SFHo EOS).
Such trends are clearly found in the  mass histogram of $Y_\mathrm{e}$, shown in the top-left panel of Fig.~\ref{fig:hist_dyn}, for the dynamical ejecta when their temperature decreases to $T=\SI{5}{GK}$.

Model SFHo125-145 has a feature between SFHo135-135 and SFHo120-150; the dynamical ejecta has a bimodal $Y_\mathrm{e}$ distribution: The distribution with the peak at a lower value of $Y_\mathrm{e}\approx 0.06$ stems from the matter ejected by the tidal interaction, while that with the peak at a higher value of $Y_\mathrm{e}\approx 0.22$ stems from the shock-driven component.
The shock-driven ejecta has somewhat lower values of $Y_\mathrm{e}$ than in model SFHo135-135 due to the lower shock heating efficiency.

The $Y_\mathrm{e}$ distribution for model SFHo130-140 is similar to that for model SFHo135-135.
However, for model SFHo130-140 the dynamical ejecta mass is smaller than those of models SFHo135-135 and SFHo125-145, as already found in~\cite{Sekiguchi2016a}.
For this model, the tidal effect is not strong enough to appreciably produce the unbound matter.
This fact is reflected  in the absence of the peak for $Y_\mathrm{e}<0.1$ seen in models SFHo125-145 and SFHo120-150.
In addition, the shock heating is not as strong as that for SFHo135-135. As a result of these facts, the ejecta mass for model SFHo130-140 becomes smaller than the other models.

For a massive asymmetric model, SFHo125-155, the $Y_\mathrm{e}$ distribution is similar to that of SFHo120-150, because the tidal interaction is the main channel of the dynamical mass ejection.
For this model, however, the fraction of the high $Y_\mathrm{e}$ component with $Y_\mathrm{e} \gtrsim 0.2$ is smaller than that for model SFHo120-150 despite the similar binary mass ratio.
This is due to the less efficient neutrino irradiation (for reprocessing $Y_\mathrm{e}$ in the ejecta) as well as the smaller amount of shock-driven ejecta (in which $Y_\mathrm{e}$ is increased by positron capture) because of the shorter lifetime of the hypermassive neutron star (see, e.g., \citealt{Sekiguchi2015a} and \citealt{Goriely2015a} for the effect of neutrino irradiation to the dynamical ejecta and \citealt{Hotokezaka2013a} for the effect of a hypermassive neutron star on the shock-driven mass ejection).

\begin{figure}
\epsscale{1.17}
\includegraphics[width=0.49\textwidth]{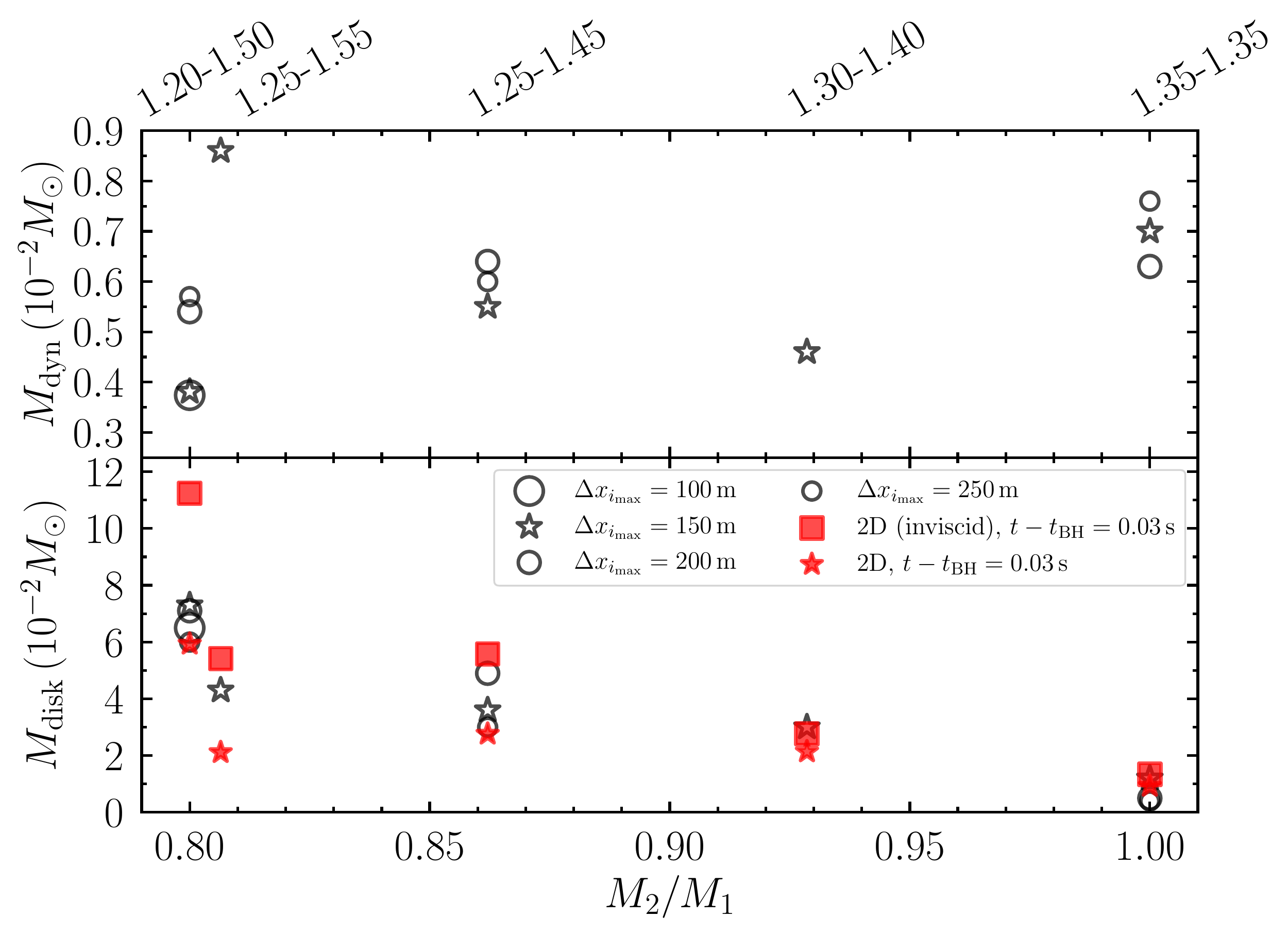}
\caption{
Masses of the dynamical ejecta (top) and the bound matter (disk; bottom) for models in the 3D simulations with different grid resolutions. The models with the fiducial resolution ($\Delta x_{i_\mathrm{max}}=\SI{150}{m}$) are marked by the open stars, while the other resolution models are marked by the open circles with different sizes.
In the bottom panel, the disk masses obtained in the 2D simulations are also plotted with the red symbols.
The filled stars and squares denote the values in the simulations with and without viscosity, respectively. All the values in this figure are obtained at \SI{30}{ms} after the black-hole formation for each model.
}
\label{fig:disk_mass}
\end{figure}

Table~\ref{tab:model-3d} summarizes the dynamical ejecta mass as a function of the mass ratio.
There is a weak trend that the dynamical ejecta mass decreases for more asymmetric merger.
This is likely due to the fact that the shock heating is the more efficient mass ejection process than the tidal interaction for this EOS~\citep{Sekiguchi2015a}, and thus, the weaker shock heating leads to the smaller dynamical ejecta mass.
It is clearly found from Table~\ref{tab:model-3d} that the average value of $Y_\mathrm{e}$ decreases with a decrease in the mass ratio, reflecting that the tidal interaction plays a more important role for the dynamical mass ejection for more asymmetric binaries.

The top-right panel of Fig.~\ref{fig:hist_dyn} displays the distribution of the maximum temperature that each tracer particle experiences.
We find that the temperature in the entire dynamical ejecta exceeds \SI{10}{GK} because of shock heating.
It is important to note that for unequal-mass cases, the (two) tidal tails of a merger also experience shock heating when the second one catches up the first one.
For SFHo135-135, the tails do not collide with each other but stem from the high-temperature shocked-surface region of merging neutron stars.

The bottom-left panel of Fig.~\ref{fig:hist_dyn} shows the cumulative mass histogram of the asymptotic specific momentum $\beta\Gamma$ of the dynamical ejecta, which is defined by $\beta\Gamma = \sqrt{(-u_t)^2-1}$ from the fact that $-u_t$ is the asymptotic Lorentz factor assuming the asymptotically stationary spacetime.
Here we ignore the contribution of the internal energy that can further accelerate the ejecta, because it only gives a minor correction for matter with $\beta\Gamma \gtrsim 0.1$.
The histogram is obtained using the snapshot at which the highest velocity for a given time slice becomes maximum in the entire time history.
After this time, the highest-velocity matter begins to be decelerated spuriously by the interaction with the artificial atmosphere.
There is a fast-moving component for all the models investigated in this work (i.e., not only for the symmetric binary but also for the asymmetric binaries).
Interestingly, it is found that the highest velocity always exceeds $0.9c$, although the matter with the highest velocity has only tiny mass and suffers from the interaction with the atmosphere (see Appendix \ref{app:resolution}).
The mass of the matter with $v_\infty/c:= \sqrt{1-(-u_t)^{-2}}>0.6$ is $\sim10^{-5}M_\odot$ irrespective of the binary mass ratio, which is consistent with another numerical-relativity result in \cite{Radice2018dec} for the SFHo EOS.
The bottom-right panel of Fig.~\ref{fig:hist_dyn} shows the cumulative kinetic energy histogram of $\beta\Gamma$.
The spectrum shape is in broad agreement with the previously reported results \citep{Hotokezaka2016nov,Hotokezaka2018nov,Hajela2022mar} apart from the extension to a higher $\beta\Gamma$ side than previously.
The dependence of $Y_\mathrm{e}$ and velocity distributions on the grid resolution is discussed in Appendix~\ref{app:resolution}.

Figure~\ref{fig:disk_mass} compares the masses of the dynamical ejecta (top) and the bound matter (or disk masses; bottom) with different grid resolutions for $M_\mathrm{tot}=2.7M_\odot$.
The ejecta and disk masses are found to converge only slowly with errors of 0.001--$0.002M_\odot$ and 0.01--$0.02M_\odot$, respectively, for the chosen range of the grid resolution.
Thus, we have to keep in mind that the masses of the dynamical ejecta and also the post-merger ejecta, the latter being a fraction of the disk mass, always have the uncertainty with such a level.

\subsection{2D simulations}

\subsubsection{Disk mass}
\label{subsubsec:diskmass}
We then turn our attention to presenting the results of the post-merger 2D simulations.
In our setting, the hypermassive neutron star collapses into a black hole at $\leq\SI{1}{ms}$ after the beginning of the 2D simulations and a disk of mass of $\sim 0.01$--$0.1M_\odot$ always remains outside the black hole. 

\begin{figure*}
\includegraphics[width=0.49\textwidth]{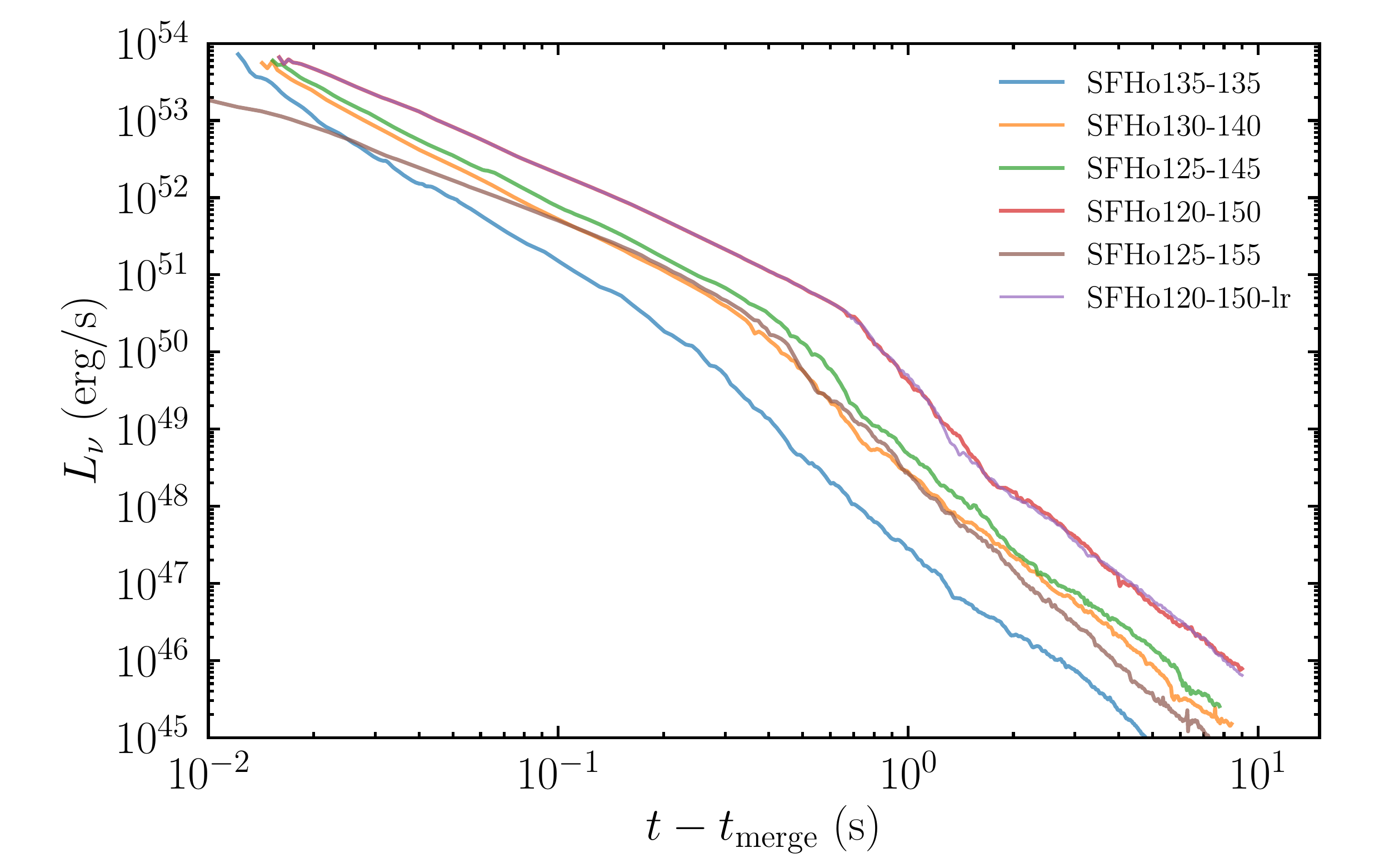}~~
\includegraphics[width=0.49\textwidth]{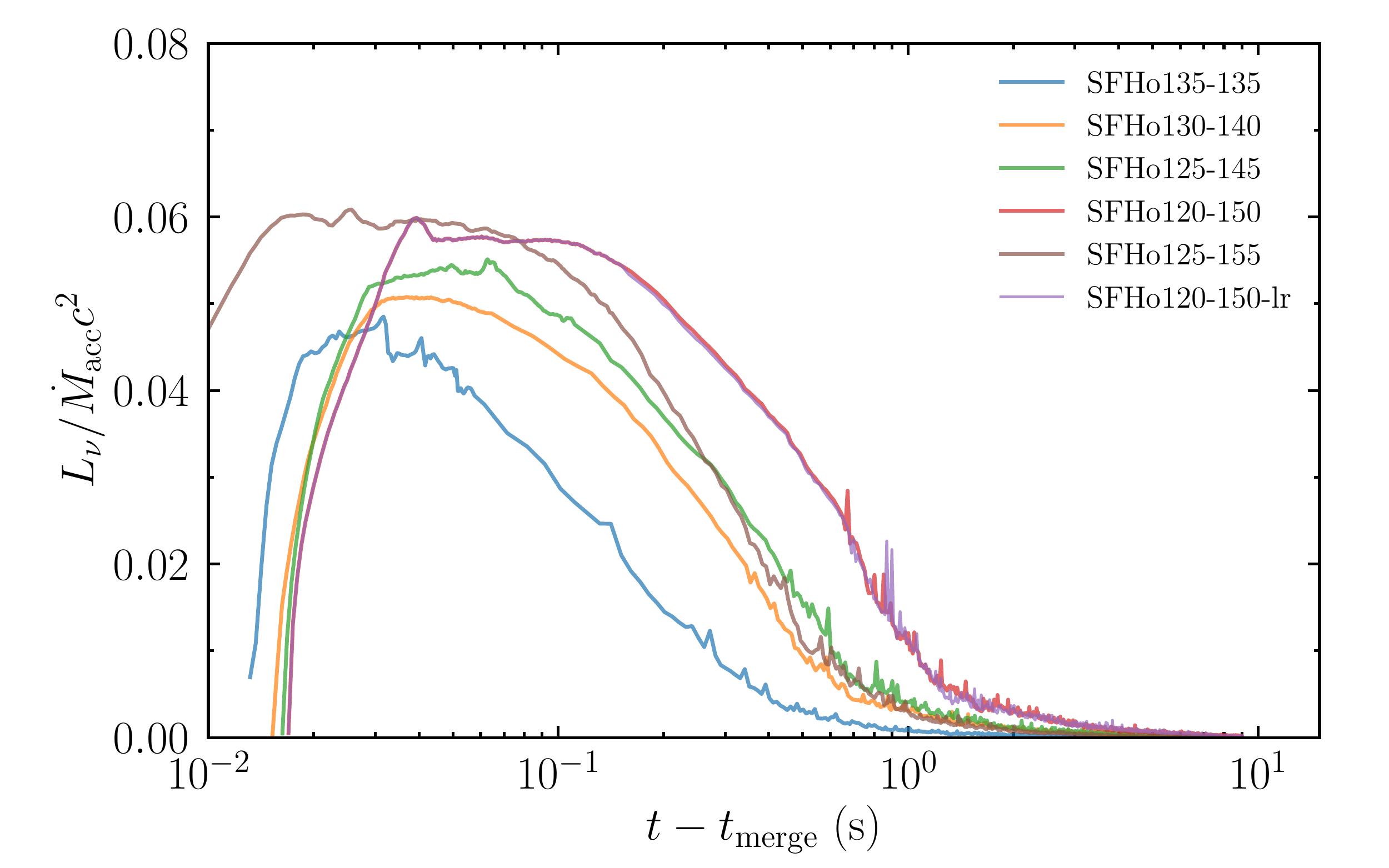}\\
\includegraphics[width=0.49\textwidth]{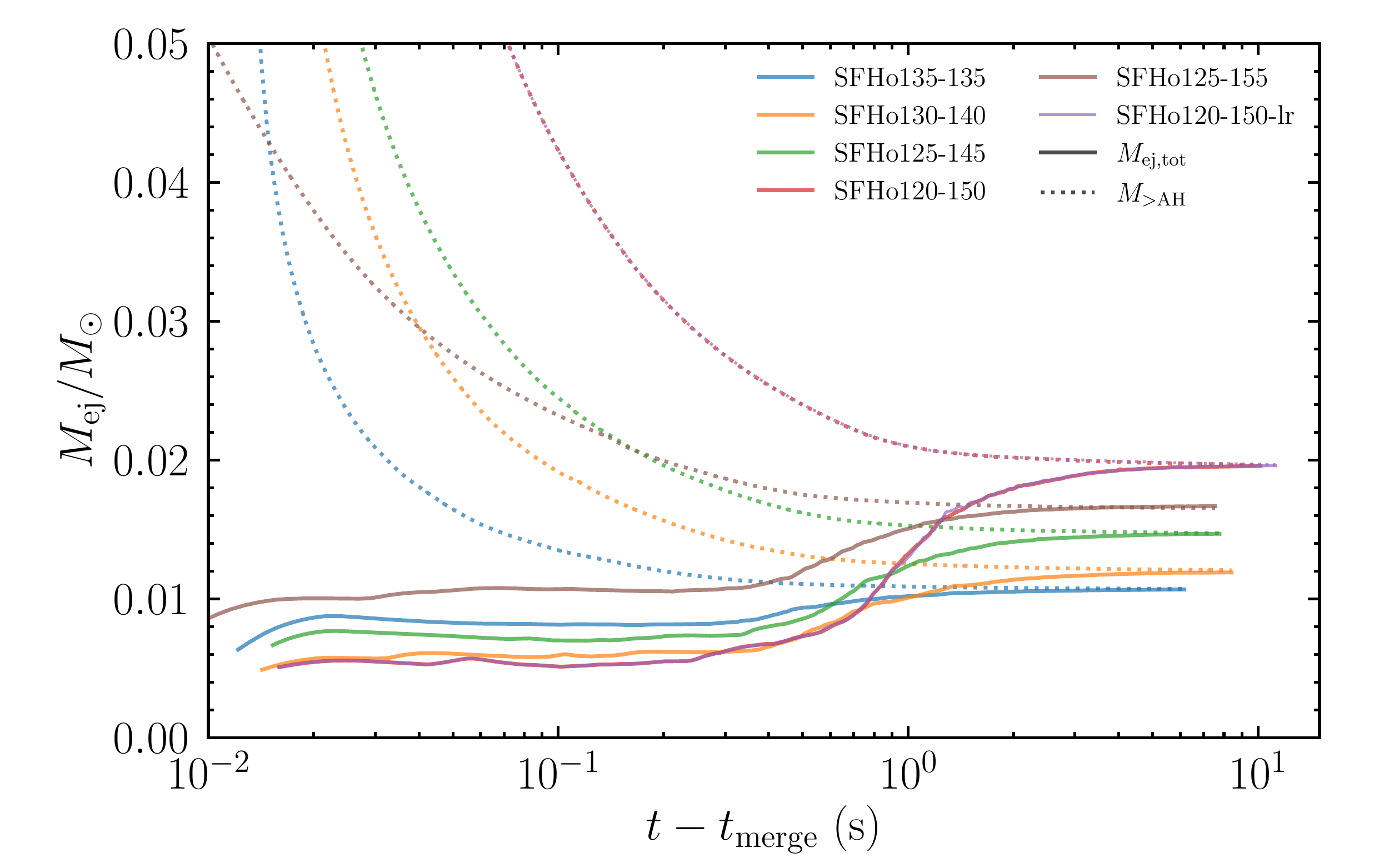}~~
\includegraphics[width=0.49\textwidth]{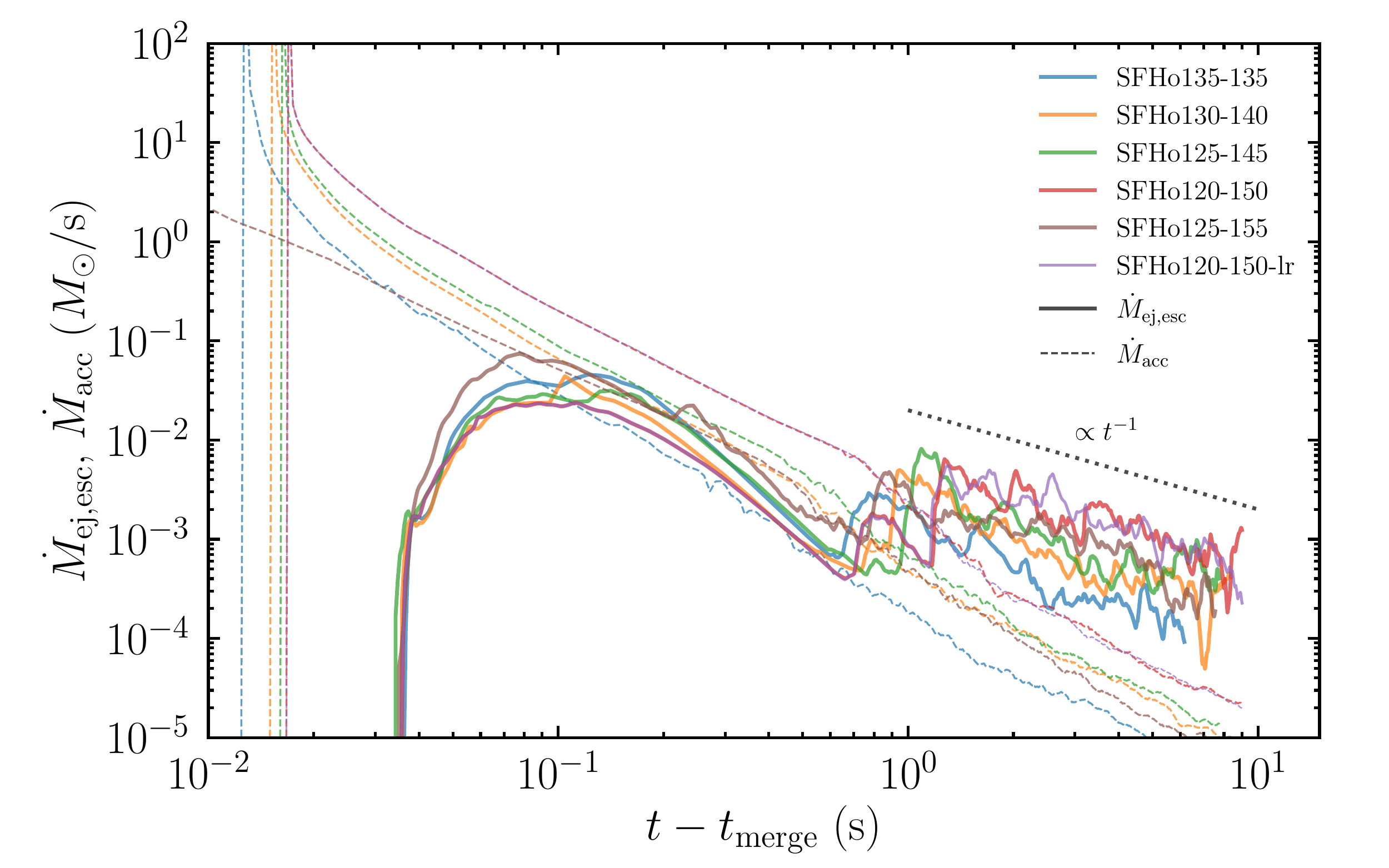}
\caption{
Top-left: Time evolution of the total neutrino luminosity.
Top-right: Time evolution of the neutrino cooling efficiency defined as the total neutrino luminosity divided by the mass accretion rate onto the black hole.
Bottom-left: Time evolution of the total ejecta mass of both outside and inside the extraction radius (solid curves) and the mass of the matter outside the apparent horizon (dotted curves). Note that for $t-t_{\rm merger} \alt 300$\,ms, the ejecta are comoposed mainly of the dynamical ejecta.
Bottom-right: Time evolution of the mass ejection rate at the extraction radius (solid curves) and the mass accretion rate onto the black hole (dashed curves).
}
\label{fig:efficiency-ejectamass}
\end{figure*}

The bottom panel of Fig.~\ref{fig:disk_mass} shows the disk mass at \SI{30}{ms} after the black-hole formation as a function of the binary mass ratio for the models with $M_\mathrm{ tot}=2.7M_\odot$.
Here, the disk mass is defined as the mass of the bound matter located outside the apparent horizon (for the criterion of the unbound matter, see the text below).
The disk mass increases with a decrease in the mass ratio of the binaries, indicating that the remnants of more asymmetric binary mergers have a possibility to eject more matter in the post-merger phase.

In the same panel, the disk mass for the 2D simulations without viscosity is also plotted, which is found to be somewhat larger than in the corresponding 3D fiducial simulations for the asymmetric merger cases (SFHo125-145 and SFHo120-150).
The reason for this is that the gravitational torque due to the non-axisymmetric matter distribution, by which the angular momentum is redistributed in the disk and the mass accretion is driven, is absent in the 2D simulations.
The effect is more significant for the more asymmetric merger cases.
By contrast, the disk mass for the 2D simulations with viscosity is consistent with that for the corresponding 3D simulations.
Thus, we consider that our 2D viscous simulations can approximately model the angular momentum transport due to the gravitational torque.

\addsf{
For a massive model SFHo125-155, however, there is still a significant non-axisymmetric structure when the mapping is performed.
The mapping of this structure may result in the artificial redistribution of angular momentum in the disk.
The disk mass at \SI{30}{ms} after the black-hole formation for the 2D simulation with viscosity is lower than that for the corresponding 3D simulation by a factor of two probably due to this effect.
Thus, the post-merger ejecta for this model may be underestimated.
}

\subsubsection{Post-merger mass ejection} \label{subsubsec:post-merger}
The mechanism of the post-merger mass ejection is the same as that described in our previous work for the case of long-lived massive neutron star formation~\citep{Fujibayashi2020c}: In the early phase of the post-merger evolution, the temperature of the disk is high ($\lesssim\SI{10}{MeV}$ at the maximum), which results in the neutrino luminosity far exceeding $10^{51}\,\mathrm{erg/s}$.
In such a phase, the internal energy generated by the viscous heating is carried away by the neutrino emission and is not efficiently used for the expansion and mass ejection of the disk matter.
However, the viscous effect transports the angular momentum inside the disk, and thus, the density and temperature of the disk decrease with time due to the outward expansion and matter infall to the black hole.
As a result of the temperature decrease, the neutrino luminosity decreases steeply with time.
The viscosity-driven mass ejection sets in when the temperature of the disk becomes sufficiently low ($k_\mathrm{B}T\lesssim2$--\SI{3}{MeV}) and the neutrino cooling becomes less efficient than the viscous heating (see \S~\ref{subsec:ye}).
In the post-merger mass ejection phase, the neutrino luminosity is below $\sim 10^{51}\,\mathrm{erg/s}$ in the present setting.

The top panels of Fig.~\ref{fig:efficiency-ejectamass} show the time evolution of the neutrino luminosity (left) and neutrino cooling efficiency (right) defined as the neutrino luminosity divided by the mass accretion rate onto the black hole.
It is indeed found that the neutrino luminosity decreases monotonically with time, and at $L_\nu \sim 10^{51}\,{\rm erg/s}$, the decreasing rate is enhanced because of the onset of the mass ejection and the resulting increase of the disk expansion rate.
It is also found that the larger the asymmetry of the binary is, the later the cooling efficiency drops.
This is partially due to the fact that the merger of a more asymmetric binary results in a higher disk mass, with which a higher mass accretion rate is achieved.
Thus, the state of high neutrino cooling efficiency is maintained for a longer timescale.
The other reason is that the typical radius of the disk, which can be calculated from the mass $M_\mathrm{ disk}$ and angular momentum $J_\mathrm{ disk}$ of the disk as $R_\mathrm{ disk}:=(J_\mathrm{ disk}/M_\mathrm{ disk})^2/(GM_\mathrm{ BH}$), is larger for a more asymmetric binary, and hence, the viscous timescale of the disk is longer.

The bottom-left panel of Fig.~\ref{fig:efficiency-ejectamass} shows the time evolution of the ejecta mass, which is defined by
\begin{align}
M_\mathrm{ej,tot} &= M_\mathrm{ej,esc}\notag\\
&+ \int_{\SI{500}{km}\le r<r_\mathrm{ext}} \sqrt{-g}\rho u^t \Theta(\Gamma_\infty - 1) d^3x, \label{eq:ejecta-mass}
\end{align}
where $\Theta$ is the Heaviside function.
In the above expression,
\begin{align}
M_\mathrm{ej,esc}(t) = \int^t \dot{M}_\mathrm{ej,esc} dt \label{eq:mej_esc}
\end{align}
is the mass of the ejecta escaped from an extraction radius $r_\mathrm{ext}$, defined by the time-integration of the mass outflow rate
\begin{align}
\dot{M}_\mathrm{ej,esc} = \int_{r=r_\mathrm{ext}} \sqrt{-g} \rho u^k \Theta(\Gamma_\infty - 1) ds_k.\label{eq:mdot-baryon-mass}
\end{align}
Here, $ds_i=\delta_{ir} {r_\mathrm{ext}}^2 \sin\theta d\theta d\phi$ is the area element of a sphere at the extraction radius.
On the other hand, the second term in Eq.~\eqref{eq:ejecta-mass} is the contribution of the unbound matter inside the extraction radius.
We note that the rest mass here is defined as the baryon number density multiplied by the atomic mass unit $m_\mathrm{u}\approx\SI{931}{MeV}/c^2$.
The criterion here for the unbound matter is the same as our previous study~\citep{Fujibayashi2020b}: The asymptotic Lorentz factor $\Gamma_\infty$ for the post-merger ejecta is defined by
\begin{align}
\Gamma_\infty = -\frac{h u_t}{h_\mathrm{min,global}}, \label{eq:gamma_inf_main}
\end{align}
where $h_\mathrm{min,global}\approx 0.9987c^2$ is the globally minimum specific enthalpy in the employed EOS table.
There are other possible criteria for the unbound matter~\citep{foucart2021dec}.
The dependence of the ejecta mass on the choice of the criteria is discussed in Appendix~\ref{app:criteria}. 
In this study we set $r_\mathrm{ext}=\SI{8000}{km}$ ($\approx 2000M_\mathrm{BH}$).

The bottom-left panel of Fig.~\ref{fig:efficiency-ejectamass} indicates the presence of unbound matter from the beginning of the 2D simulations.
This is composed mainly of the dynamical ejecta.
The approximately constant values of $M_\mathrm{ej,tot}$ for $t-t_\mathrm{merge}\alt  \SI{300}{ms}$ indicate the absence of significant mass-ejection activity in this relatively earlier post-merger phase.
For $t-t_\mathrm{merge}\gtrsim \SI{300}{ms}$, $M_\mathrm{ej,tot}$ begins to increase due to the onset of the post-merger mass ejection.
The onset time for the more asymmetric mergers becomes later because of the larger disk mass and the resultant later drop of the neutrino cooling efficiency as stated already.

\begin{figure*}
\includegraphics[width=0.49\textwidth]{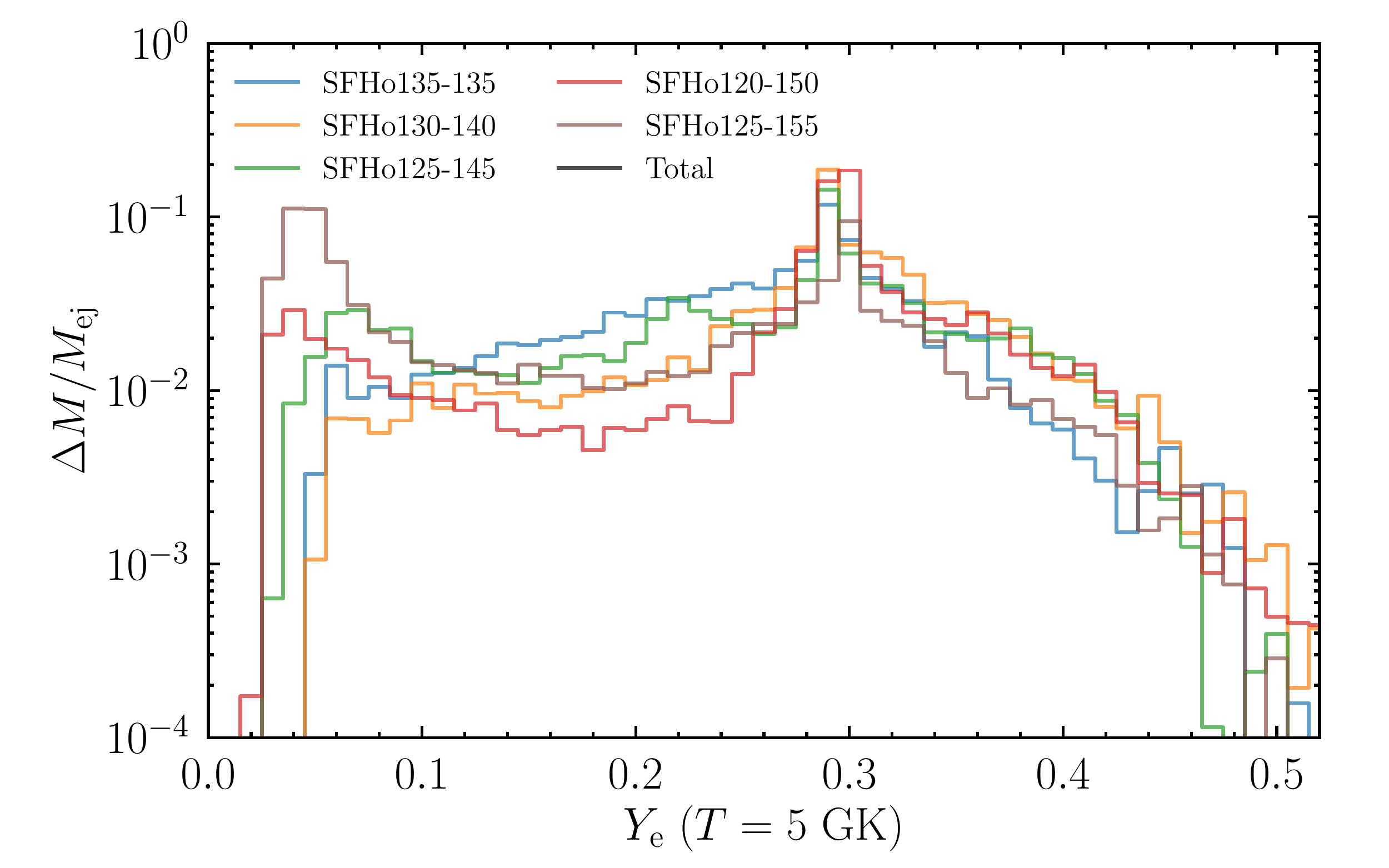}~
\includegraphics[width=0.49\textwidth]{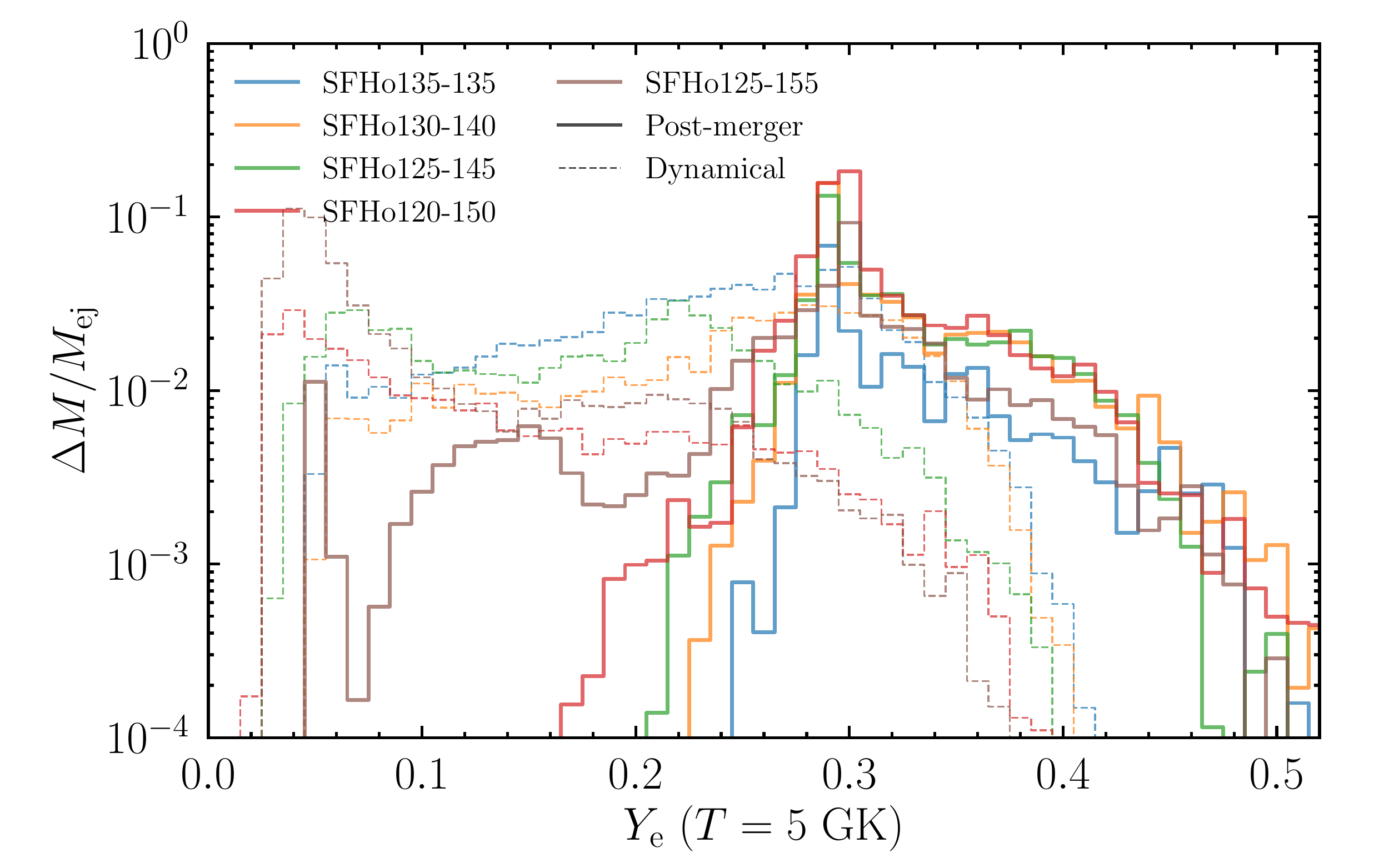}
\caption{
Mass histograms of $Y_\mathrm{e}$ at $T=\SI{5}{GK}$ for all models investigated in this work.
The histograms for the dynamical (dashed) and post-merger ejecta (solid) are shown in the right panel, while those in the left panel are the histograms for the total ejecta, which is obtained by summing up the dynamical and post-merger ejecta in 3D and 2D simulations, respectively.
See \S~\ref{subsec:tp} for the detailed procedure.
}
\label{fig:hist_summed}
\end{figure*}.

It is indeed found that $M_\mathrm{ej,tot}$ begins to increase concurrently with the steep decrease of the cooling efficiency, which saturates in several seconds after the merger.
The saturated values are very close to the total mass outside the apparent horizon, $M_\mathrm{>AH}$, irrespective of the models (see the dotted curves in the bottom-left panel of Fig.~\ref{fig:efficiency-ejectamass})

The bottom-right panel of Fig.~\ref{fig:efficiency-ejectamass} compares the mass ejection rate $\dot{M}_\mathrm{ej,esc}$ and the mass accretion rate onto the black hole $\dot{M}_\mathrm{acc}$.
After the post-merger mass ejection sets in, the mass ejection rate becomes higher than the mass accretion rate by more than one order of magnitude. 
Thus, the mass accreted onto the black hole after the onset of the post-merger mass ejection is subdominant compared to that of the post-merger ejecta.
Therefore, the total post-merger ejecta mass can be approximately estimated as the total mass located outside the apparent horizon when the cooling efficiency have already dropped sufficiently and the post-merger mass ejection have already set in.

Although we here evaluate the mass of the post-merger ejecta assuming that all the internal energy will be  converted to the kinetic energy (i.e., Bernoulli's criterion), a fraction of matter can turn around before this conversion completes and then fall back onto the black hole-disk system (e.g., \citealt{Ishizaki2021dec}).
This fall-back matter may energize late-time electromagnetic emission such as that observed in the afterglow of GRB 170817A~\citep{Ishizaki2021aug,Hajela2022mar}.
To clarify the uncertainty in the effect of fall back, we need a computation for the very long-term evolution of the outflowed matter (e.g., \citealt{Kawaguchi2021jun} and \citealt{Rosswog2014mar}).

We note that the post-merger mass ejection rate is proportional to $t^{-1}$, or $dM_\mathrm{ej,esc}/d\ln t$ is constant, approximately.
This implies that the contribution of the post-merger mass ejection in a later phase with $t\sim \SI{10}{s}$ is as important as that in an earlier phase $t\sim \SI{1}{s}$. This shows that a long-term ($\sim\SI{10}{s}$ or more) simulation is required to explore the entire mass ejection history.

\subsubsection{Resolution dependence}
Figure~\ref{fig:efficiency-ejectamass} also shows the results for model SFHo120-150-lr, in which the coarser grid spacing with $\Delta x_0 = \SI{300}{m}$ is adopted after the regridding.
The time evolution of the neutrino cooling efficiency, ejecta mass, and mass ejection and mass accretion history for the lower resolution model agree approximately with those for the higher resolution model.
Thus, we conclude that the dependence of the grid resolution on the results is weak.

\begin{figure*}
\centering

\includegraphics[width=0.325\textwidth]{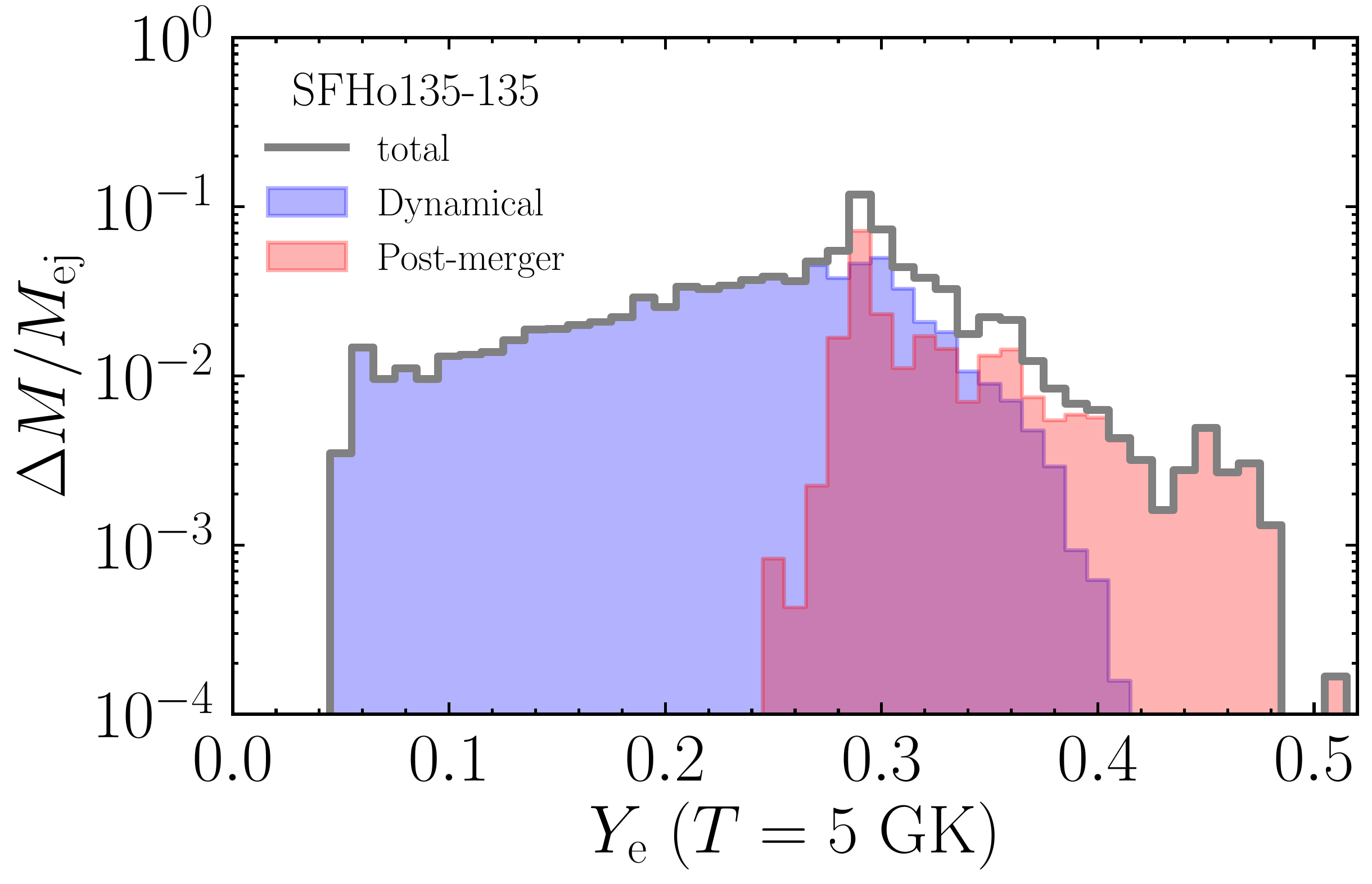}
\includegraphics[width=0.325\textwidth]{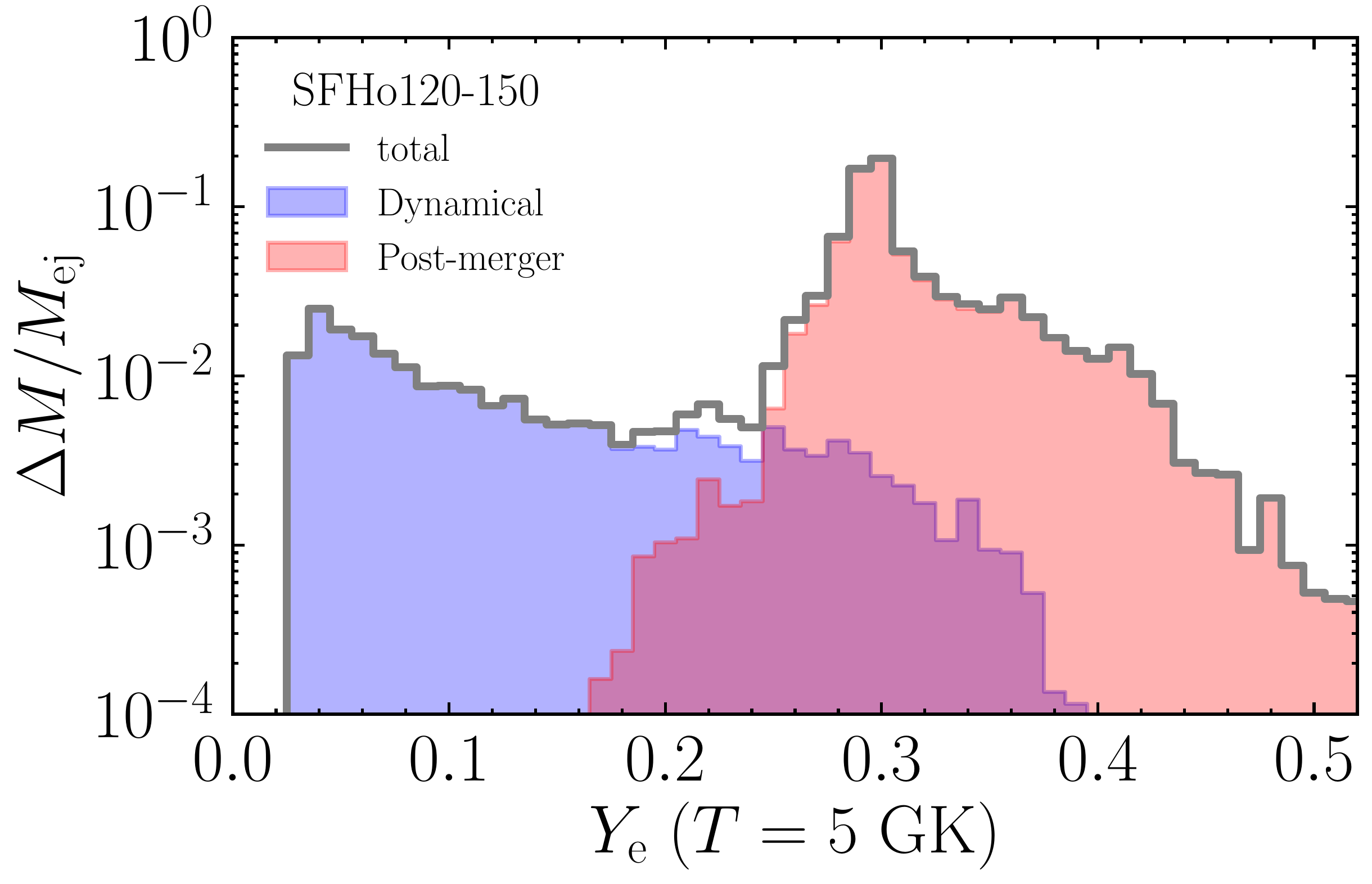}
\includegraphics[width=0.325\textwidth]{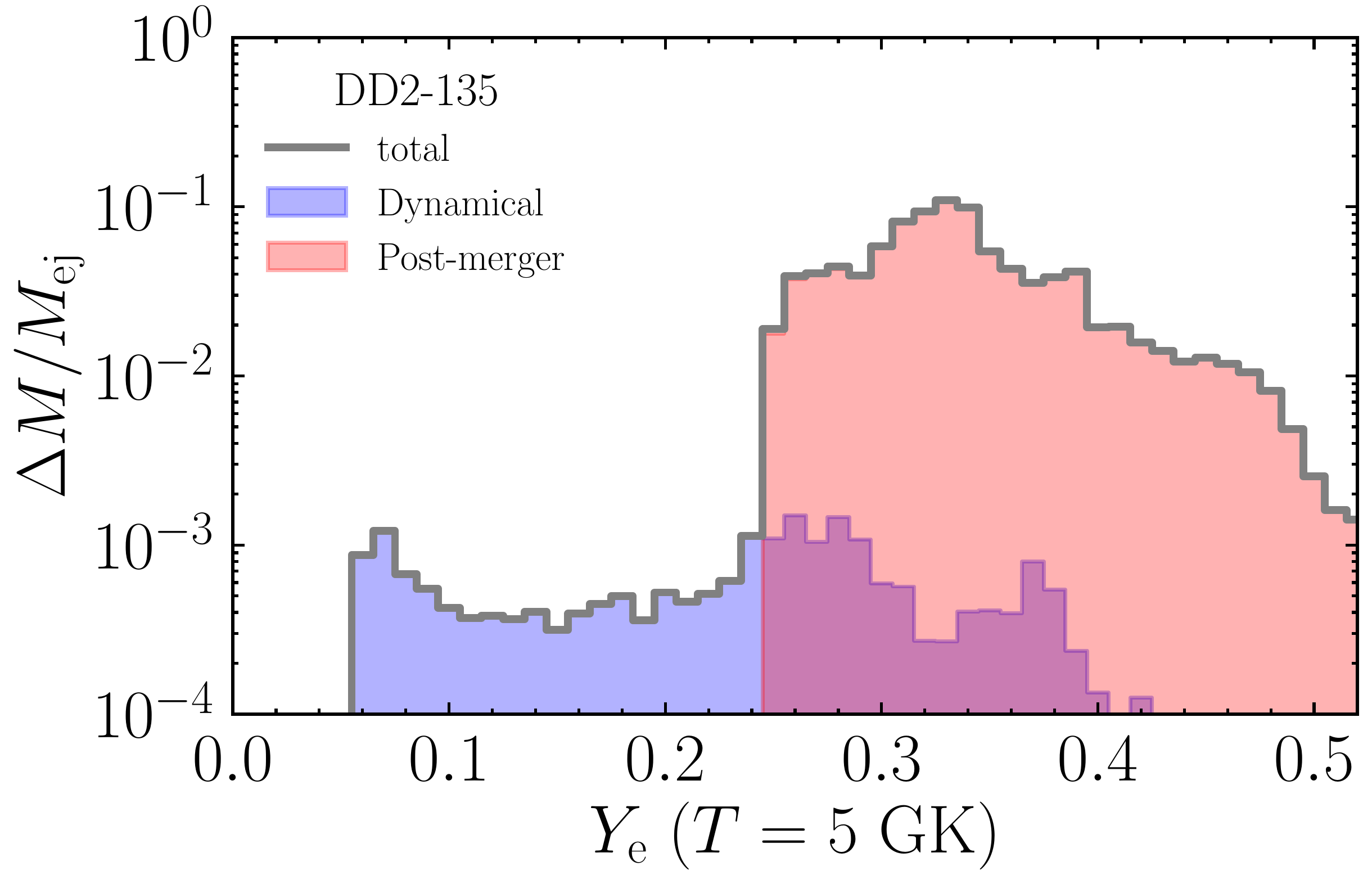}\\
\includegraphics[width=0.325\textwidth]{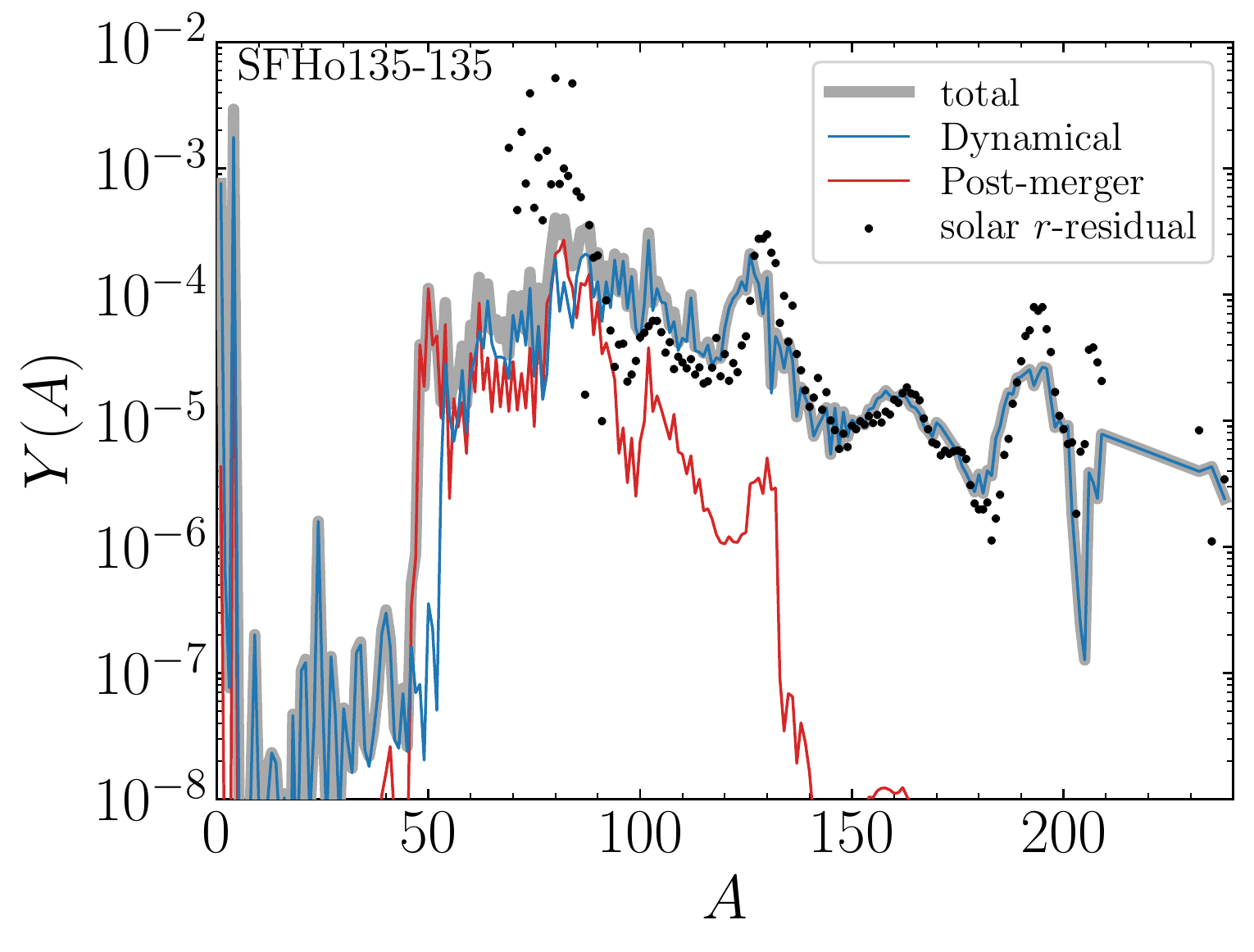}
\includegraphics[width=0.325\textwidth]{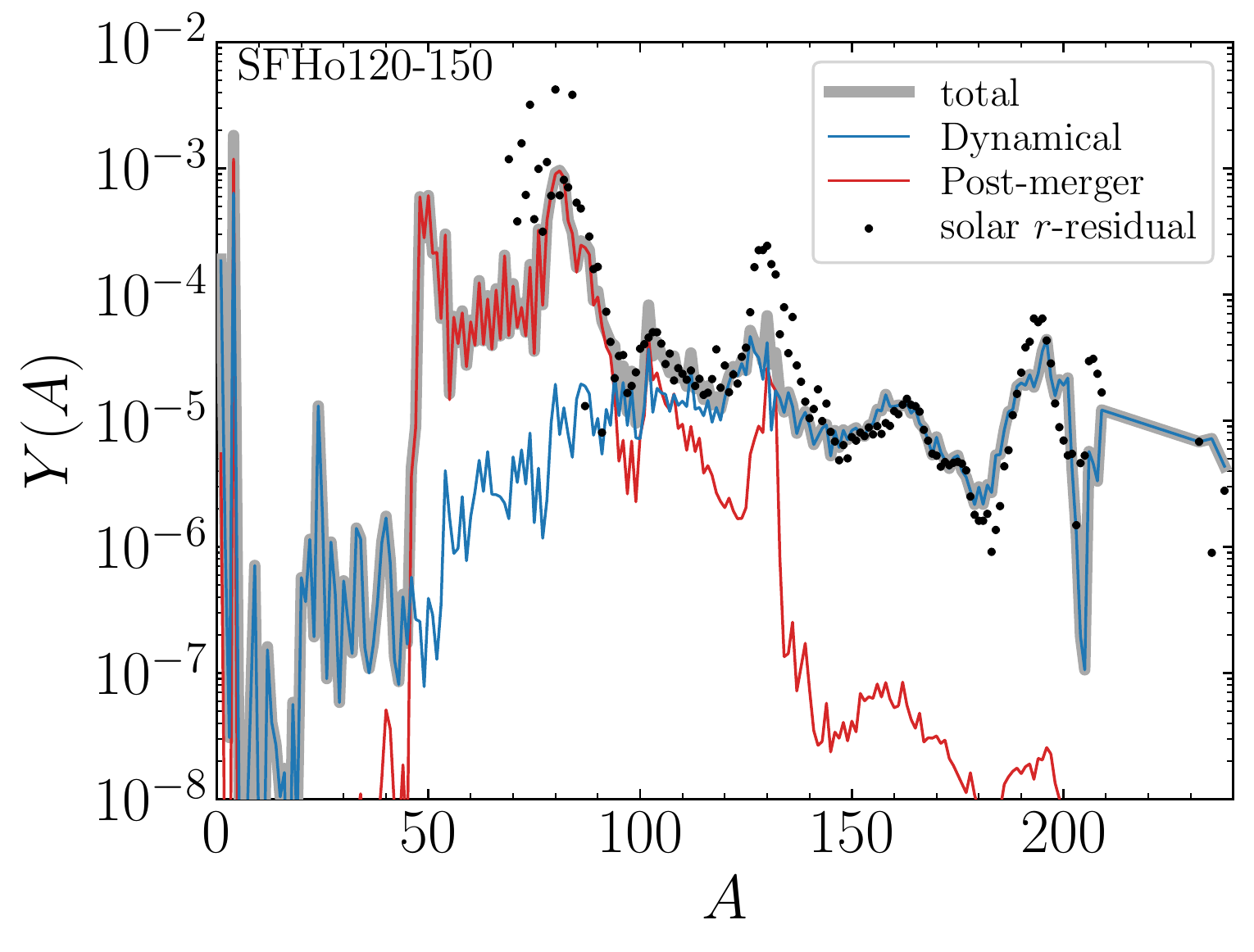}
\includegraphics[width=0.325\textwidth]{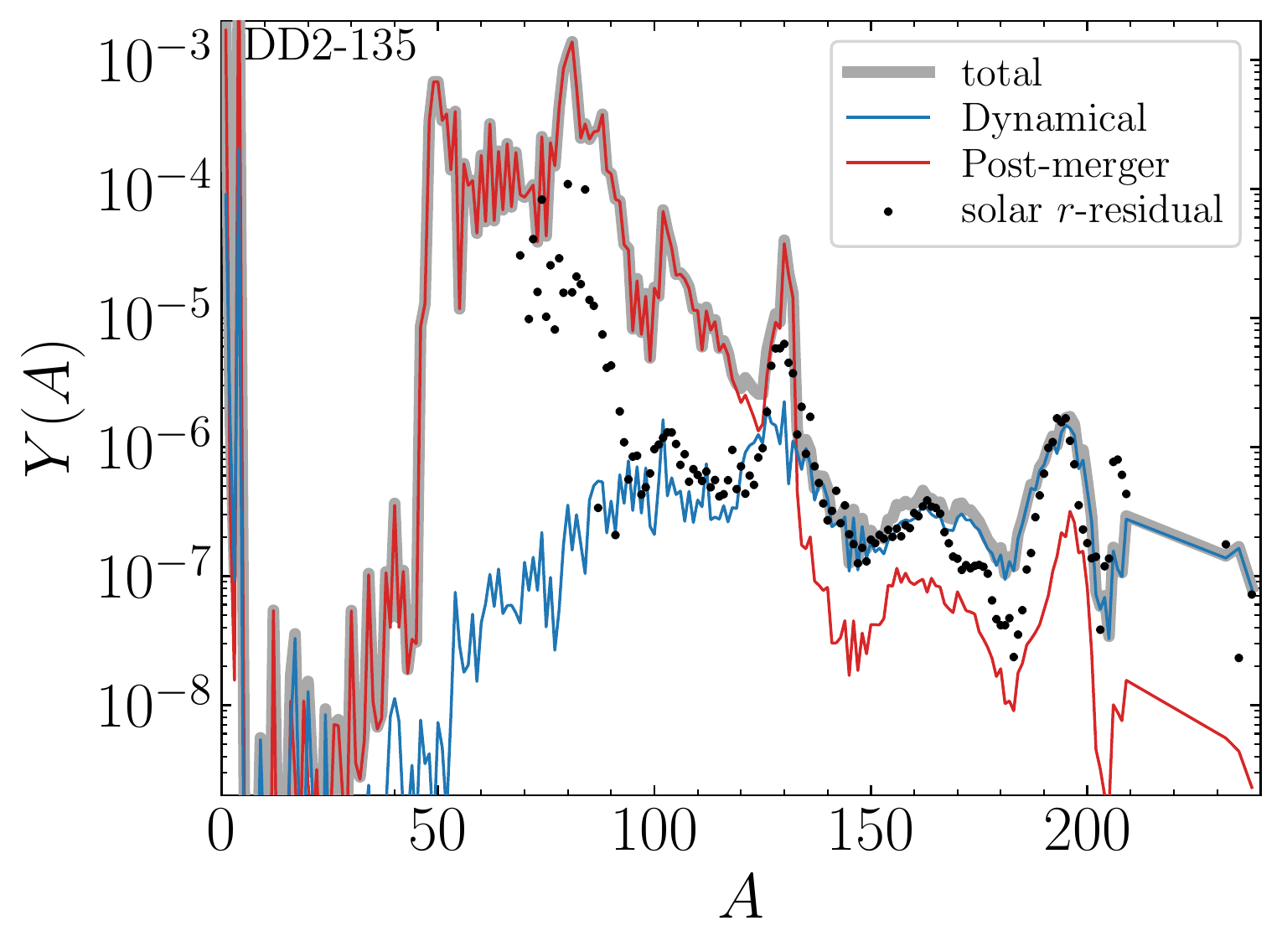}
\caption{Top panels: $Y_\mathrm{e}$ distributions for three representative cases with the total mass of $2.7\, M_\odot$; the equal-mass merger leaving a hypermassive neutron star (SFHo135-135; left), the asymmetric merger leaving a hypermassive neutron star (SFHo120-150; center), and the equal-mass merger in which a massive neutron star survives for more than $10$\,s (model DD2-135 in \citealt{Fujibayashi2020a}; right).
The blue and red shaded histograms denote those of dynamical and post-merger ejecta, respectively, and the gray lines denote the total distribution.
Bottom panels (under construction): Abundance distribution corresponding to the models in the top panels. The blue and red curves denote the contributions from dynamical and post-merger ejecta, respectively, and the gray curves denote the total nucleosynthetic yields.
}
\label{fig:ye-dist-parameter}
\end{figure*}

\subsection{Electron fraction distribution} \label{subsec:ye}

Figure~\ref{fig:hist_summed} shows the mass histograms of $Y_\mathrm{e}$ at $T=\SI{5}{GK}$ for the dynamical ejecta and post-merger ejecta (dashed and solid curves in the right panel, respectively), and for the sum of these two components (the left panel).
The post-merger ejecta have a peak at $Y_\mathrm{e}\approx 0.3$ irrespective of the models.
This value is determined approximately by the so-called freeze-out condition that the electron/positron capture timescales become comparable to the viscous expansion timescale of the matter owing to its expansion and adiabatic cooling~\citep[for more detailed description, see][and also Appendix \ref{app:process-ye}]{Fujibayashi2020c}.
We note that the peak location of $Y_\mathrm{e}$ can depend slightly on the strength of viscosity, which determines the expansion timescale.
Specifically, for a higher viscosity, the peak shifts to the lower-$Y_\mathrm{e}$ side as a result of the shorter expansion timescale~\citep{Fujibayashi2020b}.
For a plausible range of the viscous parameter (the $\alpha$-viscosity of 0.03--0.1), however, the variation of the $Y_\mathrm{e}$-peak location is not very appreciable ($\lesssim0.05$) for black-hole disk systems~\citep{Fujibayashi2020a}.
\addsf{Importantly, the neutrino-absorption timescale is always longer than the electron/positron capture timescale in the relevant physical conditions (see Appendix~\ref{app:process-ye}).
Thus, the dependence of resulting $Y_\mathrm{e}$ distribution on the adopted neutrino transfer method is expected to be small as far as the neutrino emission is reasonably modeled.}

It is found that, for the mergers of asymmetric binaries (SFHo125-145,  120-150, and 125-155), the distribution of $Y_\mathrm{e}$ has double peaks; one of which at $Y_\mathrm{e}\approx 0.3$ produced primarily by the post-merger ejecta and the other by the dynamical ejecta.
By contrast, for the mergers of (nearly) symmetric binaries (SFHo135-135 and 130-140), the peak in the lower-$Y_\mathrm{e}$ side is not as prominent as that in the higher-$Y_\mathrm{e}$ side.
This reflects the fact that for the (nearly) symmetric case, the effect of the shock heating at the onset of the merger is so important that the temperature is enhanced significantly (Fig.~\ref{fig:hist_dyn}; middle).
This leads to the efficient electron-positron pair production and thus positron capture by neutrons is appreciably enhanced~\citep{Sekiguchi2015a}.

Other properties of ejecta, such as the mass histograms of velocity and entropy per baryon, are consistent with our previous findings for the viscous hydrodynamics simulations of the disks around black holes~\citep{Fujibayashi2020a,Fujibayashi2020b} as summarized in Appendices~\ref{app:criteria} and \ref{app:properties-of-ejecta}.

\subsection{Nucleosynthesis}

\subsubsection{Nuclear reaction network}
Using the time evolution of the thermodynamical quantities in the tracer particles for both 3D and 2D models, post-process nucleosynthesis calculations are performed with the nuclear reaction network code \texttt{rNET} described in \cite{Wanajo2018a}.
The network consists of 6300 isotopes with atomic number $Z=1$--110, which are connected with a set of relevant reactions.
Experimentally evaluated rates are adopted if they are available (JINA REACLIB V2.0,\footnote{\url{https://groups.nscl.msu.edu/jina/reaclib/db/index.php}} \citealt{Cyburt2010a}; Nuclear Wallet Cards\footnote{\url{http://www.nndc.bnl.gov/wallet/}}) and otherwise theoretical ones are adopted.
The theoretical rates for neutron, proton, and alpha captures \citep[TALLYS;][]{Goriely2008a} and beta-decays \citep[GT2;][]{Tachibana1990a} are based on a microscopic nuclear mass model \citep[HFB-21;][]{Goriely2010a}.
Theoretical spontaneous, beta-delayed, and neutron-induced fission rates are those predicted from the HFB-14 mass model \citep{Goriely2007a} with the fission fragment distributions adopted from the GEF model (\citealt{Schmidt2010may}; version 2021/1.1\footnote{\url{http://www.khschmidts-nuclear-web.eu/GEF-2021-1-1.html}}).
Neutrino-induced reactions are not included in the nucleosynthesis calculations, because they are expected to play only minor roles in our present models (except for setting the values of $Y_\mathrm{e}$ for $T\gtrsim \SI{10}{GK}$; see the bottom panel of Fig.~\ref{fig:ye-timescale}).

Each nucleosynthesis calculation starts when the temperature decreases to $\SI{10}{GK}$ with the initial composition of $1-Y_\mathrm{e}$ and $Y_\mathrm{e}$ for free neutrons and protons, respectively.
Because of the high temperature, the nuclear composition immediately settles into that in NSE after the beginning of the calculation.
Such a simple choice for the initial composition is justified from the fact that almost the entire ejecta, even the tidally expelled component of the dynamical ejecta, experience higher temperature than \SI{10}{GK} as shown in the top-right panel of Fig.~\ref{fig:hist_dyn}.

\subsubsection{Nucleosynthetic yields}
The bottom-left and bottom-middle panels of Fig.~\ref{fig:ye-dist-parameter} show the calculated nucleosynthetic yields for models SFHo135-135 and SFHo120-150.
Here, $Y(A)$ indicates the abundance (number per nucleon) of the nuclei with atomic mass number $A$.
For the equal-mass merger case, the nuclear abundance of the dynamical ejecta (shown in the blue curve) is in a reasonable agreement with that of the solar $r$-residuals with a small underproduction of the first peak of $r$-process nuclei~\citep[$A \sim 80$; as also found in][]{Wanajo2014a,Radice2018dec,Kullmann2022feb}.
On the other hand, as a result of the lower typical value of $Y_\mathrm{e}$, the first peak nuclei are more severely underproduced in the dynamical ejecta of the asymmetric merger.
However, because of the similar typical electron fraction with $Y_\mathrm{e}\approx 0.3$ (see the top-left and top-middle panels of Fig.~\ref{fig:ye-dist-parameter}), the post-merger ejecta for mergers of both equal-mass and asymmetric binaries have a similar abundance pattern (red curves) with a production mainly of the first peak nuclei, which compensates the underproduced first peak nuclei in the dynamical ejecta.
The ratio of post-merger to dynamical ejecta mass is larger for the merger of more asymmetric binaries (see Table~\ref{tab:model}), resulting in more contribution to the production of the first peak nuclei.
Hence, the total nucleosynthetic yield approximately reproduces the solar pattern for both equal-mass and asymmetric merger cases.

\begin{figure}
\includegraphics[width=0.47\textwidth]{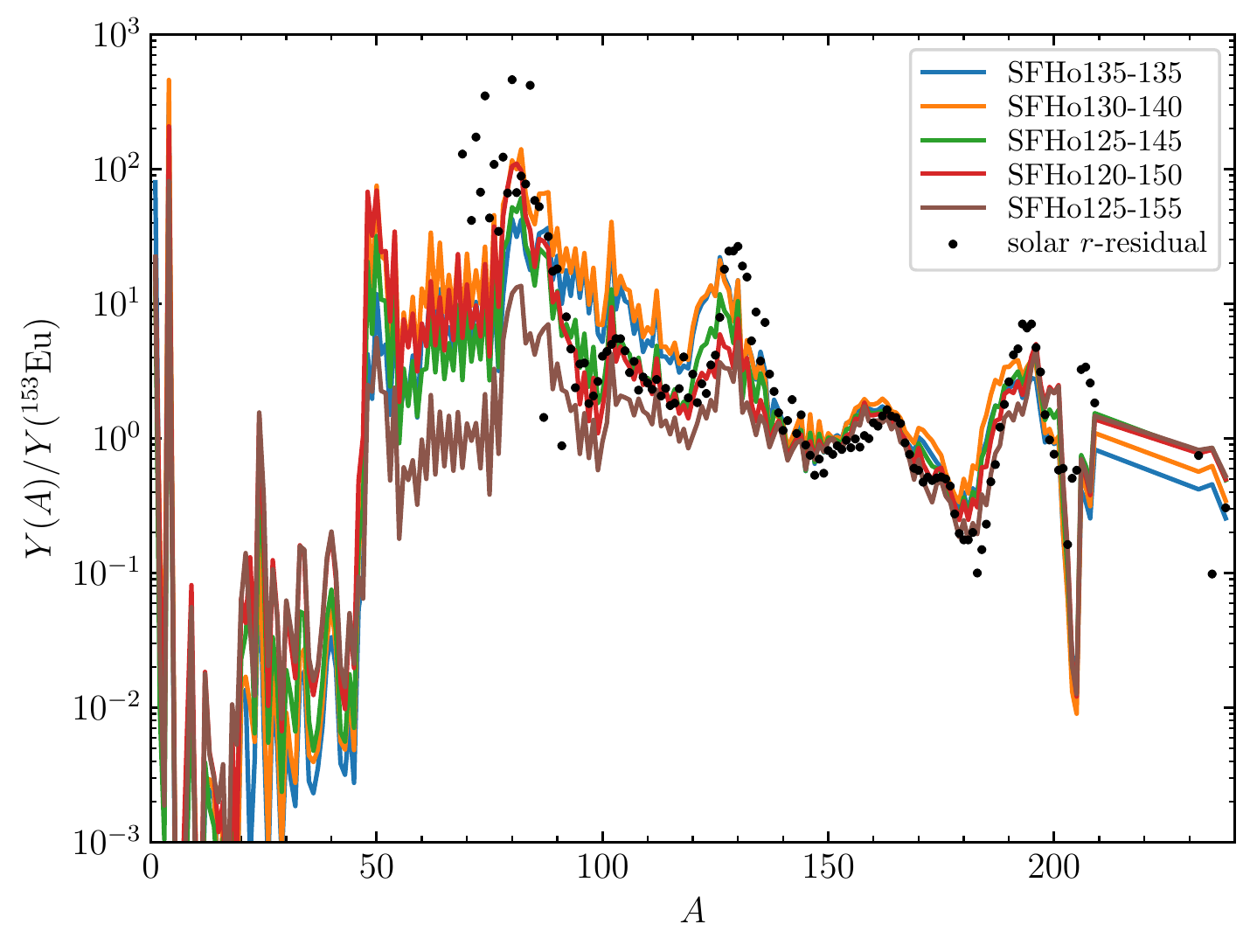}
\caption{
Total isobaric abundances for all the models, which are normalized by that of $^{153}$Eu.
The solar $r$-residuals are adopted from \citet{Prantzos2020a}.
}
\label{fig:aabun}
\end{figure}

\begin{figure}
\includegraphics[width=0.47\textwidth]{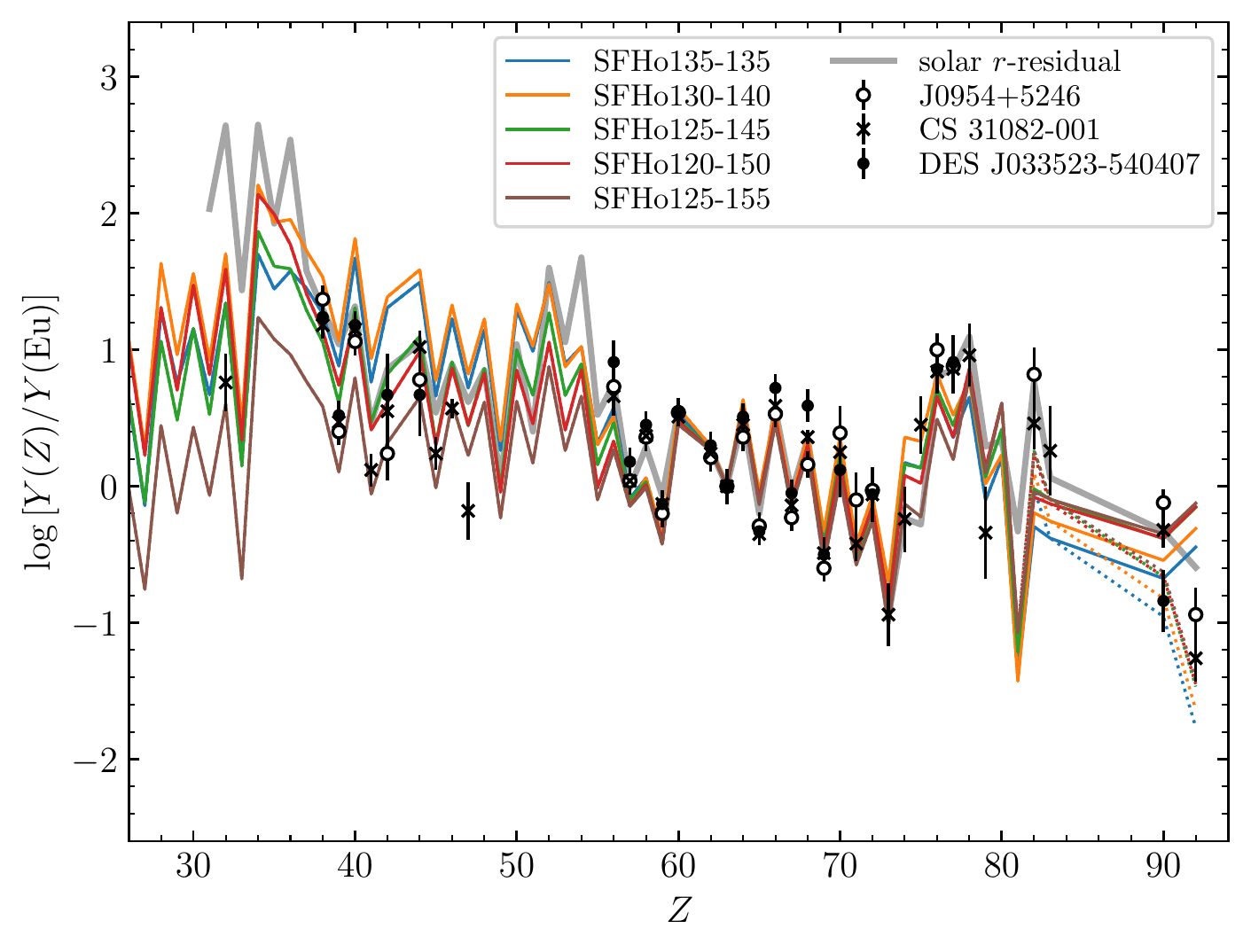}
\caption{
Total elemental abundance distributions for all the models.
The solid and dashed curves denote the distributions at the end of computation (1 yr) and at 13 Gyr, respectively (all trans-Pb nuclei except for Th and U are assumed to have decayed).
Stellar abundances of J0954+5246 (open circles; \citealt{Holmbeck2018jun}), CS~31082-001 (crosses; \citealt{Siqueira2013a}), and DES~J033523-540407 (filled circles; \citealt{Ji2018apr}) are also shown.
The grey curve denotes the solar $r$-residual pattern \citep{Prantzos2020a}.
Each distribution is normalized by that of Eu ($Z=63$).
}
\label{fig:zabun}
\end{figure}

Figure~\ref{fig:aabun} shows the total nucleosynthetic yields for all the models explored in this study.\footnote{The tables of nucleosynthetic yields are available upon request to the authors.}
It is found that the pattern of the solar $r$-residuals is reasonably reproduced irrespective of the mass ratio of the binaries \addsf{(typically within a factor of 2--3; see \S~\ref{sec:discussion})}, in particular for those between $A \sim 140$ and 200.
\addsf{This is qualitatively consistent with earlier work, e.g., \cite{Radice2018dec} and \cite{Kullmann2022feb}, although they only take into account the contribution from the dynamical ejecta.
}
However, there is a tendency that more asymmetric models lead to less production for $A < 140$ and more production for $A > 200$, respectively, in the abundances normalized by that of $^{153}$Eu (as representative of lanthanide nuclei).

In Fig~\ref{fig:zabun}, the elemental abundance distributions for all the models are compared to those measured in metal-poor stars J0954+5246 \citep[with the highest measured Th/Eu abundance ratio,][]{Holmbeck2018jun}, CS~31082-001~\citep{Siqueira2013a}, and DES~J033523-540407 \citep[with the lowest measured Th/Eu abundance ratio,][]{Ji2018apr}, which are enhanced in $r$-process elements.
Here, $Y(Z)$ is the abundance of the element with atomic number $Z$.
The calculated abundance patterns agree approximately with those for such $r$-process-enhanced metal-poor stars, in particular for the elements between $Z = 56$ and 79.
Our results exhibit \addsf{about a factor of 2--3} variation in the production of lighter elements, which can be also found in the $r$-process-enhanced stars \citep{Siqueira2014a,Holmbeck2020aug}.

Asymmetric mergers (SFHo125-145, 120-150, and 125-155) result in the higher ratio of actinide (Th and U) to Eu owing to the ejection of more matter with very low electron fraction $Y_\mathrm{e}\lesssim 0.1$.
The Th/Eu abundance ratio spans $-0.84 \le \log\, [Y(\mathrm{Th})/Y(\mathrm{Eu})] \le -0.63$ at 13 Gyr (given this being the ages of $r$-process-enhanced stars) after the merger for models investigated here.
Such a variation in the Th/Eu ratio can also be found in $r$-process-enhanced stars ($-0.95 \le \log\, [Y(\mathrm{Th})/Y(\mathrm{Eu})] \le -0.12$; see Fig.~\ref{fig:zabun}), although the enhancement of Th in our result is below the level of the so-called ``actinide-boosted" stars such as J0954+5246 and CS~31082-001.

The Th/Eu ratios for models SFHo125-145, 120-150 and 125-155 are very similar, although the fraction of the matter with $Y_\mathrm{e}<0.1$ in the dynamical ejecta for model SFHo125-155 is approximately three times larger than that for model SFHo125-145.
This implies that the Th/Eu ratio converges to $\log\, [Y(\mathrm{Th})/Y(\mathrm{Eu})] \approx -0.33$ (at \SI{1}{yr} after the merger) with a reduction of the binary mass ratio.
\addsf{Thus, this Th/Eu ratio} is likely to be the upper limit for binary neutron-star merger models with the SFHo EOS and the GEF fission-fragment distributions\addsf{, which cannot account for the ratios in actinide-boosted stars} \citep[for the  \addsf{dependence of nuclear data inputs to Th/Eu}, see][]{Holmbeck2019jan}.

\section{Discussion} \label{sec:discussion}
Here, we discuss the dependence of the nucleosynthetic outcomes of the binary neutron star mergers on the binary parameters, in particular on the lifetime of the remnant massive neutron stars (which depends on the total binary mass as well as the adopted EOS).
Figure~\ref{fig:ye-dist-parameter} (upper panels) summarizes the $Y_\mathrm{e}$ distribution of the ejecta of the binary neutron star mergers for three representative models with the total masses of $2.7\, M_\odot$; an equal-mass merger model leaving a hypermassive neutron star (model SFHo135-135; left panel), an asymmetric merger model leaving a hypermassive neutron star (model SFHo120-150; center panel), and an equal-mass merger model in which a massive neutron star survives for a timescale longer than \SI{10}{s} (model DD2-135 presented in \citealt{Fujibayashi2020c}; right panel).

For the equal-mass merger leaving a hypermassive neutron star (SFHo135-135), the dynamical ejecta has a broad $Y_\mathrm{e}$ distribution with the range between $\sim 0.05$ and $\sim 0.4$ and with a peak around $Y_\mathrm{e} \sim 0.3$.
This results in the abundance pattern in a reasonable agreement with that of the solar $r$-residuals except for a slight underproduction of the first peak nuclei ($A\approx80$).
On the other hand, for the asymmetric merger leaving a hypermassive neutron star (SFHo120-150), the distribution for the dynamical ejecta has a peak at a low value of $Y_\mathrm{e}<0.1$ and the mass with $Y_\mathrm{e} > 0.2$ is not very appreciable.
This results in the synthesis mainly of heavy $r$-process nuclei with $A > 100$ and a severe underproduction of the first peak nuclei.
By contrast, the post-merger ejecta have a peak at $Y_\mathrm{e}\approx 0.3$ irrespective of the binary parameter (although the distribution could depend weakly on the model of the angular momentum transport, i.e., prescription and parameter of the viscosity, or model with magnetohydrodynamics).
As a result, the lighter $r$-process nuclei are the main products in the post-merger ejecta.

Importantly, the order of magnitude for the mass of the post-merger ejecta is always comparable to that of the dynamical ejecta for the models leaving hypermassive neutron stars (see Table~\ref{tab:model}).
In addition, the contribution of the post-merger ejecta is larger for the more asymmetric binaries.
Hence, the post-merger ejecta supply a larger amount of light $r$-process nuclei for more asymmetric mergers.
As a consequence of the fact that the under-abundance of lighter nuclei in the dynamical ejecta is compensated by the contribution of the post-merger ejecta, \addsf{a solar-like} $r$-process nucleosynthetic distribution is reasonably achieved irrespective of the binary mass ratio.

By contrast, for the case in which a long-lived massive neutron star survives (DD2-135), the mass of the post-merger ejecta is by about two orders of magnitude larger than that of the dynamical ejecta ($M_\mathrm{post}/M_\mathrm{dyn}=O(100)$).
Hence, the lighter $r$-process nuclei are overproduced, and as a result, \addsf{a solar-like} pattern cannot be reproduced.
Such an overproduction of lighter $r$-process nuclei is found \addsf{for all binaries that leave long-lived massive neutron stars} as investigated in \citet{Fujibayashi2020c}.
\addsf{Note that the abundance variation owing to such a fundamental difference of $M_\mathrm{post}/M_\mathrm{dyn}$ (originating from the adopted EOS or the assumed total binary mass) appears more prominent than the local modification of abundance distribution due to different EOSs \citep[in dynamical ejecta,][]{Radice2018dec,Kullmann2022feb}.}

These results naturally lead to the following conclusion: Suppose that binary neutron star mergers are the main site for the $r$-process nucleosynthesis in the universe.
Then the events with the formation of long-lived remnant neutron stars must be a minority among entire binary neutron star mergers: otherwise \addsf{a solar-like $r$-process pattern} cannot be reproduced.
This suggests that the events with the formation of magnetars (long-lived magnetized neutron stars as a remnant) should be also a minority of binary neutron star mergers and that the magnetar scenario for short gamma-ray bursts \citep[][]{Usov1992jun,Metzger2008apr} might be applicable only to such a sub-class of mergers. 
\addsf{This also implies that a relatively softer EOS such as SFHo (rather than DD2) is favored to reproduce a solar-like $r$-process pattern, given that the typical total binary mass is $\sim 2.7\, M_\odot$.}

\addsf{

\begin{figure}
\includegraphics[width=0.47\textwidth]{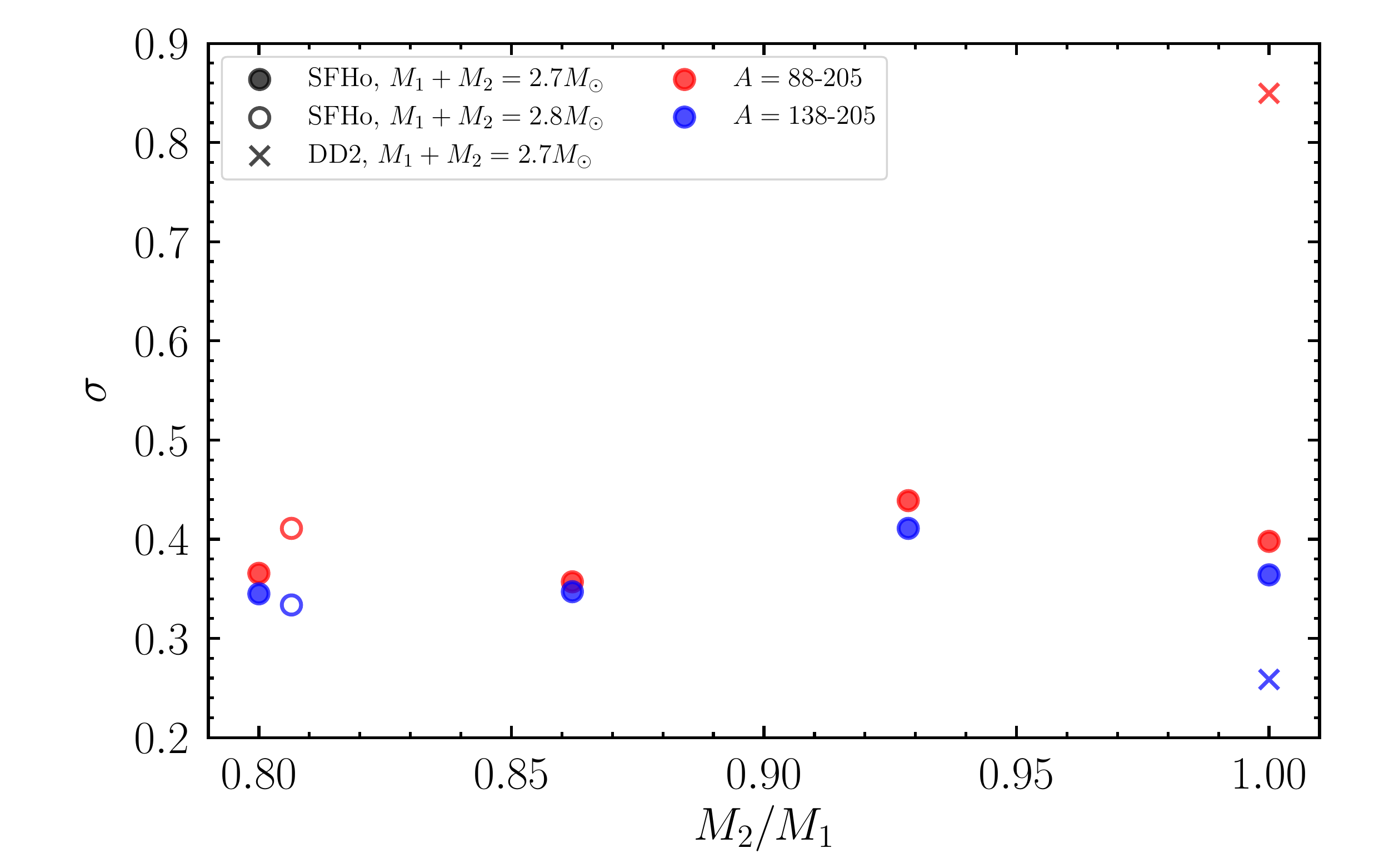}
\caption{
Average logarithmic deviations from solar $r$-residual pattern for nucleosynthetic yields of models investigated in this study as well as those for a long-lived massive neutron star \citep[DD2-135 in][]{Fujibayashi2020c}.
The filled and open markers denote the results for the total masses of 2.7 and 2.8$M_\odot$, respectively.
The results for ranges $A=88$--205 and $A=138$--205 in Eq.~\eqref{eq:dev} are shown with red and blue markers, respectively.
}
\label{fig:dev}
\end{figure}

To quantify the level of agreement between the patterns of nucleosynthetic yields and solar $r$-residuals, we define the average logarithmic deviation, $\sigma$, of a given yield $N(A)$ from the solar $r$-residual $N_\odot(A)$ by
\begin{align}
\sigma^2 &= \frac{1}{n_\mathrm{tot}}\sum_{A}\biggl(\log_{10}  N(A) - \log_{10} N_\odot(A)\biggr)^2. \label{eq:dev}
\end{align}
Figure~\ref{fig:dev} compares the deviations $\sigma$ for all the models investigated in this study (with SFHo EOS; circles) as well as the model in which a long-lived neutron star remains after the merger (with DD2 EOS; crosses).
If we limit the summation in Eq.~\eqref{eq:dev} to $A=138$--205 ($n_\mathrm{tot}=68$), for which the dynamical ejecta component is responsible (Fig.~\ref{fig:ye-dist-parameter}), the deviation ranges within $\sigma =$ 0.25--0.45 for all the models (shown in blue), irrespective of the adopted EOS or the total binary mass.
This indicates that their yields agree with the pattern of solar $r$-residuals within a factor of 2--3.
Even if we consider $A=88$--205 by including lighter nuclei produced in the post-merger ejecta ($n_\mathrm{tot}=118$), the deviation for the models explored in this study (red open and filled circles) still stay in the range of $\sigma = 0.35$--0.45.
By contrast, for the model with a long-lived neutron star, we obtain $\sigma = 0.85$ (deviation by a factor of seven).
Given a factor of 2--3 deviation being the acceptable range as a common practice, all the models explored in this study (with SFHo) may reasonably represent neutron star mergers as the dominant sources of $r$-process nuclei.
By contrast, the deviation for the model with DD2 is too large to represent typical merger events, although this can be a possible explanation for a weak $r$-process signature found in some $r$-process-deficient stars \citep{Honda2006a,Fujibayashi2020c}.

\begin{figure}
\includegraphics[width=0.47\textwidth]{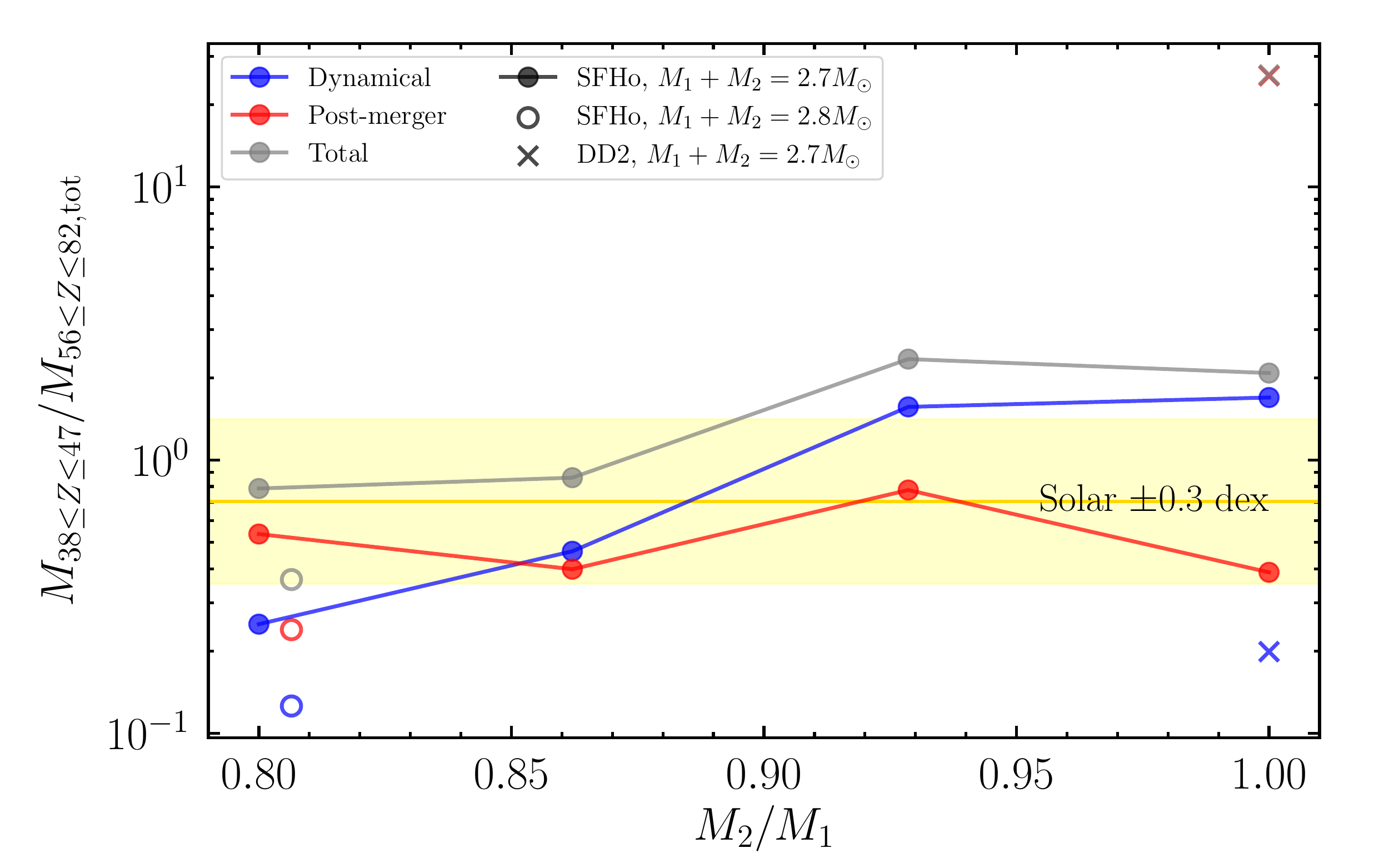}
\caption{
Mass ratios of light ($38\le Z\le 47$) to heavy ($56 \le Z \le 82$) elements for dynamical (blue), post-merger (red), and total (grey) ejecta.
The lines indicate the results for the models with the SFHo EOS and $M_1+M_2=2.7M_\odot$.
The open circles and crosses indicate the results for model SFHo125-155 and the DD2 EOS, respectively.
Note that the denominator is always the total mass of heavy elements $M_{56\le Z\le 82,\mathrm{tot}}$.
The yellow band indicates the mass ratio for the solar $r$-residuals with $\pm0.3$~dex, which is the level of scatter in the abundance ratios of light-to-heavy $r$-process elements for $r$-process-enhanced metal-poor stars (\citealt{Cowan2021jan,Holmbeck2020aug}).
}
\label{fig:light-heavy-ratio}
\end{figure}

To quantitatively discuss how lighter $r$-process nuclei are compensated by post-merger ejecta, the mass ratios of lighter (defined by $38\le Z\le47$) to heavier ($56 \le Z\le82$) $r$-process elements (for those with sufficient observational data) in the ejecta components are compared in Fig.~\ref{fig:light-heavy-ratio}.
Here, the denominator $M_{56 \le Z\le 82,\mathrm{tot}}$ is always the mass of the heavier $r$-process elements in the total ejecta, which is dominated by the dynamical ejecta.
Note that the measured light-to-heavy abundance ratios of $r$-process-enhanced stars exhibit about a factor of two variation from the solar ratio \citep{Cowan2021jan,Holmbeck2020aug}, which is indicated by the yellow band ($\pm 0.3$~dex) in Fig.~\ref{fig:light-heavy-ratio}.
For the model with $M_1+M_2=2.7M_\odot$, the ratio for the dynamical ejecta (shown in blue) steeply decreases with an increasing mass asymmetry.
On the other hand, the ratio for the post-merger ejecta (red) is relatively insensitive to mass asymmetry.
As a result, the ratio for the total ejecta (grey) mildly decreases with an increasing mass asymmetry, being in good agreement with the solar ratio for SHFo120-150 and SFHo125-145 and three times higher than that for SFHo130-140 and SFHo135-135.
For the massive model SFHo125-155 ($M_1+M_2=2.8M_\odot$), which leaves the shortest-lived massive neutron star among the models investigated in this study, the ratio for the total ejecta is a factor of two smaller than that for the solar $r$-residuals.
Thus, for the models with short-lived massive neutron stars explored in this study (the gray line and circle), the scatter level of light-to-heavy $r$-process elements are in reasonable agreement with (or slightly higher than) that observed in $r$-process-enhanced stars.
This implies that such a scatter can be attributed to a variation of the mass asymmetry as well as of the total mass.
The model with a long-lived neutron star (DD2-135) shows, on the other hand, more than one order of magnitude larger ratio than the solar (or observational) ratio.
Thus, the population of mergers leaving long-lived massive neutron stars should be subdominant in all merger events.
}

A word of caution is appropriate here, because in the present study, we do not take into account the effects of the neutrino oscillation.
The latest numerical simulations (e.g., \citealt{Li2021jun} and \citealt{Just2022mar}) indicate that the value of $Y_\mathrm{e}$ for the post-merger ejecta may be decreased by up to 0.03.
In such a case, the production of the heavy $r$-process elements are enhanced while that of the lighter one is reduced.
Such a slight decrease in $Y_\mathrm{e}$ for the post-merger ejecta does not change the conclusion in this paper.
However, in the presence of a long-lived massive neutron star as a remnant, the neutrino-oscillation effect may be significant because of a longer-term neutrino irradiation.
Moreover, the quantitative effect of the neutrino oscillation is not well understood, because the current studies are based on an approximation.
Thus, it should be kept in mind that the future more detailed study for the mass ejecion taking into account the neutrino oscillation may change the conclusion in this paper.
It is also important to note that nuclear ingredients such as the rates of neutron capture and $\beta$-decay (and thus nuclear mass models) affect local abundance distributions typically with a factor of 2--3 level \citep[e.g.,][]{Vassh2021feb}. In addition, for the cases in which neutron-rich matter ($Y_\mathrm{e} \lesssim 0.1$) dominates the dynamical ejecta (SFHo120-150 and SFHo125-155), fission recycling becomes important for determining the nucleosynthetic abundance distribution \citep{Goriely2011sep,korobkin2012nov,Eichler2015jul,Goriely2015feb,Holmbeck2019jan,Vassh2020jun,Lemaitre2021feb}. A choice of different set of nuclear data inputs may affect the agreement levels of nucleosynthetic abundances with the solar pattern to some extent.

\section{Summary} \label{sec:summary}
We performed a set of 3D numerical relativity simulations for the mergers of binary neutron stars leading to hypermassive neutron stars and subsequently axisymmetric simulations for the post-merger phase to investigate the properties of the ejecta in a self-consistent manner.
In the post-merger phase, the viscosity was taken into account to model the angular momentum transport due to the turbulent motion developed by the magnetohydrodynamical processes.
In this work, we considered binaries with the total masses of $2.7M_\odot$ and $2.8M_\odot$, which are canonical values based on the mass distribution of the Galactic binary neutron stars \citep{Tauris2017a}.
We used an EOS of nuclear matter referred to as SFHo.
In this setup, we investigated the cases in which the merger remnant is a short-lived massive neutron star that collapses into a black hole in $\sim \SI{20}{ms}$.
This is one of the differences from our previous work with another EOS (DD2), in which we considered the case that the remnant is a long-lived massive neutron star~\citep{Fujibayashi2020c}.
We considered a wide range of the mass ratio for the merging binary and investigated the dependence of the ejecta properties on the mass ratio.

We reconfirmed that the $Y_\mathrm{e}$ distribution of the dynamical ejecta depends on the mass ratio~\citep{Sekiguchi2016a,Radice2018dec}.
The equal-mass merger produces the nuclei with the abundance pattern close to the solar pattern as shown in \citet{Wanajo2014a,Radice2018dec,Kullmann2022feb}.
On the other hand, a more asymmetric binary merger results in a more neutron-rich distribution of ejecta ~\citep{Sekiguchi2016a} because of the enhanced importance of tidal interaction for the mass ejection.
As a result, heavier $r$-process nuclei are mainly synthesized and lighter $r$-process nuclei are underproduced.
However, a more asymmetric binary merger results in the formation of more massive disk, and hence, the mass of the post-merger ejecta, which are composed primarily of moderately neutron-rich matter with $Y_\mathrm{e} \sim 0.3$, can be larger.
Because the lighter $r$-process nuclei are the main products in such a condition, summing up the dynamical and post-merger ejecta naturally results in the approximate reproduction of the solar $r$-process pattern within the mean deviation of a factor of three.
The light-to-heavy abundance scatter (by a factor of a few) observed in $r$-process-enhanced stars can also be reasonably reproduced as a result of variation in the binary mass ratio and total mass.

There are rooms to improve the present study.
In particular we have to keep in mind that the model of the angular momentum transport in this work assumes the constant length scale of the turbulent eddies ($\alpha_\mathrm{vis}H_\mathrm{tur}$), and it might be too simple. For more realistic modeling for the angular momentum transport, magnetohydrodynamics instead of viscous hydrodynamics would be necessary. 
A high-resolution magnetohydrodynamics simulation for a binary neutron star merger and its remnant~\citep{kiuchi2018a} suggests that the actual angular momentum transport in the disk is more complicated, and hence, the electron fraction of the post-merger ejecta may have a different $Y_\mathrm{e}$ distribution from those obtained in this work.
Indeed, in a recent general relativistic radiation magnetohydrodynamics simulation for the merger of black hole-neutron star binaries~\citep{Hayashi2021} predicts a slightly broader $Y_\mathrm{e}$ distribution than that predicted with viscous hydrodynamics models for the post-merger ejecta, although other properties of the post-merger ejecta, such as the velocity and the mass (per the disk mass), are consistent with the present results \citep[see also][]{Just2022}. Thus, to obtain more quantitative ejecta composition, a more sophisticated modeling of angular momentum transport processes might be necessary, although the viscous hydrodynamics results for the properties of the post-merger ejecta would be in a broad agreement with the magnetohdyrodynamics ones. 
\addsf{The dynamics of ejecta may also be affected by radioactive energy released during $r$-processing, which is not taken into account in our simulations.
This effect can lead to an ejection of the marginally bound matter during both the dynamical and post-merger phases~\citep[e.g.,][]{foucart2021dec}.
Inclusion of $r$-process heating in merger simulations will be necessary in our future work (but see Appendix~\ref{app:criteria}).
}

\acknowledgements

This work was in part supported by Grant-in-Aid for Scientific Research (Grant Nos.~JP20H00158, 19K14720, 22K03617) of Japanese MEXT/JSPS. Numerical computations were performed at Sakura, Cobra, and Raven of the Max Planck Computing and Data Facility, Oakforest-PACS at Information Technology Center of the University of Tokyo, XC50 at National Astronomical Observatory of Japan, and Yukawa21 at Yukawa Institute for Theoretical Physics.

\appendix

\section{Other ejecta criteria}
\label{app:criteria}
There are several proposed criteria for determining unbound matter in general relativistic hydrodynamics simulations in the literatures. In this appendix we summarize them and compare the results among the different criteria.

In the so-called geodesic criterion, the condition for unbound matter is written by
\begin{align}
u_t + 1 < 0. \label{eq:criterion-geodesic}
\end{align}
This is appropriate in the presence of a time-like Killing vector and in the case that thermal effects of the ejecta are negligible.

The so-called Bernoulli's criterion provides another condition taking the thermal effects of the matter into account. If the stationary flow is assumed in addition to the presence of a time-like Killing vector, the Bernoulli's argument ($h u_t$ is constant along a stream line) predicts the asymptotic Lorentz factor $\Gamma_\infty = -u_{t,\infty} = -hu_t/h_\infty$, where $h_\infty$ is the asymptotic specific enthalpy of the matter. 
The condition for unbound matter can be then  written as
\begin{align}
\Gamma_\infty = -\frac{hu_t}{h_\infty} > 1 \iff h u_t + h_\infty < 0. \label{eq:criterion-bernoulli}
\end{align}

We note that the specific internal energy $\varepsilon$ inside $h$ ($=c^2 + \varepsilon + P/\rho$) includes the contribution of the nuclear binding energy, which is written as $\langle \Delta m\rangle c^2/ m_\mathrm{u}$, 
where $\langle \Delta m\rangle$ is the average mass excess per baryon.
Here, we define the atomic mass unit $m_\mathrm{u}(\approx \SI{931}{MeV}/c^2)$ as the reference mass of baryons.
Thus, the criterion in Eq.~\eqref{eq:criterion-bernoulli} depends on the final product of the nucleosynthesis through the asymptotic value of the specific internal energy $\varepsilon_\infty$.
If the effects of nuclear burning are not fully taken into account in a simulation, one should assume the final product of the nucleosynthesis and resulting $\varepsilon_\infty$.

\begin{figure}
\begin{center}
\includegraphics[width=0.47\textwidth]{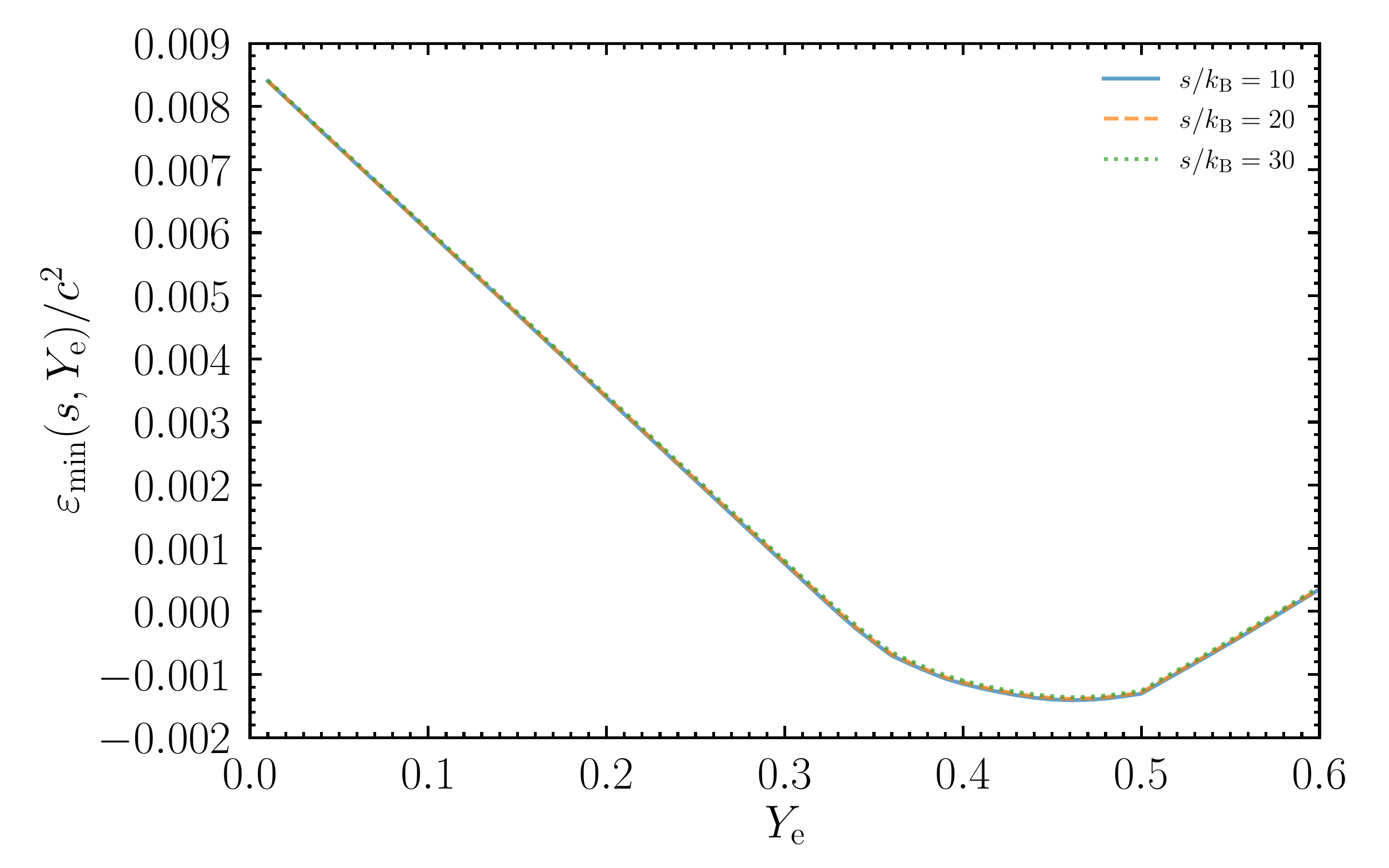}
\end{center}
\caption{
The minimum specific internal energy in the extended SFHo EOS table as a function of the electron fraction for entropies per baryon of $s/k_\mathrm{B}=10$, 20, and 30.
}
\label{fig:epsilon-min}
\end{figure}

Let us first assume that the matter is in NSE, which determines the composition. In addition, we assume that the matter expands adiabatically (i.e., the entropy per baryon, $s$, is constant) without any weak interaction processes (i.e., $Y_\mathrm{e}$ is constant).
In this assumption, the composition of the ejected matter in the asymptotic region is determined  by the NSE condition at the low-temperature limit at a given set of $s$ and $Y_\mathrm{e}$.
The assumption is consistent with our formulation of hydrodynamics, and thus, this choice for $\varepsilon_\infty$ gives the condition for the unbound matter in our simulation.
Practically, $\varepsilon_\infty$ in this case can be found as the minimum value of $\varepsilon$ in the adopted EOS table (in which the NSE is assumed in its construction) for given values of $s$ and $Y_\mathrm{e}$. We refer to it as  $\varepsilon_\mathrm{min}(s,Y_\mathrm{e})$ (we here assume that the nuclear binding energy dominates the specific internal energy at the temperature we consider).
Figure~\ref{fig:epsilon-min} shows $\varepsilon_\infty=\varepsilon_\mathrm{min}(s,Y_\mathrm{e})$ as a function of $Y_\mathrm{e}$ for selected entropy per baryon for the SFHo EOS.
We note that $\varepsilon_\mathrm{min}(s,Y_\mathrm{e})$ depends only very weakly on the entropy because the nuclear composition does not depend on the entropy at the low-temperature limit.\footnote{The free energy ($:=\varepsilon-Ts/m_\mathrm{u}$) does not depend strongly on the entropy at a low temperature ($T\ll \varepsilon m_\mathrm{u}/s)$.
Thus, the nuclear composition in NSE, which is determined to satisfy the minimum free energy, does not also depend strongly on the entropy.}

The above choice for $\varepsilon_\infty$ ignores the possible effects of non-equilibrium nuclear burning, i.e., $r$-process in our case.
The $r$-process modifies the nuclear composition and thus, $\varepsilon_\infty$.
This implies that the bound matter in the simulation can still become unbound with this additional energy.
The asymptotic value $\varepsilon_\infty$ should be determined by performing the nucleosynthesis calculation.
Otherwise it has to be assumed with a physical consideration. The effect of the nuclear burning for making matter unbound is the most significant if we assume $\varepsilon_\infty = \varepsilon_\mathrm{min,global}$, which is the globally minimum specific internal energy in the employed EOS table (for the SFHo EOS, $h_\mathrm{min,global} := c^2+\varepsilon_\mathrm{min,global} \approx 0.9987c^2$; see \S~\ref{subsubsec:post-merger}).
This is equivalent to assuming that the final product of the nucleosynthesis is always iron, which is the most stable nuclear species. This choice for $\varepsilon_\infty$ may give a too simple criterion for the unbound matter.
In reality, the final product of the $r$-process, which depends on $Y_\mathrm{e}$ of the matter, is not in general as stable as iron and $\varepsilon_\infty$ is larger than $\varepsilon_\mathrm{min,global}$.
In addition, a fraction of released rest-mass energy during the nucleosynthesis is carried away by neutrinos.
Thus, the rest-mass energy per baryon available for unbounding matter is smaller than the difference between $\varepsilon_\mathrm{min}(s,Y_\mathrm{e})$ and $\varepsilon_\mathrm{min,global}$.

A criterion of the unbound matter approximately taking the effect described above into account is proposed in \cite{foucart2021dec}, which is
\begin{align}
\Gamma_\infty = -\frac{h u_t}{h_\infty} (1-f_\mathrm{loss}) > 1, \label{eq:criterion-ff21}
\end{align}
where $f_\mathrm{loss} = \mathrm{max}(0, 0.0032 - 0.0085Y_\mathrm{e})$ is the effect of energy loss by neutrinos, which is a function of $Y_\mathrm{e}$, and $h_\infty \approx c^2$ is the asymptotic specific enthalpy for the matter that experienced the $r$-process.

\begin{figure}
\includegraphics[width=0.48\textwidth]{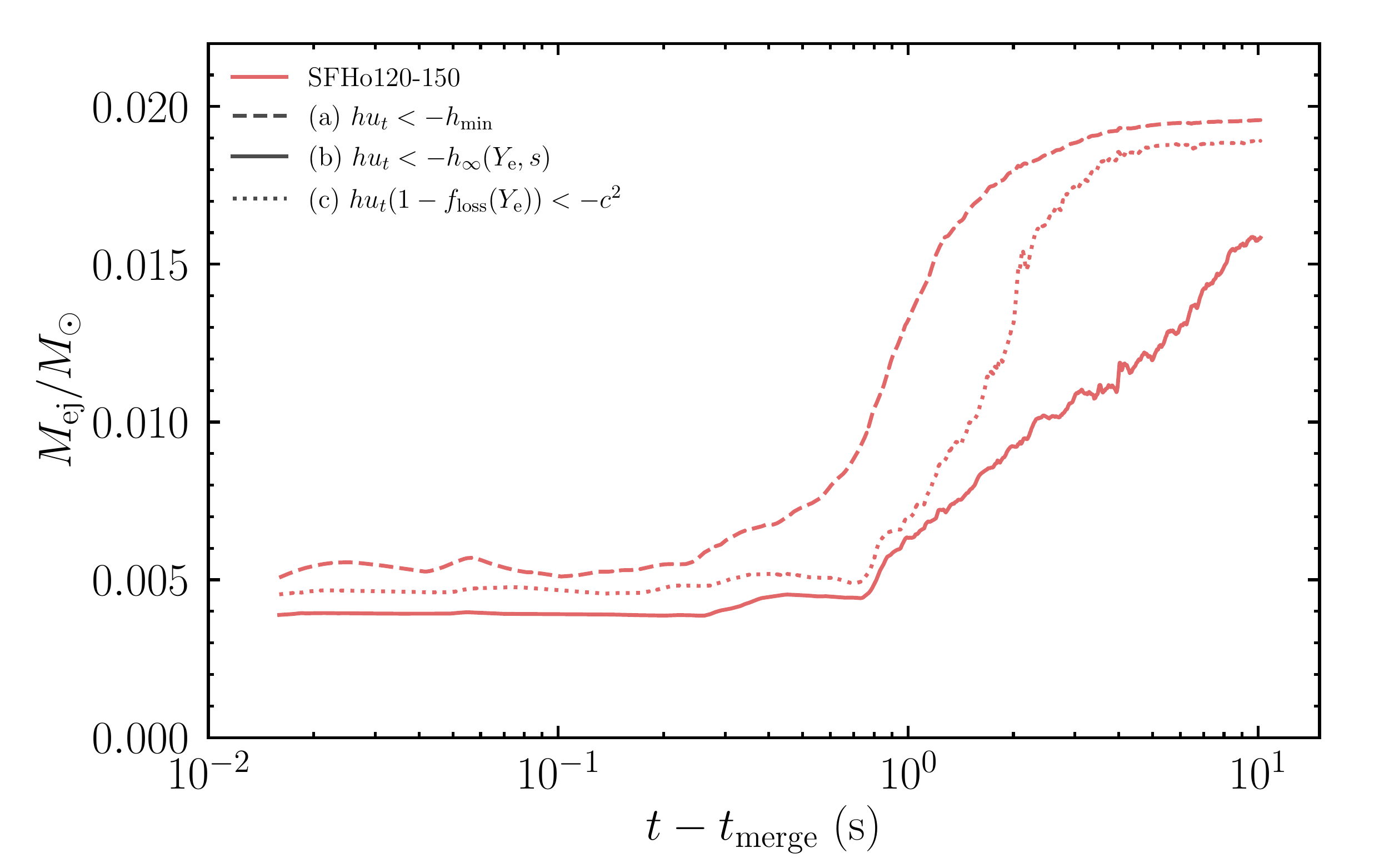}~~
\includegraphics[width=0.48\textwidth]{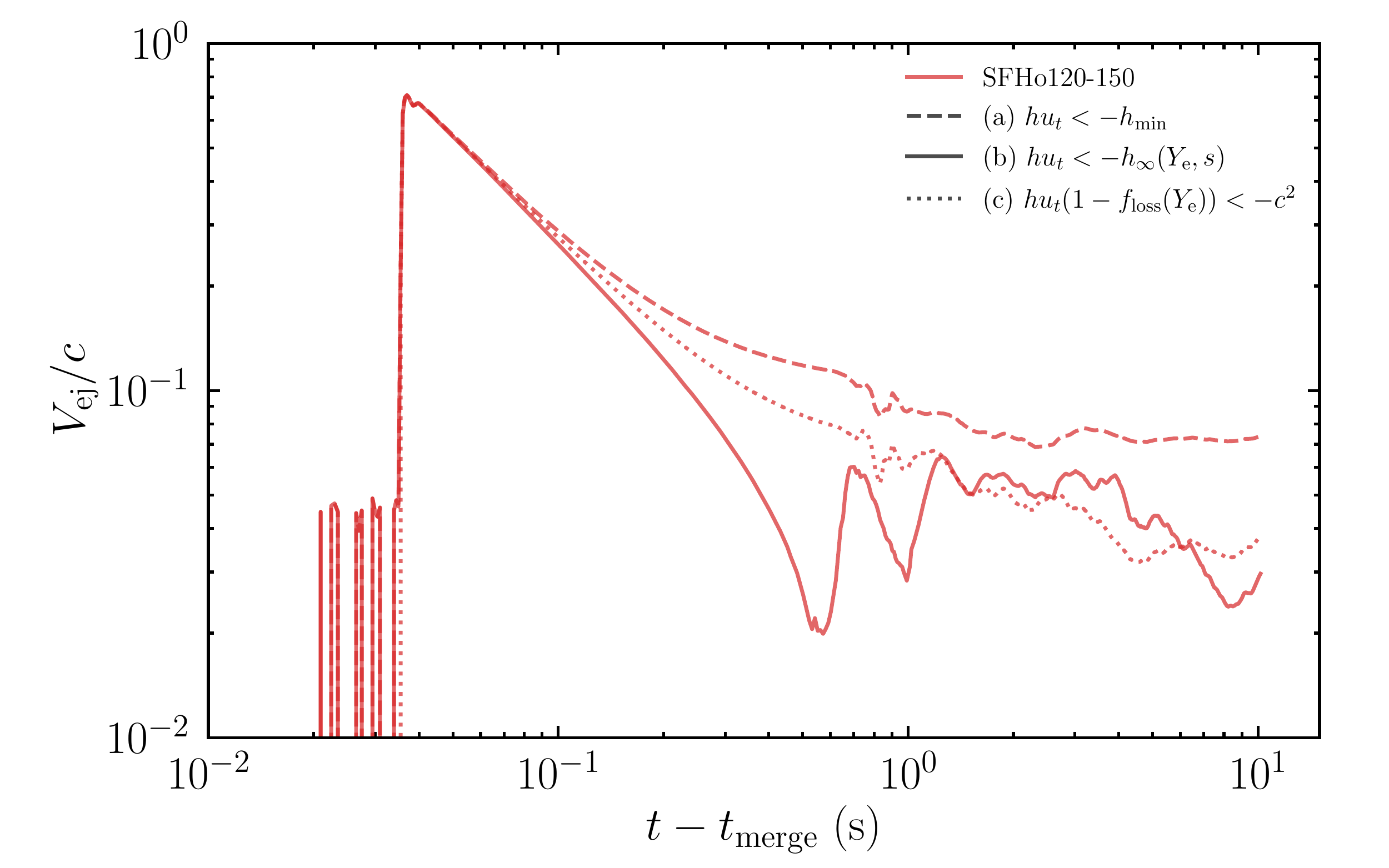}
\caption{
Left: Time evolution of the ejecta mass defined by Eq.~\eqref{eq:ejecta-mass} with different criteria for unbound matter.
Right: Asymptotic velocity of the unbound matter that passes through the extraction radius at a given time with the different criteria.
See text in Appendix~\ref{app:criteria} for the details of the criteria.
For both panels, the results of model SFHo120-150 are plotted. 
}
\label{fig:criterion}
\end{figure}

The time evolution of the ejecta masses defined by
\begin{enumerate}[\textrm{criterion} (a) : ]
\item Equation~\eqref{eq:criterion-bernoulli} with $h_\infty=h_\mathrm{min,global}$,
\item Equation~\eqref{eq:criterion-bernoulli} with $h_\infty=h_\mathrm{min}(s,Y_\mathrm{e}) := c^2+\varepsilon_\mathrm{min}(s,Y_\mathrm{e})$,
\item Equation~\eqref{eq:criterion-ff21} with $h_\infty=c^2$
\end{enumerate}
are compared in the left panel of Fig.~\ref{fig:criterion}.
Here the result of model SFHo120-150, in which the contribution of the post-merger ejecta is the largest among the models investigated in this paper, is used.
The criterion (a) gives the largest ejecta mass at any time because it maximally takes into account the effect of the possible release of nuclear binding energy.
On the other hand, the criterion (b)
gives somewhat smaller ejecta mass because it does not take the non-equilibrium nuclear burning into account.
The ejecta mass with the criterion (c) agrees approximately with that for the criterion (a), indicating that it may be acceptable to use the criterion (a) for estimating ejecta mass in the simulation.

Different criteria for the unbound matter lead to different asymptotic Lorentz factor, or velocity of the post-merger ejecta as shown below.
In the presence of a time-like Killing vector, $\xi^\nu T_\nu{}^\mu = T_t{}^\mu$ satisfies the conservation equation $\del_\mu (\sqrt{-g}T_t{}^\mu)=0$, and there exists a globally conserved energy
\begin{align}
\int_V \sqrt{-g}(-T_t{}^t) d^3x,
\end{align}
which satisfies 
\begin{align}
\del_t \int_V \sqrt{-g}(-T_t{}^t) d^3x = - \int_S \sqrt{-g}(-T_t{}^k)ds_k,
\end{align}
where $V$ is a spatial domain and $S$ is its surface.
Hereafter, the domain $V$ is assumed to be a spherical domain with a boundary of a radius $r_\mathrm{ext}$. For the perfect fluid, the outflow rate of the energy of the unbound matter from the domain is written as
\begin{align}
\dot{E}_\mathrm{tot,ej}=-\int_{r=r_\mathrm{ext}} \sqrt{-g}\rho u^k (-h u_t) \Theta(\Gamma_\infty-1) ds_k = \int_{r=r_\mathrm{ext}} \sqrt{-g}\rho u^k h_\infty \Gamma_\infty \Theta(\Gamma_\infty-1) ds_k.
\end{align}
To derive the final expression, the Bernoulli's argument is used under the assumption of the stationary flow. The mass ejection rate of the asymptotic rest mass is estimated by
\begin{align}
\dot{M}_\mathrm{ej,\infty} &= \frac{1}{c^2}\int_{r=r_\mathrm{ext}} \sqrt{-g} \rho u^k  h_\infty\Theta(\Gamma_\infty - 1) ds_k. \label{eq:rest-mass-ejecta}
\end{align}
\addsf{We note that the quantity defined by Eq.~\eqref{eq:mdot-baryon-mass} is not strictly the mass but the baryon number multiplied by the reference mass $m_\mathrm{u}$.}
The factor $h_\infty/c^2$ in Eq.~\eqref{eq:rest-mass-ejecta} corrects the difference between the reference mass $m_\mathrm{u}$ and the asymptotic mass per baryon $(1+\varepsilon_\infty/c^2) m_\mathrm{u}=h_\infty m_\mathrm{u}/c^2$.
From $\dot{E}_\mathrm{tot,ej}$ and $\dot{M}_\mathrm{ej,\infty}$, the ejection rate of the asymptotic kinetic energy of the ejecta is written as
\begin{align}
\dot{K}_\mathrm{ej} &= \dot{E}_\mathrm{tot,ej} - \dot{M}_\mathrm{ej,\infty}c^2\notag\\
&= \int \sqrt{-g} \rho u^k h_\infty (\Gamma_\infty-1) \Theta(\Gamma_\infty - 1) ds_k.\label{eq:asy-kin}
\end{align}
We define the asymptotic velocity of the ejecta that pass through a extraction radius at a given time by
\begin{align}
V_\mathrm{ej} = c\sqrt{1-\biggl(\frac{\dot{K}_\mathrm{ej,\infty}}{\dot{M}_\mathrm{ej,\infty}c^2}\biggr)^{-2}}, \label{eq:vej}
\end{align}
where $\dot{K}_\mathrm{ej,\infty}/(\dot{M}_\mathrm{ej,\infty}c^2)$ is the average asymptotic Lorentz factor of the ejecta.

It is found in Eq.~\eqref{eq:asy-kin} that the asymptotic kinetic energy depends on $\Gamma_\infty$ and hence on the criteria of unbound matter.
The right panel of Fig.~\ref{fig:criterion} shows $V_\mathrm{ej}$ for different criteria (a)--(c).
The asymptotic velocity of the dynamical ejecta, which pass through the extraction radius for $t\lesssim \SI{0.3}{s}$, has only a minor difference among the different criteria, reflecting its dominant kinetic energy over the internal energy.
On the other hand, the different criteria give different asymptotic velocities of the post-merger ejecta, which are ejected for $t\gtrsim \SI{1}{s}$.
The criterion (a) gives the largest $\approx0.07$--0.09$c$, while the other criteria (b) and (c) give smaller values $\approx0.02$--0.06$c$.
If the post-merger ejecta have a dominant contribution to the energy source by radioactive decay, the characteristics of the kilonova may be sensitive to its velocity.
Hence, there may be a systematic uncertainty of kilonova models from the estimation of the ejecta velocity. We will investigate the effect of the heating due to the non-equilibrium nuclear burning in detail in the future.

\section{Other properties of the ejecta}
\label{app:properties-of-ejecta}
\begin{figure}
\includegraphics[width=0.47\textwidth]{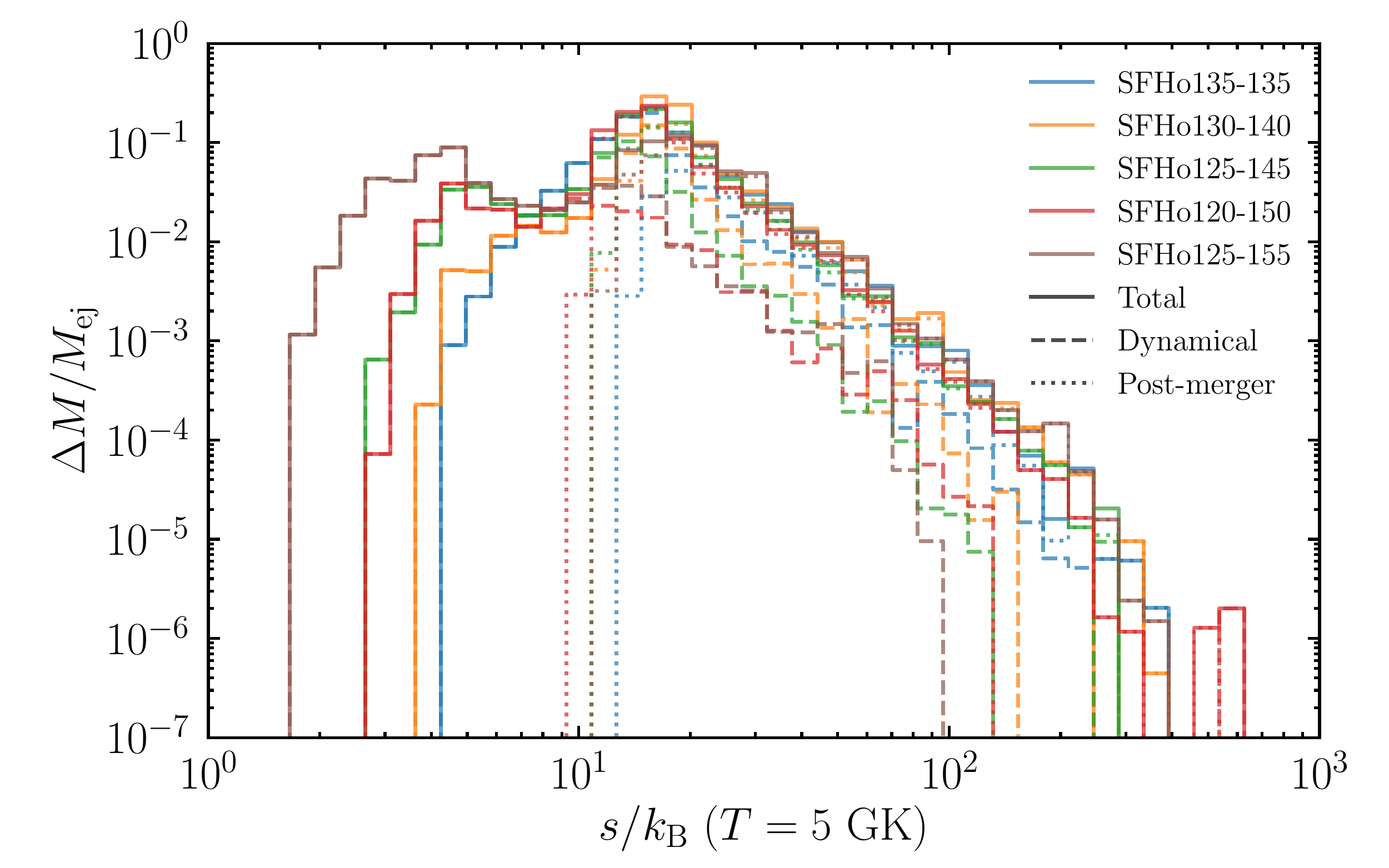}
\includegraphics[width=0.47\textwidth]{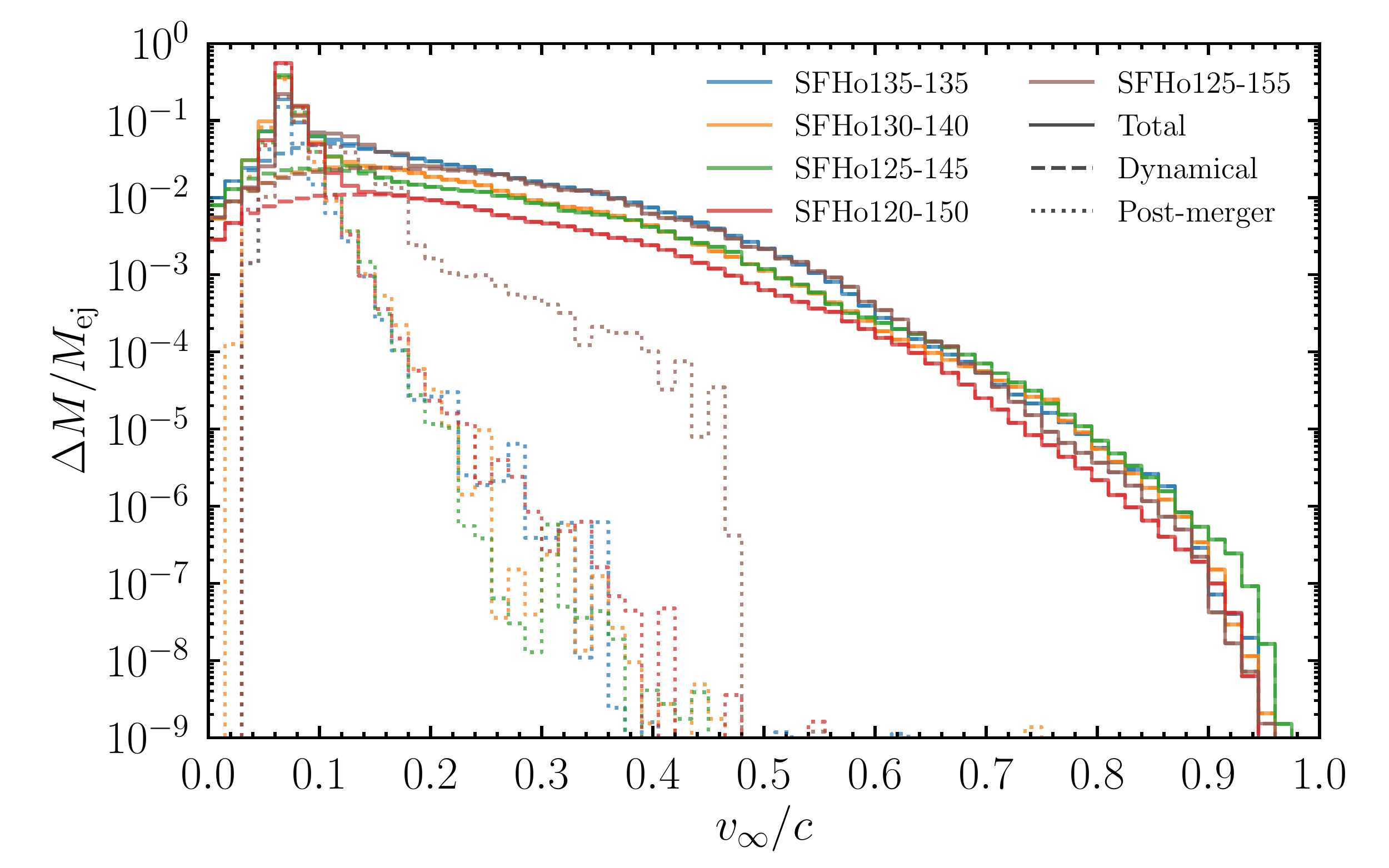}
\caption{
Mass histograms of entropy per baryon (left) and asymptotic velocity (right).
The solid, dashed, and dotted lines are histograms for total, dynamical, and post-merger ejecta, respectively.
For the velocity distributions of the dynamical ejecta, we use the same time slices as those used in the bottom panel of Fig.~\ref{fig:hist_dyn}.
}
\label{fig:VShist}
\end{figure}

The left panel of Fig.~\ref{fig:VShist} shows mass histograms of the ejecta with respect to the entropy per baryon at the time when the temperature decreases to \SI{5}{GK}.
For $s/k_\mathrm{B}\lesssim 10$, the ejecta are dominated by the dynamical ejecta for which the lowest-end of the entropy per baryon is lower for the merger of more asymmetric binaries, reflecting more contribution from the tidally ejected matter.
On the other hand, the post-merger ejecta contribute to the portion for $s/k_\mathrm{B}\gtrsim 10$ and has a similar distribution among the investigated models with a peak at $s/k_\mathrm{B}\approx 15$, reflecting the similar properties of the post-merger system, such as the mass of the central black hole and the disk (for dependence of the properties of the ejecta on them, see, e.g., \citealt{Fujibayashi2020b} and \citealt{Fernandez2020a}).
The high-entropy side ($s/k_\mathrm{B}\gtrsim 20$) has a slope approximately described by $dM/d\log s\propto s^{-3}$ for both dynamical and post-merger components.

In the right panel of Fig.~\ref{fig:VShist}, the mass histogram with respect to the asymptotic velocity is shown.
Here, in contrast to that in Fig.~\ref{fig:hist_dyn}, we use a different definition of the asymptotic velocity for the post-merger component as $v_\infty/c := \sqrt{1-(-hu_t/h_\mathrm{min})^{-2}}$, which is consistent with the criterion for unbound matter in post-merger simulations (see the definition of the asymptotic Lorentz factor in Eq.~\eqref{eq:gamma_inf_main} and also Appendix~\ref{app:criteria}).
For $v_\infty/c\lesssim 0.1$, the ejecta are dominated by the post-merger ejecta with a peak at $v_\infty/c \approx 0.06$, which again does not depend on the mass ratio significantly because of the same reason described above.

\section{Resolution- and time-dependence of electron fraction and velocity distributions of dynamical ejecta}
\label{app:resolution}

\begin{figure}
\plottwo{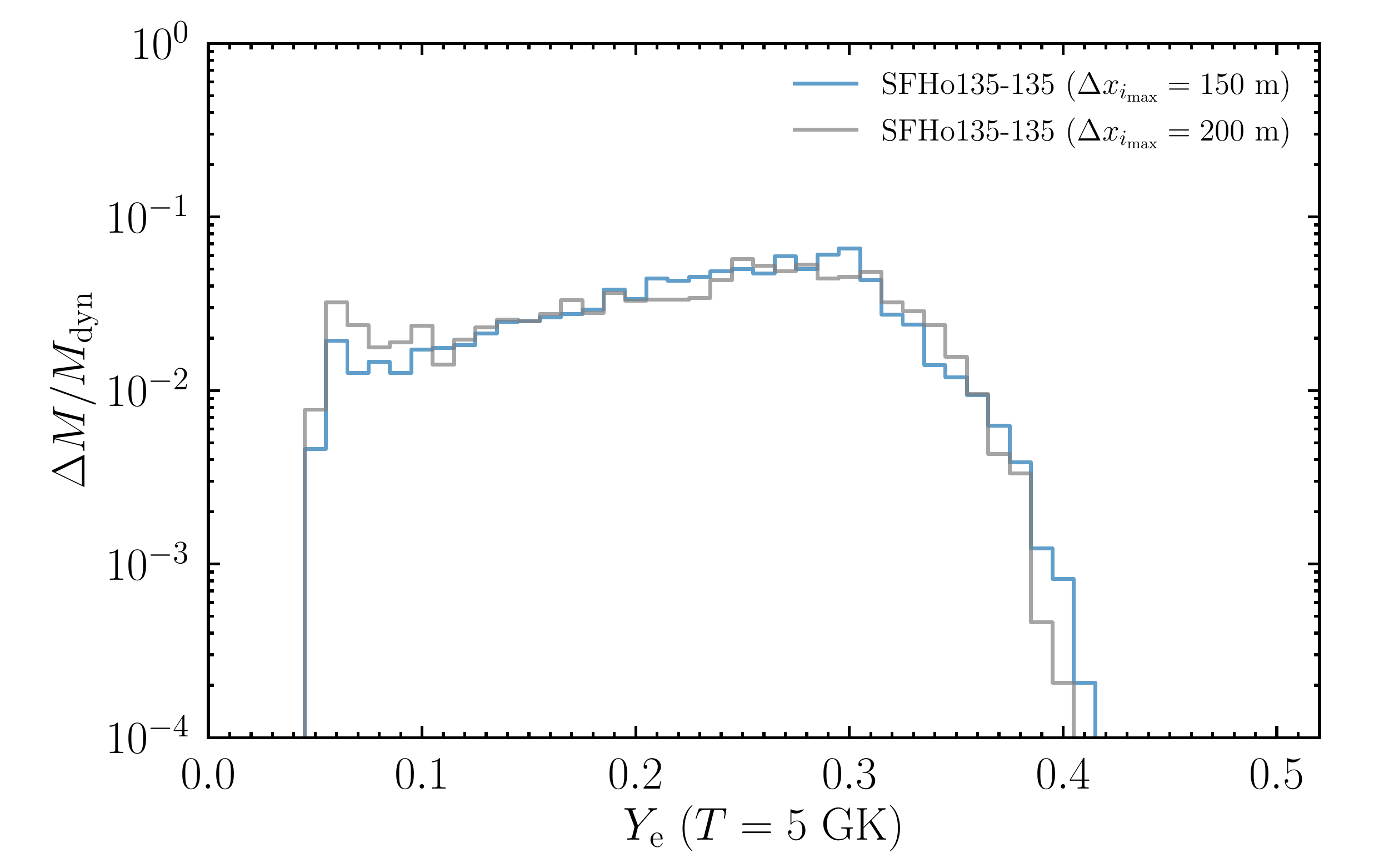}{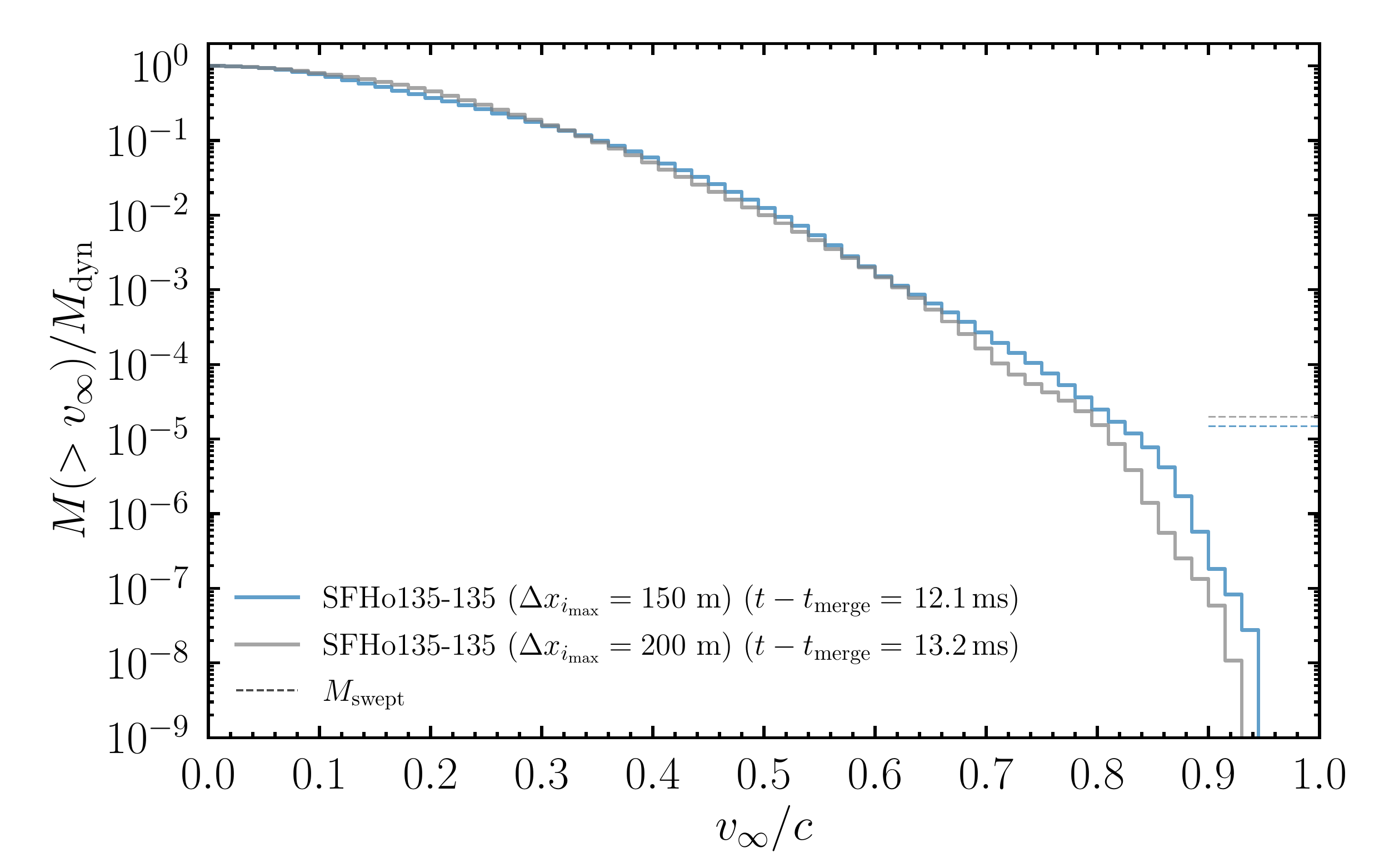}
\plottwo{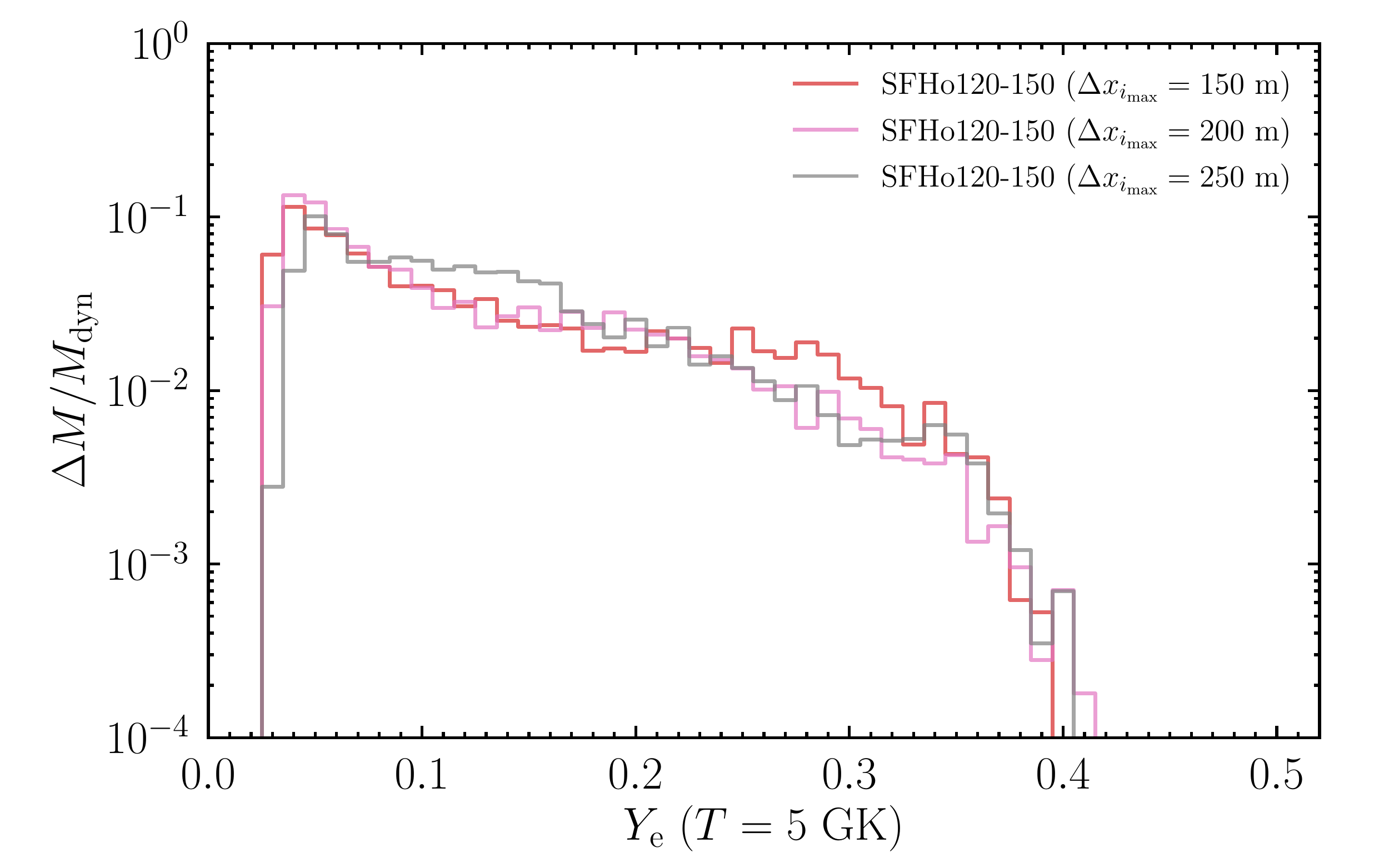}{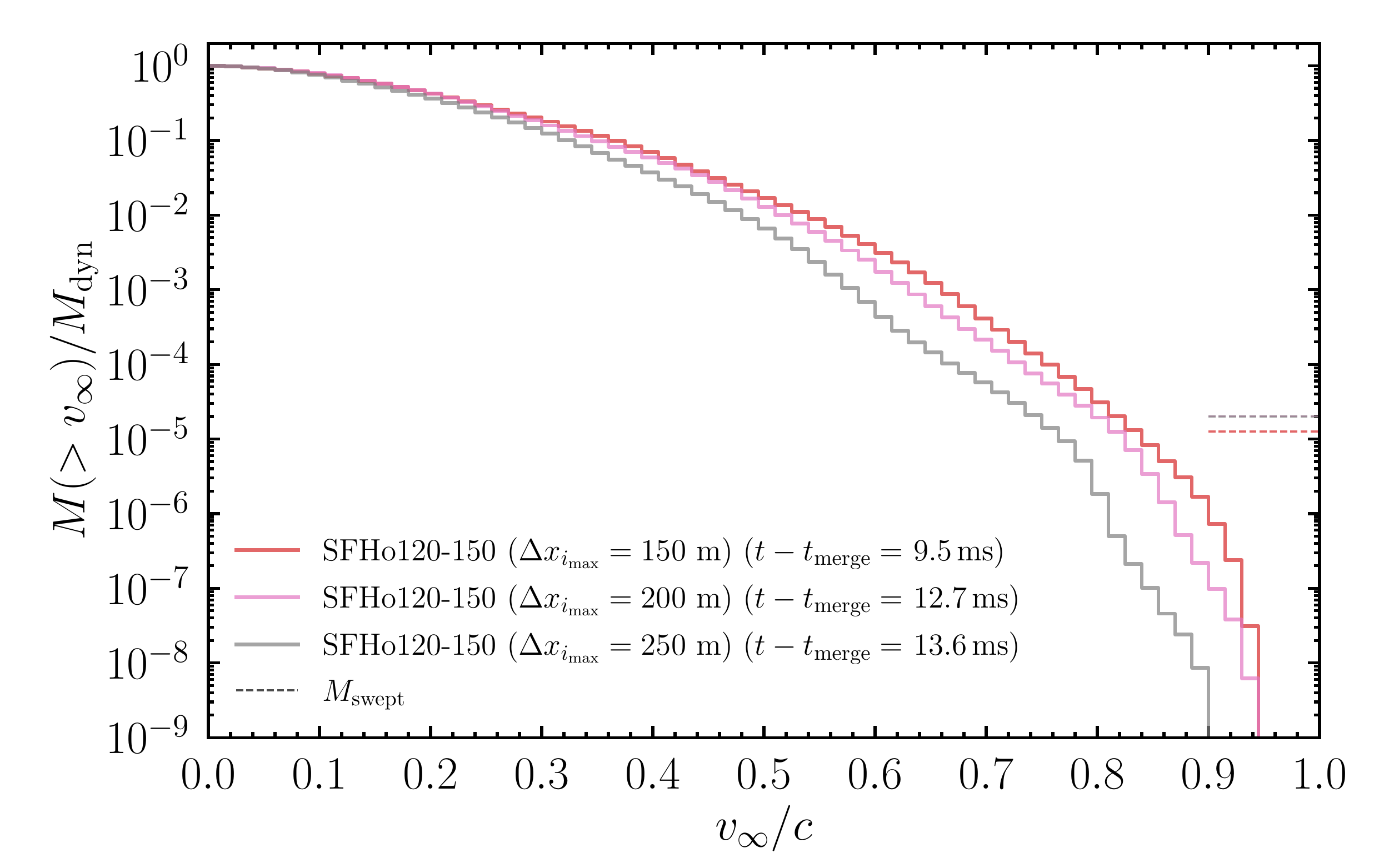}
\caption{
Comparison of mass histograms of the electron fraction (left) and cumulative mass distribution of the asymptotic velocity (right) for the dynamical ejecta.
Models SFHo135-135 (top) and SFHo120-150 (bottom) with different grid resolutions ($\Delta x_{i_\mathrm{max}}=150$ and \SI{200}{m} for 135-135, and $\Delta x_{i_\mathrm{max}}=150$, 200, and \SI{250}{m} for 120-150) are employed.
For the right panels, the distributions relative to the total mass of the dynamical ejecta are compared.
}
\label{fig:resolution}
\end{figure}

\begin{figure}
\plottwo{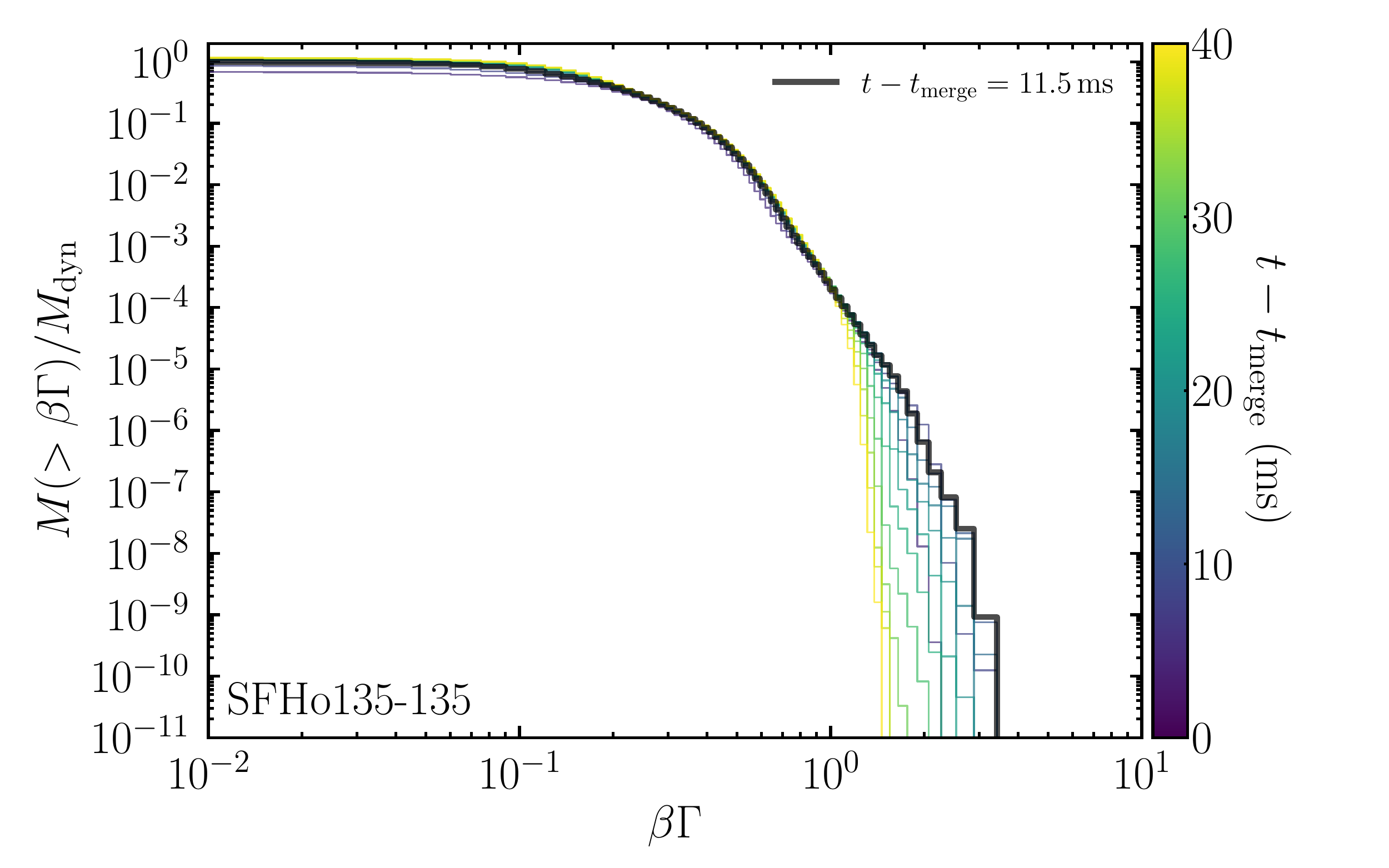}{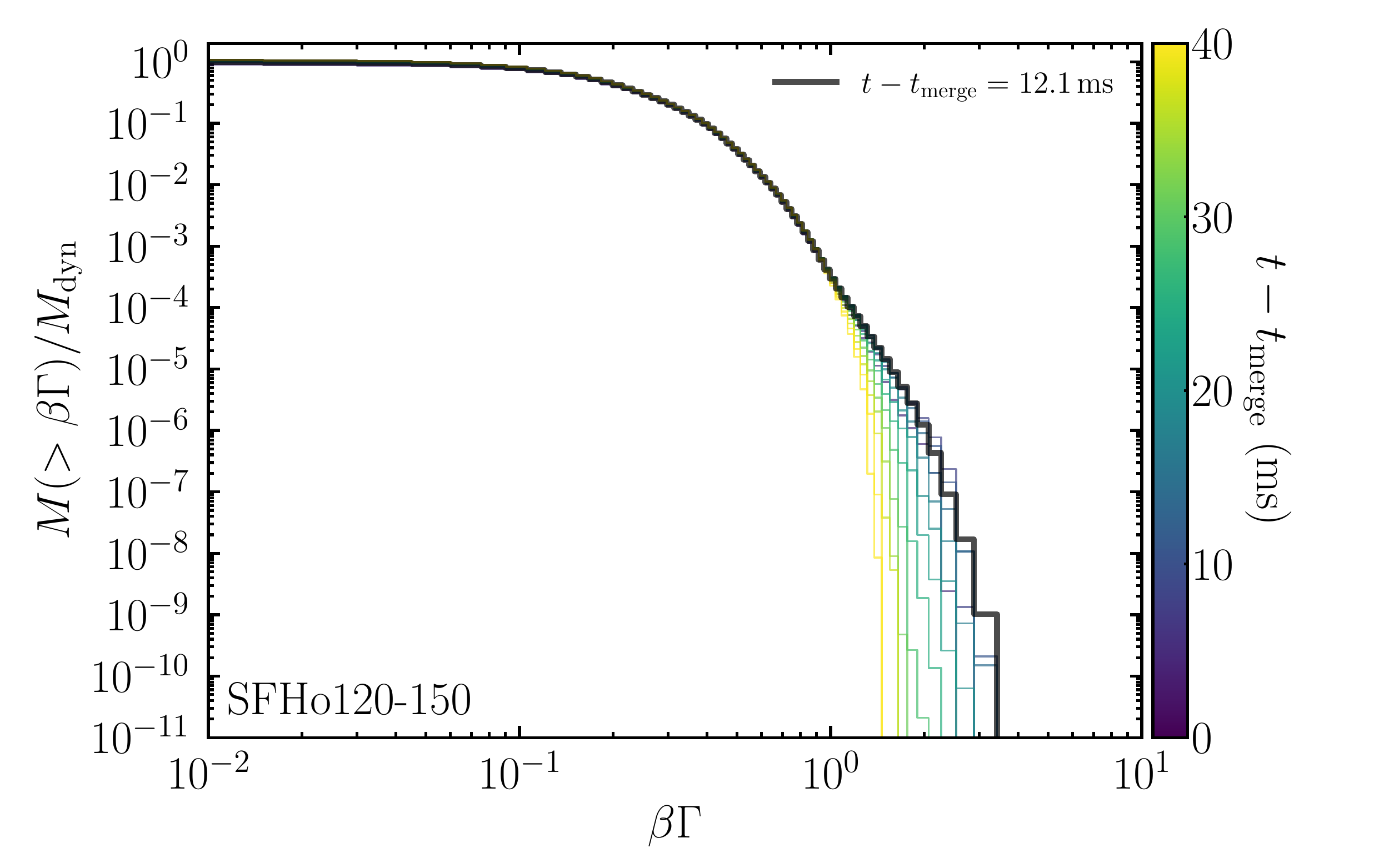}
\caption{
Cumulative mass distributions of the asymptotic specific momentum for different post-merger time are compared for models SFHo135135 (left) and 120150 (right).
The black lines correspond to the ones displayed in the bottom-left panel of Fig.~\ref{fig:hist_dyn} but normalized by the total mass of dynamical ejecta.
}
\label{fig:betagamma-time}
\end{figure}

In this section, we briefly show the dependence of the results for the dynamical ejecta in 3D simulations on the grid resolution.
\addsf{
Figure~\ref{fig:resolution} compares the mass distribution of the electron fraction and asymptotic velocity for models SFHo135-135 and SFHo120-150 with different grid resolutions ($\Delta x_{i_\mathrm{max}}=150$, \SI{200}{m} for 135-135 and $\Delta x_{i_\mathrm{max}}=150$, 200, and \SI{250}{m} for 120-150).
It is found that the $Y_\mathrm{e}$ distribution for the different resolutions agrees reasonably for both of the models.
For the lower resolutions ($\Delta x_{i_\mathrm{max}}=\SI{200}{m}$ for SFHo135-135 and $\Delta x_{i_\mathrm{max}}=200$ and \SI{250}{m} for SFHo120-150), the mass of the high-velocity ejecta are underestimated.
Specifically, the masses with $v_\infty/c\gtrsim0.7$, $\gtrsim0.5$, and $\gtrsim0.3$ are underestimated
for models SFHo135-135 with $\Delta x_{i_\mathrm{max}}=\SI{200}{m}$, SFHo120-150 with $\Delta x_{i_\mathrm{max}}=\SI{200}{m}$ and SFHo120-150 with $\Delta x_{i_\mathrm{max}}=\SI{250}{m}$, respectively. 
This reflects that the high-velocity components are severely suppressed by the numerical dissipation.

Figure~\ref{fig:betagamma-time} compares the cumulative mass distribution of the asymptotic specific momentum for selected post-merger time.
The ejecta mass with the highest velocity is the largest at $t-t_\mathrm{merge}=11.5$ and \SI{12.1}{ms} for models SFHo135-135 and 120-150, respectively (shown in black lines in each panel).
After these times, the matter with $\beta\Gamma\gtrsim 1.3$ ($v_\infty/c\gtrsim 0.8$) begins to be decelerated spuriously due to the interaction with the atmosphere.
Even at $t-t_\mathrm{merge}=11.5$ and \SI{12.1}{ms} for these models, the mass of the atmosphere swept by the highest-velocity ejecta amounts to $\sim 10^{-8} M_\odot$ (see bottom-left panel of Fig.~\ref{fig:hist_dyn}), which is comparable to the mass for $\beta\Gamma\gtrsim 2$ ($v_\infty/c\gtrsim0.9$).
This shows that the distribution of the highest-velocity end suffers from the interaction with the atmosphere and the high-velocity ejecta mass is always underestimated. 
}

\section{Processes that determine the electron fraction of the ejecta}
\label{app:process-ye}
In this section, we describe how the typical value of the electron fraction of the post-merger ejecta is determined. Specifically, we will describe how the electron fraction of the disk matter evolves toward the onset of the post-merger mass ejection. 

The evolution equation of the electron fraction in the fluid rest frame is written as
\begin{align}
\dot{Y}_\mathrm{e} = -(\lambda_\mathrm{ec}+\lambda_\mathrm{\bar{\nu}_\mathrm{e}}) X_\mathrm{p} + (\lambda_\mathrm{pc}+\lambda_\mathrm{\nu_\mathrm{e}}) X_\mathrm{n}, 
\end{align}
where $\lambda_\mathrm{ec}$ and $\lambda_\mathrm{pc}$ are the capture rates of electrons and positrons by protons and neutrons, respectively, and $\lambda_\mathrm{\nu_\mathrm{e}}$ and  $\lambda_\mathrm{\bar{\nu}_\mathrm{e}}$ are the absorption rates of electron neutrinos and antineutrinos by neutrons and protons, respectively. We note that weak interaction processes of heavy nuclei are negligible in the context of this study, and thus, ignore them in this analysis.

The reaction rates of the electron/positron capture ($\lambda_\mathrm{ec}$ and $\lambda_\mathrm{pc}$) are determined by the temperature and the chemical potential of electrons, while those of the electron (anti)neutrino absorption ($\lambda_\mathrm{\nu_\mathrm{e}}$ and  $\lambda_\mathrm{\bar{\nu}_\mathrm{e}}$) are determined by the temperature, the chemical potential of electrons, and also the number flux and its energy distribution of neutrinos \citep[see, e.g.,][in the similar context]{Just2022}.
In our method of leakage scheme-based energy-integrated neutrino transfer, the reactions are approximately calculated.
The detailed method of calculating their reaction rates is described in \cite{Sekiguchi2011} and \cite{fujibayashi2017a}.

\begin{figure}
\begin{minipage}[t]{0.48\textwidth}
\includegraphics[width=\textwidth]{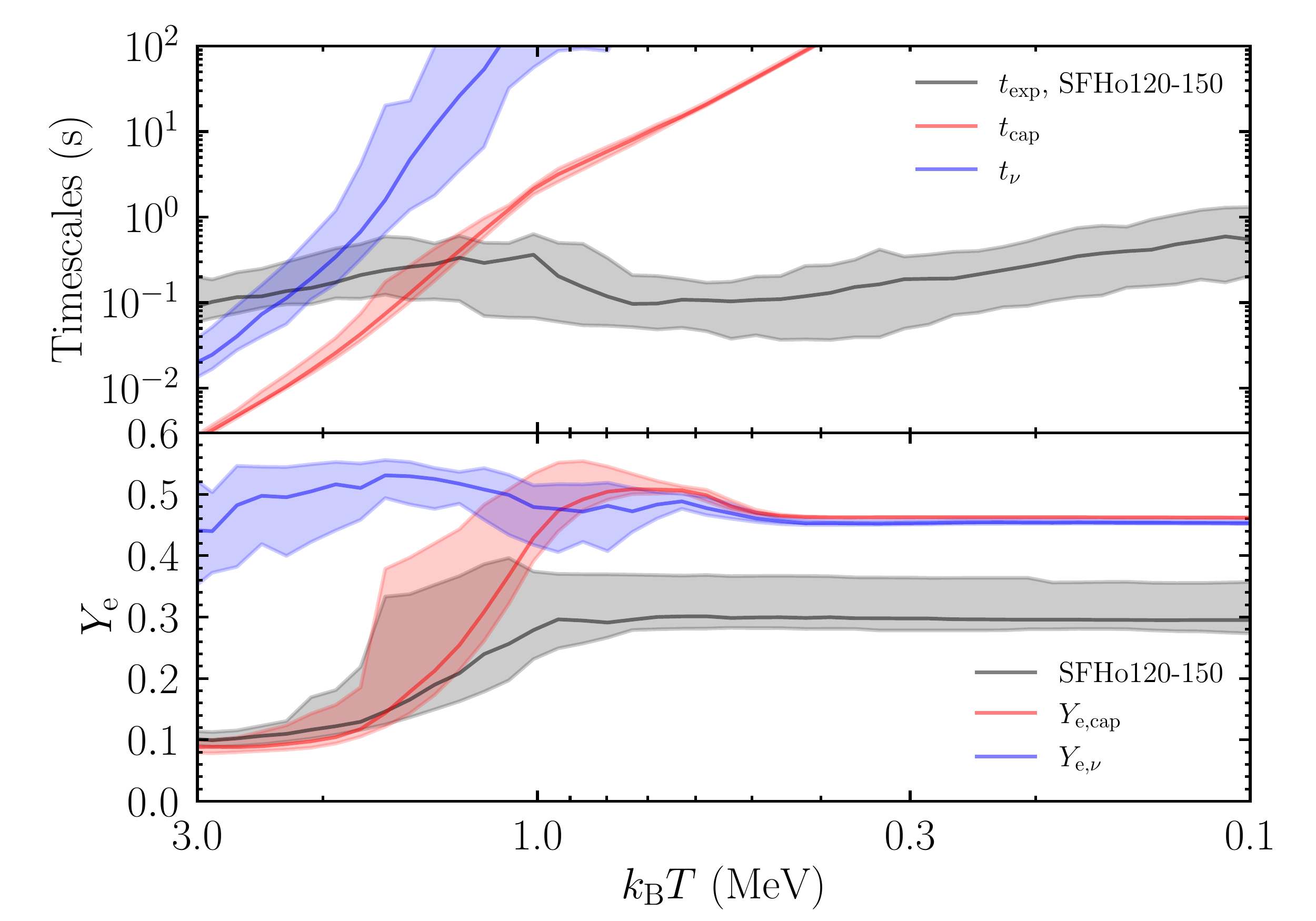}
\caption{
Top: The expansion timescale of matter (black), electron/positron capture timescale (red), and neutrino absorption timescale (blue) as functions of decreasing temperature.
Bottom: Evolution of the electron fraction of matter (black), its equilibrium value of the electron/positron capture (red), and its equilibrium value of the neutrino absorption (blue). In both panels, the solid curves denote the mass-weighted medians of the quantities, and the shaded regions contain 80\% of the mass of tracer particles around the medians.
}
\label{fig:ye-timescale}
\end{minipage}
\begin{minipage}[t]{0.48\textwidth}
\includegraphics[width=\textwidth]{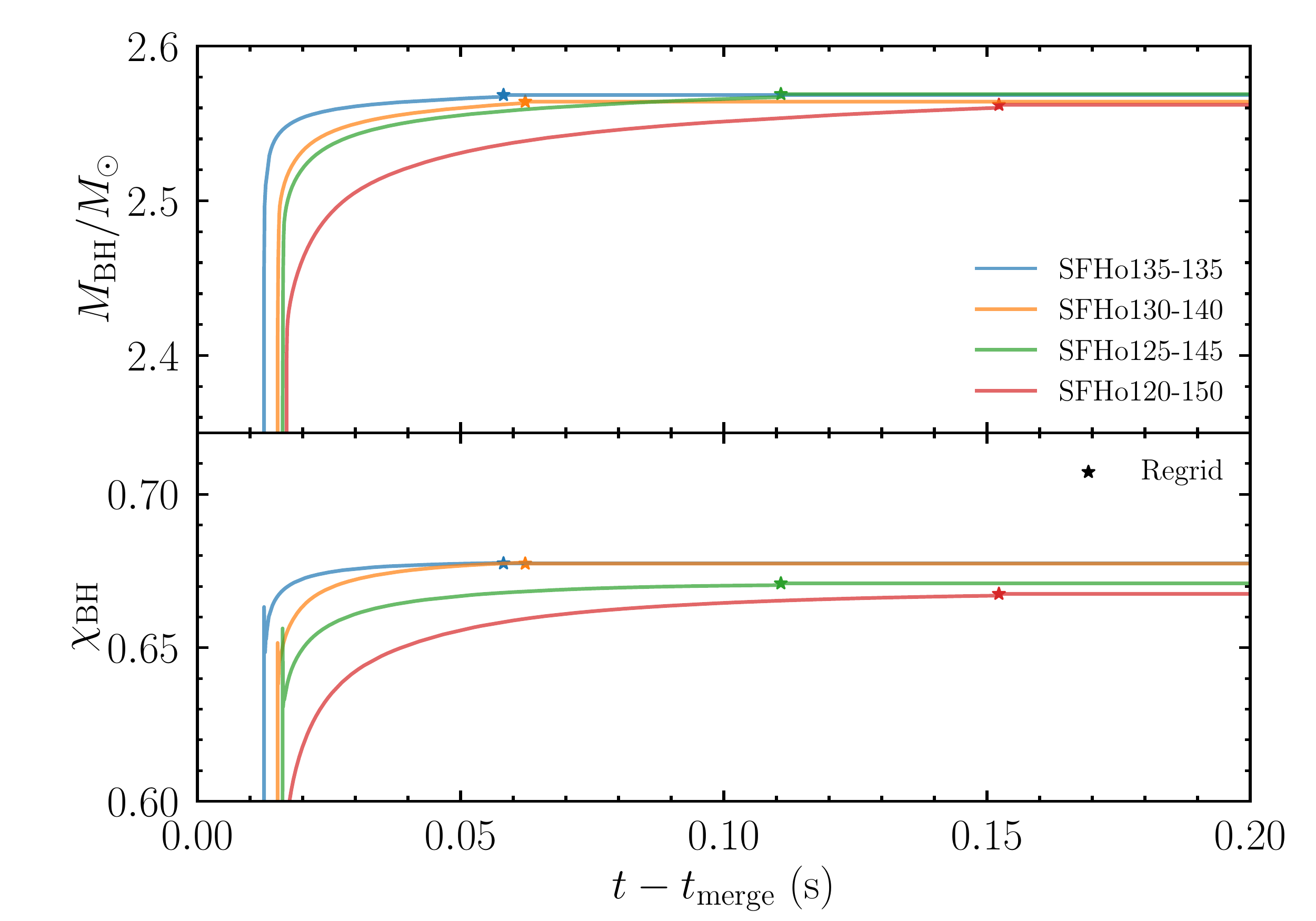}
\caption{
Time evolution of the mass and dimensionless spin of the remnant black holes.
The stars mark the time at which the time evolution of the metric is stopped and regridding is performed.
}
\label{fig:bhmass-spin}
\end{minipage}
\end{figure}

The top panel of Fig.~\ref{fig:ye-timescale} shows the timescales of the expansion of the matter, $t_\mathrm{exp}:=r/v^r$ (black), electron/positron capture, $t_\mathrm{cap}:= \mathrm{min}(1/\lambda_\mathrm{ec}, 1/\lambda_\mathrm{pc})$ (red), and absorption of neutrinos, $t_\nu = \mathrm{min}(1/\lambda_\mathrm{\nu_\mathrm{e}}, 1/\lambda_\mathrm{\bar{\nu}_\mathrm{e}})$ (blue) as functions of the temperature.
We calculate these reaction rates along each tracer particle of the ejecta.
The curves denote the mass-weighted median of these timescales for the tracer particles and the shaded regions denotes the region in which 80\% of the mass of particles are contained around the medians.

The neutrino absorption timescale is always longer than that of the electron/positron capture by at least a factor of three, indicating that the neutrino absorption does not play a significant role for determining the electron fraction.
For an early phase in which $k_\mathrm{B}T\gtrsim \SI{1.5}{MeV}$, the electron/positron capture timescale is shorter than the expansion timescale, and thus, the electron fraction is determined predominantly by these capture reactions.

The bottom panel of Fig.~\ref{fig:ye-timescale} shows the electron fraction of the matter (black), its equilibrium values of the electron and positron capture reactions,  $Y_\mathrm{e,cap}$ (red), and absorption reactions of electron neutrinos and antineutrinos, $Y_\mathrm{e,\nu}$ (blue).
The electron fraction of the matter agrees well with $Y_\mathrm{e,cap}$ for $k_\mathrm{B}T\approx \SI{3}{MeV}$ because the electron/positron capture dominates over the neutrino absorption, and its timescale is much shorter than the expansion timescale.
Because of the expansion of the matter, the degeneracy of electrons becomes weaker and the value of $Y_\mathrm{e,cap}$ becomes higher as the temperature decreases due to the expansion.
At the same time, the timescale of the electron/positron capture becomes longer as the temperature decreases.
At $k_\mathrm{B}T\approx \SI{1.5}{MeV}$, its timescale becomes comparable to that of the expansion, and for the lower temperature, the weak interaction reactions become inefficient, leading to the freeze-out of the electron fraction.
Since the post-merger ejecta have the similar expansion timescale (viscous expansion timescale), its resulting electron fraction is distributed around 0.2--$0.3$.
We note that the electron fraction can be lower (higher) if the expansion timescale of the ejecta is shorter (longer) as shown in our previous study \citep{Fujibayashi2020c}, although its typical value is not low enough for the strong $r$-process \citep[$\lesssim 0.23$, see][]{korobkin2012nov} for reasonable values of the viscosity \citep[see also][]{Fernandez2019a,Just2022}.

\section{Properties of formed black holes}

For the models employed in this paper, a black hole is always eventually formed after the merger. We here summarize the properties of the black holes formed. 

Figure~\ref{fig:bhmass-spin} shows the time evolution of the mass and dimensionless spin 
of the black holes. Here, the mass of the black hole is estimated by the equatorial circumference length of the black hole, $C_\mathrm{e}$, by (see, e.g., \cite{Shibata2016a})
\begin{align}
M_\mathrm{BH} &= \frac{C_\mathrm{e}c^2}{4\pi G}
\end{align}
and its dimensionless spin $\chi_\mathrm{BH}$ is estimated using the ratio of polar and equatorial circumferences by solving
\begin{align}
\frac{C_\mathrm{p}}{C_\mathrm{e}} &= \frac{\sqrt{2\hat{r}_+}}{\pi} \int_0^{\pi/2} \sqrt{1-\frac{\chi_\mathrm{BH}^2}{2\hat{r}_+}\sin^2\theta}d\theta
\end{align}
where $C_\mathrm{p}$ is the polar circumference length and $\hat{r}_+ = 1+\sqrt{1-\chi_\mathrm{BH}^2}$.
The mass of the black hole is shown in the top panel of Fig.~\ref{fig:bhmass-spin}.
This shows that at its formation the black-hole mass is smaller for the merger of more asymmetric binaries because of the presence of more massive disk around the black hole.
This also results in a smaller spin as shown in the bottom panel of Fig.~\ref{fig:bhmass-spin}.

It is found that the mass and dimensionless spin approach certain asymptotic values for each model, because the mass accretion onto the black hole eventually ceases. This result  indicates that the time evolution of the black hole is solved with a sufficiently high resolution in the present simulations, because
otherwise, the mass and dimensionless spin increases and decreases in time, respectively, due to the accumulation of the numerical error~\citep{Fujibayashi2020a}.

After the mass outside the apparent horizon becomes less than 1\% of the black-hole mass, we stop the evolution of the spacetime geometry and solve radiation viscous-hydrodynamics equations on a fixed background. Before we enter this computation phase, we perform a regridding to save the computation time. The time of the regridding for each model is marked by a star in Fig.~\ref{fig:bhmass-spin} and afterward the evolution of the geometrical variables is switched off. 
It shows that the regridding is performed when the mass and dimensionless spin of the black hole are approximately constant in time, indicating that the geometrical variables are also approximately constant in time and switching off their evolution does not quantitatively influence the hydrodynamics after the regridding.

\bibliography{reference}
\end{document}